\documentclass[11pt,a4paper]{article}

\pdfoutput=1
\usepackage{jheppub}

\usepackage[english]{babel}
\usepackage{amsfonts,amsmath,amssymb,amsthm,mathtools}
\usepackage{comment,enumerate,footnote,graphicx,subfloat}
\usepackage{array,tabularx,tabu,multirow,framed}
\usepackage[usenames,dvipsnames]{xcolor}
\numberwithin{equation}{section}
\usepackage{cleveref}
\crefformat{pluralequation}{#2\black{eqs.~(}#1\black{)}#3}
\Crefformat{pluralequation}{#2\black{Equations~(}#1\black{)}#3}
\crefformat{pluralfigure}{#2\black{figs.~}#1#3}
\Crefformat{pluralfigure}{#2\black{Figures~}#1#3}

\usepackage{tikz}
\usetikzlibrary{arrows,shapes}
\usetikzlibrary{trees,patterns}
\usetikzlibrary{matrix,arrows} 				
\usetikzlibrary{positioning}				  
\usetikzlibrary{calc,through}				  
\usetikzlibrary{decorations.pathreplacing}  
\usepackage{pgffor}							

\usetikzlibrary{decorations.pathmorphing}	
\usetikzlibrary{decorations.markings}
\tikzset{
	>=stealth', 
    vector/.style={decorate, decoration={snake}, draw},
	provector/.style={decorate, decoration={snake,amplitude=2.5pt}, draw},
	antivector/.style={decorate, decoration={snake,amplitude=-2.5pt}, draw},
	bigvector/.style={decorate, decoration={snake,amplitude=4pt}, draw},
    fermion/.style={draw=black, postaction={decorate},
        decoration={markings,mark=at position .55 with {\arrow[draw=black]{>}}}},
    fermionbar/.style={draw=black, postaction={decorate},
        decoration={markings,mark=at position .55 with {\arrow[draw=black]{<}}}},
    fermionnoarrow/.style={draw=black},
    gluon/.style={decorate, draw=black,
        decoration={coil,amplitude=4pt, segment length=5pt}},
    scalar/.style={dashed,draw=black, postaction={decorate},
        decoration={markings,mark=at position .55 with {\arrow[draw=black]{>}}}},
    scalarbar/.style={dashed,draw=black, postaction={decorate},
        decoration={markings,mark=at position .55 with {\arrow[draw=black]{<}}}},
    scalarnoarrow/.style={dashed,draw=black},
    momentum/.style={draw=black, postaction={decorate},
        decoration={markings,mark=at position 1 with {\arrow[draw=black]{>}}}},
    antimomentum/.style={draw=black, postaction={decorate},
        decoration={markings,mark=at position 0.1 with {\arrow[draw=black]{<}}}}
}

\tikzstyle{block} = [draw, rectangle, 
    minimum height=3em, minimum width=6em]

\newcommand{\nc}{\newcommand}

\nc{\pd}{\partial}
\nc{\bea}{\begin{eqnarray}}
\nc{\eea}{\end{eqnarray}}
\nc{\bal}{\begin{alignedat}}
\nc{\eal}{\end{alignedat}}
\nc{\beq}{\begin{equation}}
\nc{\eeq}{\end{equation}}
\nc{\bit}{\begin{itemize}}
\nc{\eit}{\end{itemize}}
\nc{\benu}{\begin{enumerate}}
\nc{\eenu}{\end{enumerate}}
\nc{\bdes}{\begin{description}}
\nc{\edes}{\end{description}}
\nc{\bma}{\begin{pmatrix}}
\nc{\ema}{\end{pmatrix}}

\newcommand{\black}[1]	{{\color{black} 	#1}}

\nc{\nn}{\nonumber}
\nc{\hc}{\text{h.c.}}
\nc{\cc}{\text{c.c.}}

\nc{\slashed}[1]{{#1}\hspace{-2mm}/}

\nc{\abs}[1]{\left| #1 \right|}
\def\[{\left[}
\def\]{\right]}
\def\({\left(}
\def\){\right)}
\def\<{\langle}
\def\>{\rangle}

\def\g5{\gamma_{5}}

\def\a{\alpha}

\def\g{\gamma}
\def\d{\delta}
\def\e{\epsilon}
\def\z{\zeta}
\def\h{\eta}

\def\k{\kappa}

\def\m{\mu}
\def\n{\nu}
\def\ks{\xi}

\def\p{\pi}
\def\r{\rho}
\def\s{\sigma}

\def\f{\phi}
\def\x{\chi}
\def\ps{\psi}

\def\vf{\varphi}

\def\G{\Gamma}

\def\W{\Omega}

\renewcommand{\vec}[1]{{\bf #1}}

\def\vrel	{v_{\rm rel}}
\def\ann		{_{\rm ann}}
\def\annC	{_{\rm ann, C}}

\def\lI	{{\ell_{\rm I}^{}}}
\def\mI	{{m_{\rm I}^{}}}
\def\lR	{{\ell_{\rm R}^{}}}
\def\mR	{{m_{\rm R}^{}}}

\def\BSF		{_{_{\rm BSF}}}
\def\BSFgr	{_{_{\rm BSF}}^{\{100\}} }
\def\BSFC	{_{_{\rm BSF, C}}}
\def\BSFCgr	{_{_{\rm BSF,C}}^{\{100\}} }

\def\M	{{\cal M}}

\title{Radiative bound-state-formation cross-sections \\ 
for dark matter interacting via a Yukawa potential}

\author[1,2]{Kalliopi Petraki,}
\author[2]{Marieke Postma}
\author[2]{and Jordy de Vries}
\affiliation[1]{LPTHE, CNRS, UMR 7589, 4 Place Jussieu, F-75252, Paris, France}
\affiliation[2]{Nikhef, Science Park 105, 1098 XG Amsterdam, The Netherlands}

\emailAdd{kpetraki@lpthe.jussieu.fr}
\emailAdd{mpostma@nikhef.nl}
\emailAdd{jordy.de.vries@nikhef.nl}

\date{\today}

\abstract{We calculate the cross-sections for the radiative formation of bound states by dark matter whose interactions are described in the non-relativistic regime by a Yukawa potential. These cross-sections are important for cosmological and phenomenological studies of dark matter with long-range interactions, residing in a hidden sector, as well as for TeV-scale WIMP dark matter. We provide the leading-order contributions to the cross-sections for the dominant capture processes occurring via emission of a vector or a scalar boson. We offer a detailed inspection of their features, including their velocity dependence within and outside the Coulomb regime, and their resonance structure. For pairs of annihilating particles, we compare bound-state formation with annihilation.
}

\begin{document}
\maketitle

\section{Introduction}

In a variety of theories, motivated on theoretical and phenomenological grounds, dark matter (DM) couples directly to light or massless force mediators, which give rise to long-range interactions. A notable example is the self-interacting DM scenario, which can currently explain the observed galactic structure better than collisionless DM. Importantly, even the Weak interactions of the Standard Model --- which have long served as the prototype of short-range interactions, and as the canonical particle-physics framework for DM --- exhibit a long-range behaviour if the interacting particles are heavier than a few TeV.

An important implication of long-range interactions is the existence of bound states. If bound states exist in the spectrum of the theory, then they may form efficiently in the early universe, and in the dense and non-relativistic environment of haloes today. This, in turn, may have dramatic consequences for the phenomenology and the detection signatures of DM. It is thus essential to accurately compute the formation of DM bound states, when long-range interactions are considered. In theories where the (effective) particle degrees of freedom are weakly coupled, the efficiency of bound-state formation (BSF) depends on the cross-sections of the relevant processes and on the thermodynamic environment. Here we are concerned with the former.

In Ref.~\cite{Petraki:2015hla}, we established a field-theoretic framework for the computation of radiative BSF cross-sections in weakly coupled theories. We expressed the amplitudes for such processes in terms of the wavefunctions of the (initial) scattering state and the (final) bound state, and an off-shell perturbative interaction involving the radiative vertex. We then reduced the fully relativistic expressions into their non-relativistic counterparts. Finally, we focused on particles interacting via a Coulomb potential, and calculated the cross-sections for BSF with the emission of a (nearly) massless vector or scalar force mediator.

In this paper, we extend these calculations to the case of a massive force mediator giving rise to a Yukawa potential,
\beq
V_Y({\bf r}) = -\frac{\a \, e^{-m_\vf r}}{r} \, ,
\label{eq:Yukawa}
\eeq
where $\a$ parametrises the interaction strength, and $m_\vf$ is the mediator mass. While no analytical expressions for the cross-sections of interest can be derived for $m_\vf >0$, our goal is to outline how to evaluate these cross-sections, and to highlight the features that are important for DM phenomenology.

The paper is organised as follows. In the next section, we offer some preliminaries that will be needed in the evaluation of the radiative BSF cross-sections. In \cref{Sec:VecMed,Sec:ScalMed}, we compute the BSF cross-sections with emission of a vector and a scalar boson, respectively. We consider particularly capture into zero and low angular momentum bound states, and inspect in detail the velocity dependence and the resonance structure of the corresponding cross-sections. For particle-antiparticle pairs, we compare BSF with direct annihilation into radiation. We conclude in \cref{Sec:Conc} with a discussion of the implications of our findings. 

Many of the formulae used in \cref{Sec:VecMed,Sec:ScalMed} are derived in the appendices. 
In \cref{App:WaveFun}, we discuss the scattering-state and bound-state wavefunctions in a Yukawa potential. In \cref{App:ConvIntegr}, we consider the integrals that convolve these wavefunctions with the radiative vertex, and which enter in the computation of the radiative BSF cross-sections. We show how to perform an expansion in powers of the coupling, for capture into bound states of any angular momentum. For capture into zero and low angular momentum bound states, we identify the leading order contributions, which we then use in our computations of \cref{Sec:VecMed,Sec:ScalMed}.
In \cref{App:Coulomb ell=0}, we derive analytical expressions in the Coulomb limit, of the convolution integrals that enter the cross-sections for capture into zero angular momentum bound states.
Throughout this work we have, in fact, applied two different methods to calculate BSF cross sections. While the main text and \cref{App:WaveFun,App:ConvIntegr,App:Coulomb ell=0} focus on a coordinate-space method, we have reproduced the calculations using a momentum-space procedure that is based on methods developed by the nuclear-physics community, for few-body problems. We outline this method in \cref{App:BSF_MomemtumSpace}.

For easy reference, we list our results and several useful formulae in \cref{tab:results}, with references to the corresponding equations and figures. In \cref{tab:notation}, we summarise the notation used throughout the paper and for succinctness, we often do not define these symbols in the text.

The radiative formation of bound states by particle-antiparticle pairs interacting via a Yukawa potential has been recently considered in Refs.~\cite{An:2016gad,An:2016kie}, where a quantum mechanical formalism has been employed from the onset. Here, we adopt the field-theoretic formalism outlined in Ref.~\cite{Petraki:2015hla}, which has a direct representation in terms of Feynman diagrams. Moreover, this formalism allows for a systematic inclusion of higher order corrections, which can be particularly important when the lowest order contributions to a specific process cancel. This, in fact, occurs in the radiative capture of a pair of identical particles or a particle-antiparticle pair, via emission of a scalar force mediator. The cancellation of the lowest-order terms implies, among else, that the BSF cross-sections may be different for bosonic and fermionic pairs of particles (cf.~\cref{sec:Convol Integr}). In \cref{sec:ScalarMed_Degen}, we calculate the BSF cross-sections for bosonic particle-antiparticle pairs, via emission of a scalar current. This computation is complementary to Ref.~\cite{An:2016kie}, which has considered fermionic DM; indeed, our results show that the two cases are different.

Besides the different formalism, the present work has broader applicability, in the following ways: (i) Our formulae are valid in the entire parameter range where bound states exist, rather than in the more limited parameter space where bound states are kinematically allowed to form with emission of the same force mediator that is responsible for their existence. Our results can therefore be easily adapted to describe BSF with emission a (lighter) vector or scalar boson that is not (primarily) responsible for the Yukawa interaction~\eqref{eq:Yukawa} (cf.~\cref{sec:xi explanations}).
This is, in fact, relevant to DM coupled to the Weak interactions of the Standard Model (WIMPs), as has been recently discussed in Ref.~\cite{Asadi:2016ybp}. 
(ii) We consider BSF by pairs of particles that do not necessarily belong to the same species. For capture via vector current emission, the BSF cross-sections computed for particle-antiparticle pairs can be easily adapted to describe the capture of particles belonging to different species, by an appropriate use of the reduced mass of the interacting pair.
However, for capture via scalar current emission, the results are markedly different depending on whether the two interacting particles have the same or different masses and couplings to the scalar current. 
Lastly, we note that in our computations, we employ a minimal parametrisation that we believe facilitates phenomenological studies (cf.~\cref{sec:param}); using this parametrisation, we perform a rather thorough inspection of the features of the BSF cross-sections.

\clearpage
\begin{table}[p]
\newcommand{\DivOne}		{{\raisebox{.3ex}{{~\tiny $\blacksquare$~~}}}}
\newcommand{\DivTwo}		{{\raisebox{.3ex}{{\hspace{16pt}$\bullet$~}}}}
\newcommand{\DivThree}	{{\raisebox{.3ex}{{\hspace{26pt}$-$~}}}}

\renewcommand{\arraystretch}{1.45}
\hspace{-0.7cm}
\begin{tabular}{|l|c|c|} 
\hline
\parbox[c]{11cm}{\medskip \bf \boldmath
Cross-sections for the radiative formation of bound states \\
$\{n\ell m\}$, by particles interacting via a Yukawa potential} 
& \parbox[c]{3cm}{\centering \bf Equations} 
& {\bf Figs}
\\[3mm] \hline \hline 


{\bf Vector force mediator}
& 
& 
\\
\DivOne
Capture into $\ell=0$ bound states
& \eqref{eqs:VecMed_BSF_n00}, \eqref{eqs:VecMed_BSF_n00_Coul}
& \ref{fig:VecMed_Coulomb} -- \ref{figs:VecMed_Sregularized}
\\[1pt]
\DivTwo
Coulomb limit and comparison with annihilation
& {\centering \eqref{eqs:VecMed_BSF_n00_Coul}, \eqref{eqs:VecMed_Ann}, \eqref{eq:VecMed_BSFoverANN}}
& \ref{fig:VecMed_Coulomb}
\\[-1pt]
\DivTwo
\parbox[t]{9cm}{Capture into the ground state $\{100\}$, and comparison \\ 
with annihilation}
& {\centering \eqref{eqs:VecMed_BSF_n00}, \eqref{eq:VecMed_S100_Coul}, \eqref{eqs:VecMed_Ann}}
& \ref{figs:VecMed_n=1}
\\[-2pt]
\DivThree
Resonant structure
& 
& \ref{fig:VecMed_n=1_Xi}
\\[-5pt]
\DivThree
Velocity dependence off-resonance
& 
& \ref{fig:VecMed_n=1_Zeta_NonRes}
\\[-5pt]
\DivThree
Velocity dependence on-resonance
& 
& \ref{fig:VecMed_n=1_Zeta_Res}
\\[-5pt]
\DivThree
Velocity dependence near threshold
& 
& \ref{fig:VecMed_n=1_Zeta_Threshold}
\\[-5pt]
\DivTwo
Capture into the first excited state $n=2, \, \ell=0$
& \eqref{eqs:VecMed_BSF_n00}, \eqref{eqs:VecMed_BSF_n00_Coul}
& \ref{fig:VecMed_n=2_l=0}
\\[-5pt]
\DivTwo
\parbox[t]{9cm}{Comparison: the resonant structure of bound-state \\ 
formation and $p$-wave annihilation}
& 
& \ref{figs:VecMed_Sregularized}
\\[-2pt] 
\DivOne
Capture into $\ell=1$ bound states
& \eqref{eqs:VecMed_BSF_n1}
& \ref{fig:VecMed_Coulomb}, \ref{fig:VecMed_n=2_l=1}
\\ \hline 


{\bf Scalar force mediator: Non-degenerate particles} 
&
& 
\\
\DivOne
Capture into $\ell=0$ states (including Coulomb limit)
& \eqref{eqs:ScalMed_NonDegen_BSF}, \eqref{eqs:ScalMed_NonDegen_BSF_Coul} 
& \ref{fig:ScalarMed_NonDegen_100}
\\[-5pt]
\DivOne
Capture into $\ell=1$ states 
& \eqref{eqs:ScalMed_NonDegen_BSF_l=1}
& \ref{fig:ScalarMed_NonDeg_n=2_l=1}
\\ \hline 


\parbox[t]{11cm}{\bf Scalar force mediator: \\ 
Non-self-conjugate bosonic particle-antiparticle pairs} 
& 
& 
\\[3pt]
\DivOne
Capture into $\ell=0$ states
& \eqref{eqs:ScalMed_Degen_BSF}, \eqref{eqs:ScalMed_Degen_BSF_Coul}
& \ref{fig:ScalarMed_Degen_Coulomb} -- \ref{figs:ScalarMed_Degen}
\\[-4pt]
\DivThree
Coulomb limit and comparison with annihilation
& \eqref{eqs:ScalMed_Degen_BSF_Coul}, \eqref{eq:ScalMed_BSFoverANN}
& \ref{fig:ScalarMed_Degen_Coulomb}
\\[-5pt]
\DivThree
Resonant structure
& 
& \ref{fig:ScalarMed_Degen_Xi}
\\[-5pt]
\DivThree
Velocity dependence off- and on-resonance
& 
& \ref{fig:ScalarMed_Degen_Zeta}
\\[-5pt]
\DivOne
Capture into $\ell=1$ states
& \eqref{eqs:ScalMed_Degen_BSF_l=1}
& \ref{fig:ScalarMed_Degen_Xi_l=1}
\\ \hline

{\bf Scalar force mediator: Identical particles} 
& \eqref{eq:IdenticalBosons} --
\eqref{eq:ScalMed_BSFoverANN}
& \ref{fig:ScalarMed_Degen_Coulomb} -- \ref{fig:ScalarMed_Degen_Xi_l=1}
\\ \hline 
\end{tabular}

\bigskip

\renewcommand{\arraystretch}{1.2}
\begin{tabular}{|l|l|} 
\hline
{\bf Convolution integrals ${\cal I}_{\vec k, n\ell m} (\vec b)$, 
{\boldmath${\cal J}$}${}_{\vec k, n\ell m} (\vec b)$, 
${\cal K}_{\vec k, n\ell m} (\vec b)$} 
&   
{\bf Equations}
\\ 
\hline \hline
Definition 
& \eqref{eqs:ConvIntegrals_Def}
\\
Expansion in powers of radiated momentum $|\vec b|$: Validity of approximation
&  \eqref{eq:expansion condition}
\\
Capture into $\{n\ell m\}$ bound states, expansion in $|\vec b|$
& \eqref{eqs:IJK_nlm_expansion}
\\
Capture into $\ell=0$ bound states 
& 
\\
~--~
Expansion in $|\vec b|$, leading-order terms 
& \eqref{eqs:IJK n00}
\\
~--~
Coulomb limit (no expansion)
&  \eqref{eqs:Conv Integrals n00 Coul}
\\
~--~
Coulomb limit: expansion in $|\vec b|$, leading-order terms 
&  \eqref{eqs:ConvIntegr_n00_Coul_expansion}
\\ 
Capture into $\ell=1$ bound states: expansion in $|\vec b|$, leading-order terms  
& \eqref{eqs:Jn1m}
\\ \hline 
\end{tabular}

\caption{References to the main results and formulae.}

\label{tab:results}
\end{table}

\clearpage
\begin{table}[p]

\centering
\renewcommand{\arraystretch}{1.7}

\begin{tabular}{|c|c|} 
\hline
{\bf Particles and masses} &   {\bf Symbols}
\\ \hline  \hline 
Interacting particles & $X_1$, $X_2$
\\ \hline
Masses of interacting particles & $m_1$, $m_2$
\\ \hline
Total mass of interacting particles & $M \equiv m_1 + m_2$
\\ \hline
Reduced mass of interacting particles & $\m \equiv \dfrac{m_1 m_2}{m_1+m_2}$
\\[1.5mm] \hline
Force mediator (scalar or vector) & $\vf$
\\ \hline
Mass of force mediator & $m_\vf$
\\ \hline
\end{tabular}

\medskip

\renewcommand{\arraystretch}{2}
\begin{tabular}{|l|l|} 
\hline
{\bf Description} &   {\bf Symbol}
\\ \hline  \hline 
Dark fine structure constant  & $\a$
\\ \hline 
\parbox[c]{7cm}{\medskip
Expectation value of relative velocity of \\
interacting particles in the scattering state} 
& \parbox[l]{2cm}{$\vec{v}_{\rm rel}$, \\ $\vrel = |\vec{v}_{\rm rel}|$}
\\[2mm] \hline
Bohr momentum & $\k  \equiv \m\a$
\\ \hline 
\parbox[c]{7cm}
{\smallskip Momentum of reduced system of \\ 
interacting particles in the scattering state} 
& 
\parbox[l]{2cm}{$\vec{k}  \equiv \m \vec{v}_{\rm rel}$, \\ $k \equiv |\vec{k}|$}
\\[2mm] \hline
Wavefunction of $n\ell m$ bound state
&$\psi_{n\ell m} (\vec r) = \k^{3/2} \left[\dfrac{\x_{n\ell}^{}(\k r)}{\k r}\right] Y_{\ell m}(\W_{\vec r})$
\\[2.5mm] \hline 
Wavefunction of scattering state
&$\phi_{\vec k} (\vec r) = \displaystyle \sum_{\ell=0}^\infty (2\ell+1) 
\left[ \dfrac{\x_{|\vec k|,\ell}^{}(\k r)}{\k r} \right] P_{\ell}(\hat{\vec k}\cdot\hat{\vec r})$
\\[3mm] \hline
Binding energy of $n\ell m$ bound state
&${\cal E}_{n\ell} = -\dfrac{\kappa^2}{2\mu} \: \gamma_{n\ell}^2(\xi)$
\\[1mm] \hline
Kinetic energy of scattering state
&${\cal E}_{\vec k} = \dfrac{\vec k^2}{2\mu}$
\\[1mm] \hline
\parbox[c]{7.5cm}{\medskip 
Phase-space suppression due to emission of \\ 
a massive force mediator, during capture \\ 
into the $n\ell m$ bound states} 
& $pss_{n\ell}$, see \cref{eq:pss}
\\[5mm] \hline
\end{tabular}

\medskip

\renewcommand{\arraystretch}{1.5}
\begin{tabular}{|c|} 
\hline
{\bf Dimensionless parameters}
\\ \hline \hline
$\h_{1,2} \equiv \dfrac{m_{1,2}}{m_1+m_2}$
\\[1.5mm] \hline
$\z  \equiv \kappa / k = \a/\vrel$
\\ \hline
$\ks \equiv \k/m_\vf = \mu \a/m_\vf$
\\ \hline
\end{tabular}

\caption{Notation}


\label{tab:notation}
\end{table}

\clearpage
\section{Preliminaries \label{Sec:Prelim}}

We shall consider two particles $X_1$ and $X_2$ that may in general belong to different species, and interact via a vector or scalar force mediator $\vf$ (cf.~\cref{fig:OBE}). The interaction Lagrangians will be specified in the following sections. In the non-relativistic regime, the interaction between $X_1$ and $X_2$ is described by the static Yukawa potential of \cref{eq:Yukawa}, which admits bound state solutions roughly if the mediator mass is less than the inverse Bohr radius, $m_\vf \lesssim \mu \a \equiv \kappa$, where $\mu$ is the $X_1 - X_2$ reduced mass.\footnote{
More precise conditions for the existence of bound states are discussed in \cref{app:BoundState WF num}.}  
Under this condition, we want to compute the cross-sections for the radiative formation of bound states, with emission of the force mediator $\vf$,
\beq
{\cal U}_{\vec k}(X_1 + X_2) \ \to \ 
{\cal B}_{n\ell m}(X_1 X_2) \ + \ \vf \, .
\label{eq:BSF process}
\eeq
Here, ${\cal U}_{\vec k}(X_1 + X_2)$ stands for a two-particle scattering (unbound) state, characterised by the continuous vector quantum number $\vec k = \mu \vec \vrel$, with $\vec \vrel$ being the expectation value of the relative velocity.  Because of their long-range interaction, $X_1$ and $X_2$ cannot be approximated by plane waves. This gives rise to the well-known Sommerfeld effect~\cite{Sommerfeld:1931}. The ${\cal U}_{\vec k}(X_1 + X_2)$ state is instead described by a wavefunction $\f_{\vec k}(\vec r)$ that obeys the Schr\"odinger equation with the Yukawa potential of \cref{eq:Yukawa} and a positive energy eigenvalue ${\cal E}_{\vec k}$, parametrised by $\vec k$. Further, ${\cal B}_{n\ell m}(X_1 X_2)$ is a bound state, whose wavefunction $\ps_{n\ell m}(\vec r)$, parametrised by the familiar principal and angular-momentum discrete quantum numbers $\{n,\ell,m\}$, obeys the Schr\"odinger equation with the same potential and a negative energy eigenvalue ${\cal E}_{n\ell}$. We discuss the numerical computation of $\ps_{n\ell m}(\vec r)$ and $\f_{\vec k}(\vec r)$ in \cref{App:WaveFun}.

As is well-known, the Yukawa potential arises from the one-boson-exchange diagram, shown in \cref{fig:OBE} (left). This is the lowest-order 2-particle-irreducible diagram contributing to the 4-point Green's function of the $X_1 - X_2$ pair. The (infinite) repetition of one-boson-exchange diagrams gives rise to the ladder diagrams of \cref{fig:OBE} (right), whose resummation amounts to solving the Schr\"odinger equation. Indeed, the wavefunctions obeying the Schr\"odinger equation appear as multiplicative factors at (and determine the strength of) the singularities of the 4-point function: the poles corresponding to bound states, and the branch cuts corresponding to scattering states (see e.g.~\cite{Petraki:2015hla}). 

It is now reasonable to wonder how do non-perturbative effects -- the Sommerfeld effect and the existence of bound states -- arise from the resummation of a perturbative Dyson series? Indeed, in the ladder diagrams of \cref{fig:OBE}, the two vertices introduced by each boson exchange imply a suppression by one power of the coupling $\a$. However, this suppression is cancelled by the loop momentum exchange, which also scales with $\a$. In particular, the average momentum exchange along each virtual boson scales as $|\vec q| \sim \mu \a$, and the off-shellness of the $X_1, \, X_2$ propagators scales as $q^0 \propto |\vec q|^2 \propto \a^2$; the integration over the loop energy and momentum also yield factors of $\a^2$ and $\a$, respectively. It is then straightforward to see that each additional loop does \emph{not} increase the order of the diagram (see e.g.~\cite{Hoyer:2014gna}); instead, the ladder diagrams add up coherently. 

The radiative BSF process \eqref{eq:BSF process} arises from the diagrams of \cref{fig:BSF}. The initial-state ladder corresponds to a scattering state and is evaluated at center-of-momentum (CM) energy $E = m_1 + m_2 + {\cal E}_{\vec k} \geqslant m_1+m_2$, while the final-state ladder corresponds to a bound state and is evaluated at CM energy $E = m_1 + m_2 + {\cal E}_{n\ell} < m_1+m_2$.
For concreteness, we assume that the energy difference is dissipated via emission of the same particle that is responsible for the attractive interaction. However, it is straightforward to adapt our results to BSF occurring via radiation of any vector or scalar current that couples either to one or both of the interacting particles (see also \cref{sec:xi explanations}).

\begin{subfigures}

\label{fig:FeynmanDiagrams}
\label[pluralfigure]{figs:FeynmanDiagrams}

\begin{figure}[t!]
\centering
\begin{tikzpicture}[line width=1.5 pt, scale=1.6]
\begin{scope}
\node at (-1.2, 0.5) {$X_1$};
\node at (-1.2,-0.5) {$X_2$};
\node at (-.5, 0.70) {$k_1$};
\node at (-.5,-0.75) {$k_2$};
\draw[fermion]   (-1, 0.5) -- (0, 0.5);
\draw[fermion]   ( 0, 0.5) -- (1, 0.5);
\draw[fermion] (-1,-0.5) -- (0,-0.5);
\draw[fermion] ( 0,-0.5) -- (1,-0.5);
\node at (.5, 0.70) {$k_1'$};
\node at (.5,-0.75) {$k_2'$};
\draw[vector] (0,-0.5) -- (0,0.5);
\node at (-0.25,0) {$\vf$};
\end{scope}
\begin{scope}[shift={(4,0)}]
\node at (-1.2, 0.5) {$X_1$};
\node at (-1.2,-0.5) {$X_2$};
\draw   (-1, 0.5) -- (1, 0.5);
\draw (-1,-0.5) -- (1,-0.5);
\draw[fermion]   (-0.7, 0.5) -- (-0.6, 0.5);
\draw[fermion] (-0.7,-0.5) -- (-0.6,-0.5);
\draw[fermion]   (0.8, 0.5) -- (0.85, 0.5);
\draw[fermion] (0.8,-0.5) -- (0.85,-0.5);
\draw[vector] (-0.5,-0.5) -- (-0.5,0.5);
\draw[vector] (-0.2,-0.5) -- (-0.2,0.5);
\draw[vector] ( 0.5,-0.5) -- ( 0.5,0.5);
\node at (-0.75,0) {$\vf$};
\node at (0.15,0) {$\cdots$};
\end{scope}
\end{tikzpicture}
\caption{
\emph{Left:} In the weak-coupling regime, the one-boson (either scalar or vector) exchange is the dominant contribution to the non-relativistic $X_1 - X_2$ potential.
\emph{Right:} The ladder diagrams -- arising from the infinite repetition of the one-boson-exchange diagrams -- give rise to non-perturbative effects: the Sommerfeld effect and the existence of bound states.
}
\label{fig:OBE}

\bigskip

\centering
\begin{tikzpicture}[line width=1.5 pt, scale=1.3]
\begin{scope}
\node at (-2.4,1) {$X_1$};
\node at (-2.4,0) {$X_2$};
\draw   (-2.2,1) -- (2.2,1);
\draw (-2.2,0) -- (2.2,0);
\node at (-1.9,0.5) {$\cdots$};
\draw[vector] (-1.4,0) -- (-1.4,1);
\draw[vector] (-1.0,0) -- (-1.0,1);
\draw[vector] (-0.6,0) -- (-0.6,1);
\node at (-0.35,0.5) {$\vf$};
\draw[vector] (0,1) -- (0.8,1.5);
\node at (0.15,1.35) {$\vf$};
\node at (0.35,0.5) {$\vf$};
\draw[vector] (0.6,0) -- (0.6,1);
\draw[vector] (1.0,0) -- (1.0,1);
\draw[vector] (1.4,0) -- (1.4,1);
\node at (1.9,0.5) {$\cdots$};
\end{scope}
\node at (2.8,0.5) {$+$};
\begin{scope}[shift={(5.8,0)}]
\node at (-2.4,1) {$X_1$};
\node at (-2.4,0) {$X_2$};
\draw (-2.2,1) -- (2.2,1);
\draw (-2.2,0) -- (2.2,0);
\node at (-1.9,0.5) {$\cdots$};
\draw[vector] (-1.4,0) -- (-1.4,1);
\draw[vector] (-1.0,0) -- (-1.0,1);
\draw[vector] (-0.6,0) -- (-0.6,1);
\node at (-0.35,0.5) {$\vf$};
\draw[vector] (0,0) -- (0.8,-0.5);
\node at (0.15,-0.4) {$\vf$};
\node at (0.35,0.5) {$\vf$};
\draw[vector] (0.6,0) -- (0.6,1);
\draw[vector] (1.0,0) -- (1.0,1);
\draw[vector] (1.4,0) -- (1.4,1);
\node at (1.9,0.5) {$\cdots$};
\end{scope}
\end{tikzpicture}
\caption{The diagrams contributing in leading order, to the formation of bound states with emission of a force mediator. For this transition, the ladder to the left of the radiative vertex corresponds to the initial scattering state, while the ladder to the right corresponds to a bound state. The mediator can be either a scalar or a vector boson.}
\label{fig:BSF}
\end{figure}
\end{subfigures}

\subsection{Parametrisation \label{sec:param}}
 
We formulate our computations and present our results in terms of two dimensionless parameters, 
\beq 
\z \equiv \frac{\mu \a}{\mu \vrel} = \frac{\a}{\vrel} 
\qquad \text{ and } \qquad 
\ks \equiv \frac{\mu \a }{m_\vf},
\label{eq:zeta,xi def}
\eeq 
where $\vrel$ is the relative velocity of the interacting particles and $\mu$ is their reduced mass. 
$\z$ and $\ks$ suffice to characterise the solutions of the Schr\"odinger equation for the Yukawa potential \eqref{eq:Yukawa}, for bound and scattering states (cf.~\cref{App:WaveFun}), and to parametrise the non-analytical parts of our computations that arise in the BSF cross-sections.

The $\z$ parameter compares the average momentum transfer $\sim \mu \vrel$ between two unbound particles, with the momentum transfer $\sim \mu \a$ that occurs in the exchange of virtual force mediators, as explained above. In practice, $\zeta$ parametrises the velocity dependence of the cross-sections. 

On the other hand, $\ks$ parametrises the model dependence of the cross-sections. It compares the two physical scales involved: the Bohr momentum $\k \equiv \mu \a$, which determines the size of the bound states and the momentum transfer along the virtual force mediators in the ladder diagrams, with the mediator mass $m_\vf$, which determines the range of the interaction. The interaction manifests as long-range roughly if $\ks \gtrsim 1$; this is the regime where non-perturbative phenomena, such as the Sommerfeld effect and the existence of bound states, emerge. 

We believe that this minimal parametrisation, in terms of $\zeta$ and $\ks$, exposes the physical significance of the features of the cross-sections, and can greatly facilitate phenomenological studies.

\subsection[The range of the $\ks$ parameter]{The range of the $\boldsymbol{\ks}$ parameter \label{sec:xi explanations}}

Bound states can form with emission of the force mediator that is responsible for their existence, only if the available energy from the transition to a lower energy state suffices, $m_\vf < {\cal E}_{\vec k} - {\cal E}_{n\ell}$, where ${\cal E}_{\vec k} = {\vec k}^2/(2\mu)$ is the kinetic energy of the scattering state in the CM frame, and ${\cal E}_{n\ell} = -\g_{n\ell}^2 \times \kappa^2/(2\mu)$ is the binding energy of the bound state [cf.~\cref{eq:E_nl,eq:E_k}]. 
In most phenomenological applications related to DM, the formation of bound states becomes important in the regime where the kinetic energy is lower than the binding energy, ${\cal E}_{\vec k} \lesssim |{\cal E}_{n\ell}|$ (which roughly implies $\vrel \lesssim \a/n$). Then, the condition for BSF with emission of a force mediator becomes roughly $m_\vf \lesssim \mu \a^2/(2n^2)$, or, in terms of the $\ks$ parameter,
\beq
\ks > \frac{2}{\a \, \g_{n\ell}^2(\ks)} \gg n^2 \, .
\label{eq:BSF threshold}
\eeq
(Note that $\g_{n\ell}(\ks) \leqslant 1/n$, with the equality realised in the Coulomb limit.)
The condition \eqref{eq:BSF threshold} is significantly more stringent than the condition for the existence of bound states, $\g_{n\ell}(\ks) >0$, which implies roughly $\ks = \mu \alpha / m_\vf \gtrsim n^2$ (cf.~\cref{app:BoundState WF num}). In fact, the condition \eqref{eq:BSF threshold} asserts that in its regime of validity, the bound-state wavefunctions can be well approximated by their Coulomb limit\footnote{This approximation was employed in Refs.~\cite{An:2016gad,An:2016kie}.}. 
This is not necessarily the case for the scattering state wavefunctions though, which depend on both $\ks$ and $\zeta$. As we shall see in the following, the Coulomb limit, which formally corresponds to $\ks \to \infty$, is essentially attained for 
\beq \ks \gtrsim \zeta, \label{eq:CoulombRegime} \eeq
i.e.~when the average momentum transfer between the interacting particles exceeds the mediator mass, $\mu \vrel \gtrsim m_\vf$.

However, in many models of interest, BSF may occur via emission of a lighter species than the force mediator that is (primarily) responsible for their existence. This species may couple only to one of the particles participating in the bound state, and would therefore not mediate a long-range interaction between them. 
It is also possible that a light bosonic species couples to both of the interacting particles, albeit more weakly than the heavier force mediator, whose contribution then dominates the interaction between the two particles. In either case, BSF can occur for values of $\ks$ that do not satisfy the condition \eqref{eq:BSF threshold}. In the following, we shall thus consider the entire range of $\ks$ values for which bound states exist in the spectrum of a theory, even if they cannot form with emission of the force mediator responsible for the potential \eqref{eq:Yukawa}. To the extent possible, we will separate the phase-space suppression due to the emission of a massive particle [cf.~\cref{eq:pss}], from the effect of the non-zero mediator mass on the bound-state and scattering-state wavefunctions, and through them, on the amplitude of the process. This renders it possible to adapt our results for BSF processes that occur via emission of any vector or scalar boson.

\subsection{The cross-section} 

The differential cross-section times relative velocity for the 2-to-2 process \eqref{eq:BSF process} is
\beq
\vrel \: \frac{d\s\BSF^{\{n \ell m\}} }{d\W} 
= \frac{|\vec P_\vf|}{64 \p^2 M^2\mu } 
\ |\M_{\vec k \to n \ell m}|^2 \, ,
\nn
\eeq
where $\M_{\vec k \to n\ell m}$ is the transition amplitude, and $|\vec P_\vf|$ is the momentum of the emitted mediator. In the CM frame, the energy dissipated during BSF is the sum of the kinetic and binding energies, which implies
\beq
\sqrt{|\vec{P}_\vf|^2  + m_\vf^2} = {\cal E}_{\vec k} - {\cal E}_{n\ell} 
= \frac{\vec{k}^2}{2\mu} + \frac{\kappa^2}{2\mu} 
\gamma_{n\ell}^2(\ks) \, ,
\nn 
\eeq
where we used \cref{eq:E_nl,eq:E_k}. Then
\begin{subequations} 
\label{eq:P vf}
\label[pluralequation]{eqs:P vf}
\begin{align}
\frac{|\vec{P}_\vf|}{\kappa} &= 
\frac{\a}{2} \[ \frac{1 + \z^2 \gamma_{n\ell}^2(\ks)}{\z^2} \] 
\times \ pss_{n\ell}^{1/2} \, ,
\label{eq:ForceMed_momentum}
\\
pss_{n\ell} &= 1 - \frac{4 \, \z^4}{\a^2\ks^2 [1 + \z^2 \gamma_{n\ell}^2(\ks)]^2}  \, ,
\label{eq:pss}
\end{align}
\end{subequations}
where $pss_{n\ell}$ is the phase-space suppression factor for the emission of a force mediator, during capture into the $n\ell m$ bound state; in the Coulomb limit $\ks \to \infty$ and $pss_{n\ell}^{}=1$. Putting everything together, we obtain
\beq
\vrel \: \frac{d\s\BSF^{\{n \ell m\}} }{d\W} 
= \ \frac{\a^2}{2^7 \p^2 M^2} \ pss_{n\ell}^{1/2} \
\(\frac{1+\z^2 \g_{n\ell}^2(\ks)}{\z^2}\) \, |\M_{\vec k \to  n \ell m}|^2
\, . 
\label{eq:DiffSigma}
\eeq

The general expression for the amplitude $\M_{\vec k \to  n \ell m}$ in terms of the initial and final state wavefunctions and the radiative vertex can be found in  Ref.~\cite[section 3]{Petraki:2015hla}. In the following, we shall reproduce only the expressions that are relevant for the interactions we are considering.

\subsection{Initial and final state convolution integrals \label{sec:Convol Integr}}

The transition amplitudes $\M_{\vec k \to n \ell m}$ depend on the scattering-state and bound-state wavefunctions, $\psi_{n\ell m}$ and $\f_{\vec k}$, and on the radiative vertex, via the convolution integrals~\cite{Petraki:2015hla}
\begin{subequations} 
\label{eq:ConvIntegrals_Def}
\label[pluralequation]{eqs:ConvIntegrals_Def}
\begin{align}
{\cal I}_{\vec k, n\ell m} (\vec b) 
&\equiv 
\int \frac{d^3p}{(2\p)^3} 
\: \tilde \psi_{n\ell m}^* (\vec p) 
\: \tilde \f_{\vec k}  (\vec p + \vec b)
= \int d^3 r 
\: \psi_{n\ell m}^* (\vec r) 
\: \f_{\vec k}  (\vec r) 
\: e^{-i \vec b \cdot \vec r} \: ,
\label{eq:Ical_def}
\\
\boldsymbol{\cal J}_{\vec k, n\ell m} (\vec b) 
&\equiv 
\int \frac{d^3p}{(2\p)^3} \: \vec p 
\: \tilde \psi_{n\ell m}^* (\vec p) 
\: \tilde \f_{\vec k}  (\vec p +\vec b) 
= i \int d^3 r 
\: [\nabla \psi_{n\ell m}^* (\vec r)] 
\:  \f_{\vec k}  (\vec r) 
\: e^{-i \vec b \cdot \vec r}
\: ,
\label{eq:Jcal_def}
\\
{\cal K}_{\vec k, n\ell m} (\vec b) 
&\equiv 
\int \frac{d^3p}{(2\p)^3} \: \vec p^2 
\: \tilde \psi_{n\ell m}^* (\vec p) 
\: \tilde \f_{\vec k}  (\vec p + \vec b) 
= - \int d^3 r 
\: [\nabla^2 \psi_{n\ell m}^* (\vec r)] 
\: \f_{\vec k}  (\vec r) 
\: e^{-i \vec b \cdot \vec r}
\: .
\label{eq:Kcal_def}
\end{align}
\end{subequations}
Here $\tilde{\psi}_{n\ell m}$ and $\tilde{\phi}_{\vec k}$ are the Fourier transforms of $\psi_{n\ell m}$ and $\phi_{\vec k}$ respectively. The momentum  $\vec{b}$ is proportional to the momentum of the radiated particle, $\vec b \propto \vec P_\vf$, as we shall see in the following. 
In \cref{App:ConvIntegr}, we expand the integrals~\eqref{eqs:ConvIntegrals_Def}  in powers of $|\vec b|/\kappa \propto |\vec P_\vf|/\kappa \propto \alpha$ [cf.~\cref{eq:P vf}], and identify the leading-order contributions for the transitions of interest. The range of validity of this approximation is given by the condition \eqref{eq:expansion condition}.

We note that the formalism developed in Ref.~\cite{Petraki:2015hla} assumed that the interacting particles have zero spin. However, the lowest-order computations using the expressions of Ref.~\cite{Petraki:2015hla} for the radiative BSF cross-sections, are  applicable to both fermionic and bosonic interacting species. Indeed, in the non-relativistic regime, and to lowest order in the coupling, the spin of each of the interacting particles is conserved in the capture process, and the BSF cross-sections do not depend on the spin (or the spin configuration) of the incoming particles. The computations of \cref{Sec:VecMed,sec:ScalarMed_NonDegen} fall in this category. However, in the radiative capture of a particle-antiparticle pair via emission of scalar force mediator, the lowest-order contributions cancel each other. As discussed in \cref{sec:ScalarMed_Degen}, in this case, we are forced to consider higher-order terms, which may in general be different for fermions and bosons.

\clearpage
\section{Vector force mediator  \label{Sec:VecMed}}

\subsection{Radiative capture into a bound state}

We assume $X_1$ and $X_2$ to be coupled to a gauged U(1) force, 
\begin{subequations} 
\label{eq:VecMed_L}
\label[pluralequation]{eqs:VecMed_L}
\begin{align}
{\cal L}
&= (D_\m X_1)^\dagger (D^\m X_1) + (D_\m X_2)^\dagger (D^\m X_2)
- \frac{1}{4} F_{\m\n} F^{\m\n}
- m_1^2 |X_1|^2 - m_2^2 |X_2|^2 
+ \frac{1}{2} \, m_\vf^2 \vf_\mu \vf^\mu
\: ,
\label{eq:VecMed_L_ScalarX}
\\
{\cal L}
&= \bar{X}_1 i\slashed{D} X_1 + \bar{X}_2 i\slashed{D} X_2 
- m_1 \bar{X}_1 X_1 - m_2 \bar{X}_2 X_2 
- \frac{1}{4} F_{\m\n} F^{\m\n}
+ \frac{1}{2} \, m_\vf^2 \vf_\mu \vf^\mu
\: ,
\label{eq:VecMed_L_FermionX} 
\end{align}
\end{subequations}
where $F^{\m\n} = \partial^\m \vf^\n - \partial^\n \vf^\m$ and $D^\m = \partial^\m - i c_j g \vf^\m$, with $c_1, \: c_2$ being the charges of $X_1, \: X_2$. The mass $m_\vf$ of $\vf^\mu$ may have arisen either via the Higgs or the St\"{u}ckelberg mechanisms. While the details of the local U(1) breaking are not important for our purposes, we do assume in the Lagrangians~\eqref{eqs:VecMed_L} that the individual global U(1) symmetries associated with $X_1$ and $X_2$ remain unbroken.\footnote{In the Higgs mechanism, this can be ensured by an appropriate choice of the charge of the scalar field breaking the gauged U(1) symmetry. In the St\"{u}ckelberg mechanism, the massive gauge boson couples to a conserved current.}
In the non-relativistic regime, the $\vf^\mu$ exchange between $X_1$ and $X_2$ gives rise to the Yukawa potential of \cref{eq:Yukawa}, with
\beq
\a = -\frac{c_1 c_2 g^2}{4\p} \: .
\label{eq:VecMed_alpha}
\eeq
The interaction is attractive if $c_1 c_2 < 0$.

Since BSF involves gauge interactions with conserved currents, as seen from \cref{eqs:VecMed_L}, the Ward identity ensures that $P_\vf^\mu \M_\mu = 0$. This implies $\M^0 = P_\vf^j \M^j / P_\vf^0$. Then, the unpolarised amplitude is
\begin{align}
\sum_\e |\M_{\vec k \to n \ell m}|^2 
\ &= \ -\(g_{\m\n} - \frac{P_{\vf , \mu} P_{\vf , \nu}}{m_\vf^2} \) 
\M_{\vec k \to n \ell m}^\mu \M_{\vec k \to n \ell m}^{\nu*}
\nn \\
\ &= \ \M_{\vec k \to n\ell m}^j \M_{\vec k \to n\ell m}^{j*} 
- \frac{|P_\vf^j \M_{\vec k \to n \ell m}^j|^2 }{ P_\vf^2 + m_\vf^2 }\, . 
\label{eq:VecMed_M_PolarSum}
\end{align}
We thus need only the spatial components of the BSF amplitude, which are~\cite{Petraki:2015hla}
\begin{multline}
\M_{\vec k \to n \ell m}^j = 
- 2g\sqrt{2\mu} \ \left\{
 \frac{c_1}{\h_1}\, {\cal J}_{\vec k, n\ell m}^j ( \h_2 \vec P_\vf) 
-\frac{c_2}{\h_2}\, {\cal J}_{\vec k, n\ell m}^j (-\h_1 \vec P_\vf) 
\right. \\  \left.
+ \[c_1 \(K^j - \frac{\h_1 -\h_2}{2\h_1} P_\vf^j \) \: {\cal I}_{\vec k, n\ell m} (\h_2 \vec P_\vf)
+c_2 \(K^j + \frac{\h_1 -\h_2}{2\h_2} P_\vf^j \) \: {\cal I}_{\vec k, n\ell m} (-\h_1 \vec P_\vf) \]
\right\} \: ,
\label{eq:VecMed_M}
\end{multline}
where ${\cal I}_{\vec k, n\ell m}$ and ${\cal J}^j_{\vec k, n \ell m}$ are defined in \cref{eqs:ConvIntegrals_Def}. 
It is evident from the momentum factor in the integrand of ${\cal J}_{\vec k, n\ell m}^j$ [cf.~\cref{eq:Jcal_def}] and the momentum dependence of the ${\cal I}_{\vec k, n\ell m}$ contribution to $\M_{\vec k \to n \ell m}$, that \cref{eq:VecMed_M} describes a radiative process that proceeds via a derivative current interaction. Moreover, both charged particles contribute to the emitted radiation.

\subsection[Capture into $\ell=0$ bound states]{Capture into $\boldsymbol{\ell=0}$ bound states}

From \cref{eqs:IJK n00}, we find that the lowest-order contributions to the amplitude \eqref{eq:VecMed_M} arise from the $\boldsymbol{\cal J}_{\vec k, n00}^{}$ integrals and are of zeroth order in $|\vec{P}_\vf|/\kappa$. The unpolarised squared amplitude \eqref{eq:VecMed_M_PolarSum} is
\beq
\sum_\e |\M_{\vec k \to n00}|^2 \simeq 
\frac{2^5\pi \, \mu\alpha}{\h_1^2 \h_2^2} \[\frac{(\h_2 c_1 - \h_1 c_2)^2}{-c_1c_2}\] \(
|\boldsymbol{\cal J}_{\vec k, n00}^{}|^2 - pss_{n,0}^{} \, 
|\hat{\vec{P}}_\vf \cdot \boldsymbol{{\cal J}}_{\vec k, n00}^{}|^2 \) \, .
\label{eq:VecMed_Mn00}
\eeq
(Since we are now interested only in the zero-th order terms in $|\vec{P}_\vf|/\kappa$, we have dropped the arguments of the $\boldsymbol{\cal J}_{\vec k, n00}^{}$ functions.) Substituting \cref{eq:Jn00 expansion} into \cref{eq:VecMed_Mn00} yields
\begin{multline}
\sum_{\rm \e} |\M_{\vec{k} \to n00}|^2 \simeq 
\frac{2^7\p^2 M^2}{\mu^2} \[\frac{(\h_2 c_1 - \h_1 c_2)^2}{-c_1 c_2}\] 
\ (1-pss_{n,0}^{} \, \cos^2\theta) \, \times
\\ 
\times \left|   \int_0^\infty dx 
\[\frac{d\x_{n,0}^*(x)}{dx} - \frac{\x_{n,0}^*(x)}{x}\] 
\x_{|\vec k|,1}^{}(x)   \right|^2 \, ,
\label{eq:VecMed_Mn00_final}
\end{multline}
where $\x_{n,\ell}(x)^{}$ and $\x_{|\vec k|,\ell}^{}(x)$ are related to the bound-state and scattering-state wavefunctions as described in \cref{App:ConvIntegr}. From \cref{eq:DiffSigma} and \eqref{eq:VecMed_Mn00_final}, we find the BSF cross-section to be\footnote{\label{foot:SBSF factorisation}
Here and in the following, we typically choose to factorise the various BSF cross-sections (i.e.~separate out the $S\BSF$ factors) in a way that facilitates the comparison between BSF and annihilation processes for particle-antiparticle pairs. However, we note that the $S\BSF$ factors do not carry any physical significance on their own. This is in contrast to the $S\ann$ factors appearing in the various annihilation cross-sections [cf.~\cref{sec:VectorMed_Ann,sec:ScalarMed_Ann}], which represent the enhancement of the corresponding processes due to the non-perturbative Sommerfeld effect, and are often referred to as ``Sommerfeld enhancement factors". At $\z,\ks \ll 1$, the non-perturbative effects switch off, and the annihilation cross-sections reduce to their perturbative values ($S\ann \simeq 1$). In the same regime, $S\BSF \to 0$, as the very existence of bound states is a non-perturbative effect. The BSF processes do not have a perturbative limit, and the $S\BSF$ factors are not enhancement factors of otherwise perturbative processes, by a non-perturbative effect.}
\begin{subequations} 
\label{eq:VecMed_BSF_n00}
\label[pluralequation]{eqs:VecMed_BSF_n00}
\beq
\s\BSF^{\{n00\}} \vrel \ \simeq \
\frac{\p\a^2}{4\mu^2} \: \[\frac{(c_1 \h_2-c_2\h_1)^2}{-c_1 c_2}\] 
\ pss_{n,0}^{1/2} \: \(\frac{3-pss_{n,0}^{}}{2}\) 
\times S\BSF^{\{n00\}} (\z,\ks) \, , 
\label{eq:VecMed_sigma_n00}
\eeq
where
\beq
S\BSF^{\{n00\}} (\z,\ks) =  \frac{2^5}{3}
\ \(\frac{1+\z^2\g_{n,0}^2(\ks)}{\z^2}\)
\left|   \int_0^\infty dx 
\[\frac{d\x_{n,0}^*(x)}{dx} - \frac{\x_{n,0}^*(x)}{x}\] 
\x_{|\vec k|,1}^{}(x)   \right|^2  \, .
\label{eq:VecMed_Sn00}
\eeq
\end{subequations}
\subsubsection*{Coulomb limit}
At $m_\vf \to 0$, the above becomes
\begin{subequations} 
\label{eq:VecMed_BSF_n00_Coul}
\label[pluralequation]{eqs:VecMed_BSF_n00_Coul}
\beq
\lim_{\ks\to\infty} \s\BSF^{\{n00\}} \vrel \simeq
\frac{\p\a^2}{4\mu^2} 
\: \[\frac{(c_1 \h_2-c_2\h_1)^2}{-c_1 c_2}\] 
S\BSFC^{\{n00\}} (\z) \, , 
\label{eq:VecMed_sigma_n00_Coul}
\eeq
where, using \cref{eq:Jn00 Coul expansion}, we find
\begin{multline}
S\BSFC^{\{n00\}} (\z) 
= \( \frac{2\p \z}{1-e^{-2\p\z}} \) 
\ \frac{2^9}{3n^7} 
\ \z^4 \, (1+\z^2) \, (1+\z^2/n^2) \ \times
\\ \times
\abs{\sum_{s=0}^{n-1} \frac{n!(2n-s)}{(n-s-1)!} \frac{2^s \z_n^s}{(s+2)!\,s!}
\ \frac{d^s}{d\z_n^s} 
\[\frac{e^{-2\z {\rm arccot}\,\z_n}}{(1+\z_n^2)^2}\]
}^2_{\z_n = \z/n}   \,.
\label{eq:VecMed_Sn00_Coul_sum}
\end{multline}
Expanding the sum, $S\BSFC^{\{n00\}}$ takes the form
\beq
S\BSFC^{\{n00\}} (\z) =
\( \frac{2\p \z}{1-e^{-2\p\z}} \) 
\ \frac{2^9}{3n^3} 
\ \frac{\z^4 \, (1+\z^2)}{(1+\z^2/n^2)^{2n-1}} 
\ [\varrho_n(\z)]^2 
\ e^{-4\z\,{\rm arccot}(\z/n)} \, ,
\label{eq:VecMed_Sn00_Coul}
\eeq
where $\varrho_n(\z)$ is a rational function of $\z^2$ (and specifically, a polynomial of degree $n-2$, for $n \geqslant 2$), with $\lim_{\z\to 0}\varrho_n(\z) = 1$. We give the explicit expressions of $\varrho_n(\z)$ for $1\leqslant n \leqslant 5$ in \cref{tab:varrho n00 vector & scalar non-degen}. For capture into the ground state in particular,
\beq
S\BSFC^{\{100\}} (\z) =
\( \frac{2\p \z}{1-e^{-2\p\z}} \) 
\ \frac{2^9}{3^{~}} 
\, \frac{\z^4}{(1+\z^2)^2} 
\ e^{-4\z\,{\rm arccot} \,\z} \, .
\label{eq:VecMed_S100_Coul}
\eeq
\end{subequations}
We illustrate in detail the features of $S\BSF^{\{n00\}} (\z,\ks)$ and $S\BSFC^{\{n00\}} (\z)$ in \cref{fig:VecMed_Coulomb,fig:VecMed_n=1,fig:VecMed_n=2_l=0,fig:VecMed_Sregularized}, and we discuss them in \cref{sec:VecMed_Discussion}.

\begin{table}
\centering
\renewcommand{\arraystretch}{2.1}
\begin{tabular}{|l|l|}
\hline
$\boldsymbol{n}$ & $\boldsymbol{\varrho_n(\z)}$
\\ \hline  
1& $\dfrac{1}{1+\z^2}$
\\ \hline
2& 1
\\ \hline
3& $1 + \dfrac{7\,\z^2}{3^3}$
\\ \hline
4& $1 + \dfrac{3\,\z^2}{2^3} + \dfrac{23\,\z^4}{2^8 3}$
\\ \hline
5& $1 + \dfrac{11\,\z^2}{5^2} + \dfrac{509\,\z^4}{3 \cdot 5^5} 
+ \dfrac{7 \cdot 13\,\z^6}{3 \cdot 5^6}$
\\ \hline
\end{tabular}

\caption{\label{tab:varrho n00 vector & scalar non-degen}
The functions $\varrho_n(\zeta)$ appearing in the Coulomb limit of the cross-sections for the radiative formation of zero angular momentum states of principal quantum number $n$, (i) with emission of a vector force mediator [cf.~\cref{eq:VecMed_Sn00_Coul}], and (ii) with emission of a scalar mediator, by particles of different species [cf.~\cref{eq:ScalMed_NonDegen_Sn00_Coul}].}
\end{table}

\bigskip
Let us now recount the origin of the various factors in \cref{eq:VecMed_sigma_n00,eq:VecMed_Sn00_Coul}. In \cref{eq:VecMed_sigma_n00}, the factor $(c_1 \h_2 - c_2 \h_1)^2/(-c_1c_2) = (c_1/\h_1-c_2/\h_2)^2 \times [\h_1^2\h_2^2/(-c_1c_2)]$ emanates from the vertices of the radiated gauge boson in the Feynman diagrams of \cref{fig:BSF}, and asserts that lighter particles radiate more easily; it becomes 1 for $c_1 = -c_2$. The phase-space suppression factor $pss_{n,0}^{1/2}$  is due to the limited energy available for the radiation of a massive force mediator. The factor $(3-pss_{n,0})$ accounts for the contribution of the three polarizations of the emitted massive gauge boson; it reduces to 2 for a massless mediator.

The function $S\BSF(\z,\xi)$ measures the overlap of the initial and final states. Its Coulomb limit \eqref{eq:VecMed_Sn00_Coul} is illustrative. The first factor, $S_0(\z) = 2\p\z/(1-e^{-2\p\z})$, is an overall multiplicative constant in the scattering-state wavefunction [cf.~\cref{eqs:WFs Coul}], and is responsible for the characteristic scaling of long-range inelastic processes, $\s\BSF \vrel \propto 1/\vrel$ at $\z \gg 1$.  The other factors in \cref{eq:VecMed_Sn00_Coul} depend on the details (space/momentum dependence) of the scattering-state and bound-state wavefunctions, as well as the radiative vertex. At $\z < 1$, $S\BSF (\z,\xi)$ is very small and BSF is typically inefficient; however, at $\z \gtrsim 1$, $S\BSFC (\z)\simeq 3.13 \times 2\p\z$. This point will become important in \cref{sec:VectorMed_Ann}, where we compare BSF and annihilation for particle-antiparticle pairs.

The BSF cross-sections we calculate in the following sections have similar structure to the one described above.

\subsection[Capture into $\ell=1$ bound states]{Capture into $\boldsymbol{\ell=1}$ bound states}

Similarly to before, the zeroth order terms in $|\vec{P}_\vf|/\kappa$, of the $\boldsymbol{\cal J}_{\vec k, n1m}^{}$ integrals yield the dominant contribution to the amplitude \eqref{eq:VecMed_M} for capture into $\ell=1$ bound states, with the unpolarised squared amplitude \eqref{eq:VecMed_M_PolarSum} being
\beq
\sum_\e |\M_{\vec k \to n1m}|^2 \simeq 
\frac{2^5\pi \, \mu\alpha}{\h_1^2 \h_2^2} \[\frac{(\h_2 c_1 - \h_1 c_2)^2}{-c_1c_2}\] \(
|\boldsymbol{\cal J}_{\vec k, n1m}^{}|^2 - pss_{n,0}^{} \, 
|\hat{\vec{P}}_\vf \cdot \boldsymbol{{\cal J}}_{\vec k, n1m}^{}|^2 \) \, .
\eeq
For convenience, we define
\begin{subequations} 
\label{eq:VecMed_A0&A2}
\label[pluralequation]{eqs:VecMed_A0&A2}
\begin{align}
A_0(\z,\ks) &\equiv \int_0^\infty dx 
\[\chi_{n,1}'(x) + \frac{\chi_{n,1}^{}(x)}{x}\]^* 
\chi_{|\vec k|,0}^{}(x) \, ,
\label{eq:VecMed_A0}
\\
A_2(\z,\ks) &\equiv \int_0^\infty dx 
\[\chi_{n,1}'(x) - \frac{2\chi_{n,1}^{}(x)}{x}\]^* 
\chi_{|\vec k|,2}^{}(x) \, .
\label{eq:VecMed_A2}
\end{align}
\end{subequations}
Then, from \cref{eqs:Jn1m} we find
\begin{subequations} \label[pluralequation]{eqs:Jn1m squared}
\begin{align}
|\boldsymbol{\cal J}_{\vec k, n10}^{}|^2 
&= \frac{4\pi}{\kappa} 
\[\frac{1}{3}(|A_0|^2 + |A_2|^2) + |A_2|^2 \cos^2\theta_{\vec k} 
+ (A_0 A_2^* + A_0^* A_2) \( \cos^2\theta_{\vec k}-\frac{1}{3} \) \] \, ,
\\
|\boldsymbol{\cal J}_{\vec k, n1\pm1}^{}|^2 
&= \frac{2\pi}{\kappa} 
\[\frac{2}{3}(|A_0|^2 + |A_2|^2) + |A_2|^2 \sin^2\theta_{\vec k} 
+ (A_0 A_2^* + A_0^* A_2) \(\sin^2\theta_{\vec k}-\frac{2}{3} \)\] ,
\end{align}
and
\begin{multline}
|\boldsymbol{\cal J}_{\vec k, n10}^{}\cdot \hat{\vec{P}}_\vf|^2 
= \frac{12\pi}{\kappa} \[
\frac{|A_0-A_2|^2}{9} \, \cos^2 \theta_{\vec{P}_\vf}
+ (\hat{\vec k}\cdot \hat{\vec{P}}_\vf)^2 \, |A_2|^2 \cos^2\theta_{\vec k}
\right. \\ \left.
+ (\hat{\vec k}\cdot \hat{\vec{P}}_\vf) 
\, \cos\theta_{\vec k} \, \cos\theta_{\vec P_\vf} 
\: \frac{A_2 (A_0^*-A_2^*) + \cc}{3}
\] ,
\end{multline}
\begin{multline}
|\boldsymbol{\cal J}_{\vec k, n1\pm1}^{}\cdot \hat{\vec{P}}_\vf|^2 
= \frac{6\pi}{\kappa} \[
\frac{|A_0-A_2|^2}{9} \, \sin^2 \theta_{\vec{P}_\vf}
+ (\hat{\vec k}\cdot \hat{\vec{P}}_\vf)^2 \, |A_2|^2 \sin^2\theta_{\vec k}
+ 
\right. \\  \left.
+ 
(\hat{\vec k}\cdot \hat{\vec{P}}_\vf) 
\, \sin\theta_{\vec k} \, \sin\theta_{\vec P_\vf} 
\frac{A_2 (A_0^*-A_2^*) \exp[i(\phi_{\vec k}- \phi _{\vec P_\vf})] + \cc}{3}
\] .
\end{multline}
\end{subequations}
To calculate the total cross-section for capture to any $\ell=1$ state, we sum over all values of $m$.
\begin{multline}
\sum_{m=-1}^1  \sum_\e |\M_{\vec k \to n1m}|^2 \simeq
\frac{2^7\pi^2}{\h_1^2 \h_2^2} \[\frac{(\h_2 c_1 - \h_1 c_2)^2}{-c_1c_2}\]   \times
\\
\( |A_0|^2 + 2 |A_2|^2  - pss_{n,1} 
\[ \frac{|A_0-A_2|^2}{3} + (|A_2|^2 + A_2 A_0^* + A_2^* A_0) 
\, (\hat{\vec k} \cdot \hat{\vec P}_\vf)^2 
\right. \right. \\ \left. \left.
- 2\,{\rm Im}(A_2 A_0^*)
\sin\theta_{\vec k}\sin\theta_{\vec P_\vf} \sin(\phi_{\vec k} - \phi_{\vec P_\vf}) 
\, (\hat{\vec k} \cdot \hat{\vec P}_\vf)  
\]  \) \, .
\label{eq:VecMed_Mn1m}
\end{multline}
Note that 
$\hat{\vec k} \cdot \hat{\vec P}_\vf = 
\cos\theta_{\vec k}\cos\theta_{\vec P_\vf} + 
\sin\theta_{\vec k}\sin\theta_{\vec P_\vf}\cos(\phi_{\vec k}-\phi_{\vec P_\vf})$, 
and the term proportional to ${\rm Im}(A_2 A_0^*)$ in \cref{eq:VecMed_Mn1m} gives a vanishing contribution when integrated over $d\W_{\vec P_\vf}$. 
Using the above and \cref{eq:DiffSigma}, we find the total cross-section for capture into an $\ell=1$ state (for fixed $n$),
\begin{subequations} \label[pluralequation]{eqs:VecMed_BSF_n1}
\beq
\sum_{m=-1}^1 \s\BSF^{\{n1m\}}\vrel = 
\frac{\p\a^2}{4\mu^2} \: \[\frac{(c_1 \h_2-c_2\h_1)^2}{-c_1 c_2}\] 
\ pss_{n,1}^{1/2} \: \(\frac{3-pss_{n,1}^{}}{2}\) 
\times S\BSF^{\{n1\}} (\z,\ks) \, , 
\label{eq:VecMed_sigma_n1}
\eeq
where
\beq
S\BSF^{\{n1\}} (\z,\ks) =  \frac{2^5}{3}
\ \(\frac{1+\z^2\g_{n,1}^2(\ks)}{\z^2}\)
\[ |A_0(\z,\ks)|^2 + 2|A_2(\z,\ks)|^2 \]
\, ,
\label{eq:VecMed_Sn1}
\eeq
with $A_0$ and $A_2$ defined in \cref{eqs:VecMed_A0&A2}.
\end{subequations}

\subsection{Annihilation of particle-antiparticle pairs  \label{sec:VectorMed_Ann}}

Bound states of particle-antiparticle pairs are unstable and decay into radiation. This effectively provides an extra annihilation channel. It is then instructive to compare BSF with the direct annihilation into radiation.

The (spin-averaged) annihilation cross-section times relative velocity of a particle-antiparticle pair of scalars and fermions respectively, is
\begin{subequations} 
\label{eq:VecMed_Ann}
\label[pluralequation]{eqs:VecMed_Ann}
\begin{align}
\s\ann^s \vrel &= \frac{\p\a^2}{2\mu^2} \times S\ann^{(0)}(\z,\ks) \, ,
\label{eq:VecMed_Ann_Scalars}
\\
\s\ann^f \vrel &= \frac{\p\a^2}{4\mu^2} \times S\ann^{(0)}(\z,\ks) \, ,
\label{eq:VecMed_Ann_Fermions}
\end{align}
where $\mu = m_X/2 = M/4$ and 
\beq 
S\ann^{(0)} (\z,\ks) \equiv |\f_{\vec k}(\vec r=0)|^2 
= \lim_{x\to 0} \[ \frac{\x_{|\vec k|,\ell=0}^{} (x) }{x} \] \, .
\label{eq:Sann0}
\eeq
The Coulomb limit of \cref{eq:Sann0} is [cf.~\cref{eqs:WFs Coul}]
\beq 
S\annC^{(0)} (\z) = \frac{2\pi \zeta}{1- e^{-2\pi \zeta}} \equiv S_0(\zeta) \, .
\label{eq:Sann0_Coul}
\eeq
\end{subequations}
Note that the perturbative value of the annihilation cross-section (recovered in the limit $S\ann^{(0)} \to 1$) is different for bosons and fermions. In contrast, the leading order BSF cross-section does not depend on the spin of the particles, as already discussed in \cref{sec:Convol Integr}. 

We compare annihilation and BSF in \cref{fig:VecMed_Coulomb,fig:VecMed_n=1}, and offer our comments in the next section.

\subsection{Discussion \label{sec:VecMed_Discussion}}

We now discuss the main features of the BSF cross-sections computed in this section.

\subsubsection*{Coulomb limit} 
For a massless force mediator, and at large enough $\z$, the BSF cross-sections scale as $\s\vrel \propto \z$. This behaviour is realised at $\z \gtrsim n$ for capture into an $\{n\ell m\}$ bound state, as evident in \cref{fig:VecMed_Coulomb}; for capture into $\ell = 0$ bound states, it can also be confirmed analytically using \cref{eq:VecMed_Sn00_Coul}. Importantly, the $1/\vrel$ scaling is anticipated due to the upper bound on the inelastic cross-sections imposed by unitarity~\cite{vonHarling:2014kha, Petraki:2015hla, Baldes:2017gzw}, which we discuss below.

The radiative capture into the ground state dominates over capture into excited levels, $n>1$. On the other hand, for $n>1$, the formation of $\ell \neq 0$ states (summed over $-\ell \leqslant m \leqslant \ell$) dominates over capture into the $\ell=0$ state, in part due to the larger multiplicity of the former.

The above points imply that for a given value of $\z$, the total BSF cross-section is dominated by capture into the $n<\z$ levels, with the $\ell > 0$ states yielding a significant contribution. Indeed, the capture into excited states gives rise to a logarithmic enhancement of the total cross-section for radiative BSF~\cite{Belotsky:2015osa,An:2016gad}, described by Kramer's formula, 
\beq
\s\BSF^{\rm tot} \vrel \simeq \frac{\p \a^2}{4\mu^2} \times 
\frac{2^7}{3\sqrt{3}} \, \z \, \[\ln \z + 0.16 + {\cal O}(\z^{-1})\] \, ,
\label{eq:KramersLaw}
\eeq
which is valid for $\z \gtrsim 2$. The enhancement with respect to capture into the ground state [cf.~\cref{eq:VecMed_sigma_n00_Coul,eq:VecMed_S100_Coul}], is  
$\s\BSF^{\rm tot}/\s\BSF^{\{100\}} \simeq 1+ 1.25 \, \ln (\z/1.89)$.

However, the enhanced BSF rate implied by \cref{eq:KramersLaw} is not always relevant for the phenomenology of DM, since the observable implications of bound states do not depend only on the total rate at which bound states form, but also on their features. For example, the formation of unstable (particle-antiparticle) bound states in the early universe, and their subsequent decay into radiation, can reduce the DM relic density~\cite{vonHarling:2014kha}; however, the efficiency with which DM is depleted depends on the balance between BSF, ionisation and decay, which in turn depends sensitively on the quantum numbers of the bound states that form. Similarly, the cosmological formation of stable bound states by asymmetric DM~\cite{Petraki:2013wwa}, and their survival until today, depends typically on a rather complex interplay between formation, ionisation, excitation and de-excitation processes~\cite{CyrRacine:2012fz}. Moreover, the detectability of the radiation emitted inside halos today during the formation of stable bound states of asymmetric DM~\cite{Pearce:2015zca,Pearce:2013ola,Cline:2014eaa}, obviously depends on the energy release in the specific transition that takes place. In all these cases, the phenomenological importance of BSF cannot be assessed based solely on \cref{eq:KramersLaw}, even in the Coulomb regime. On the other hand, the high-energy signals arising from the decay of unstable bound states of symmetric DM inside halos may reflect the logarithmic enhancement of \cref{eq:KramersLaw}, provided that the excited states get de-excited or decay into radiation promptly enough in astrophysical timescales~\cite{An:2016gad}. Note though that away from the Coulomb regime, this enhancement is curtailed due the phase-space suppression [cf.~\cref{eq:pss}] that becomes more severe for capture into excited states~\cite{An:2016gad}.

The Coulomb limit is attained at $\ks \gtrsim \zeta$, as can be observed in \cref{fig:VecMed_n=1,fig:VecMed_n=2_l=0,fig:VecMed_n=2_l=1,fig:VecMed_Sregularized}, and was already noted in \cref{sec:param}. The physical interpretation of this condition is that the momentum transfer between the two incoming particles exceeds the mediator mass, $\mu \vrel \gtrsim m_\vf$.

\subsubsection*{Resonance structure} 
\begin{figure}[h!]
\centering

{\boldmath \bf Vector mediator: Capture into first excited state $\{200\}$}

\includegraphics[height=7cm]{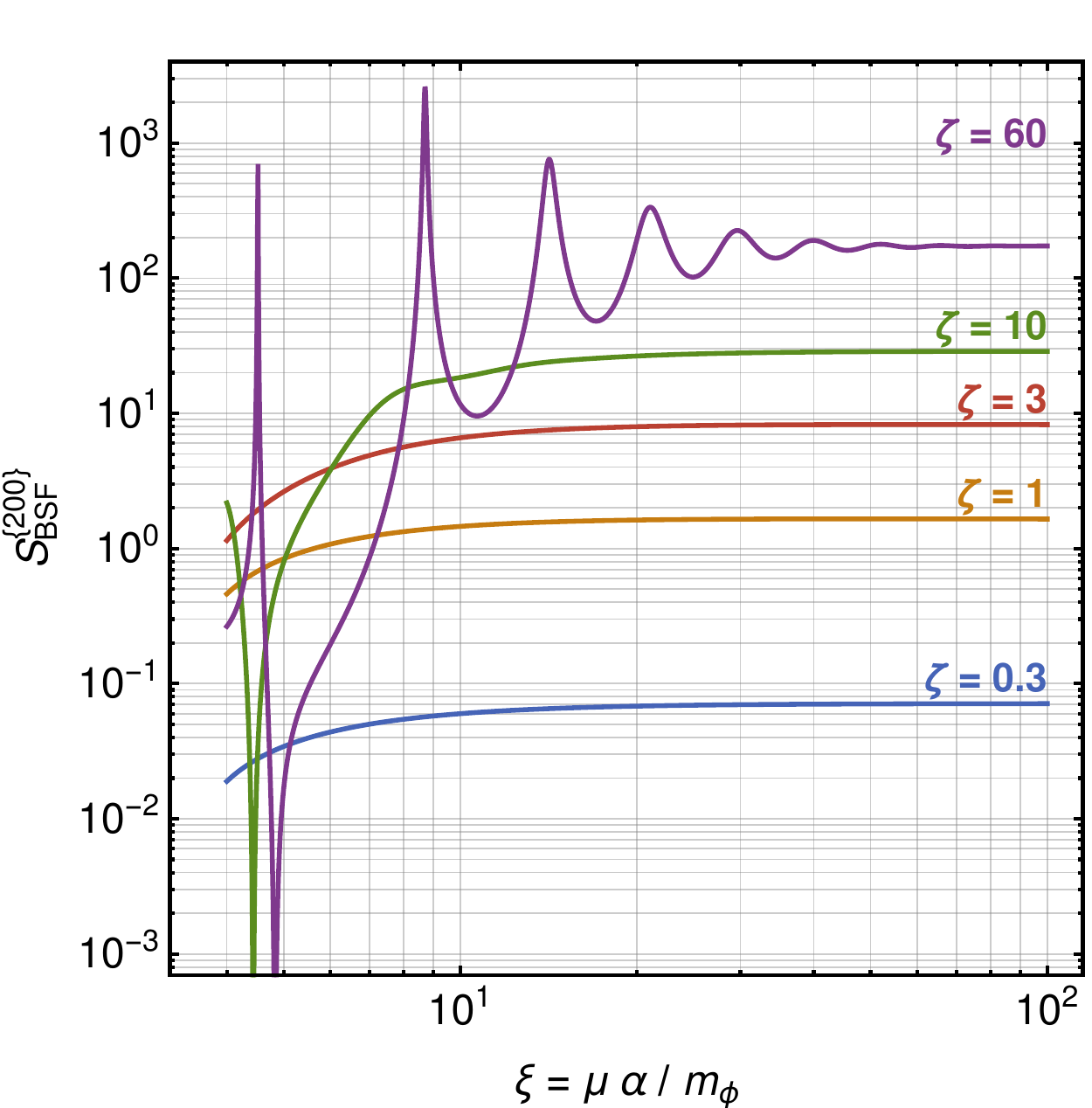}
~~~~~
\includegraphics[height=7cm]{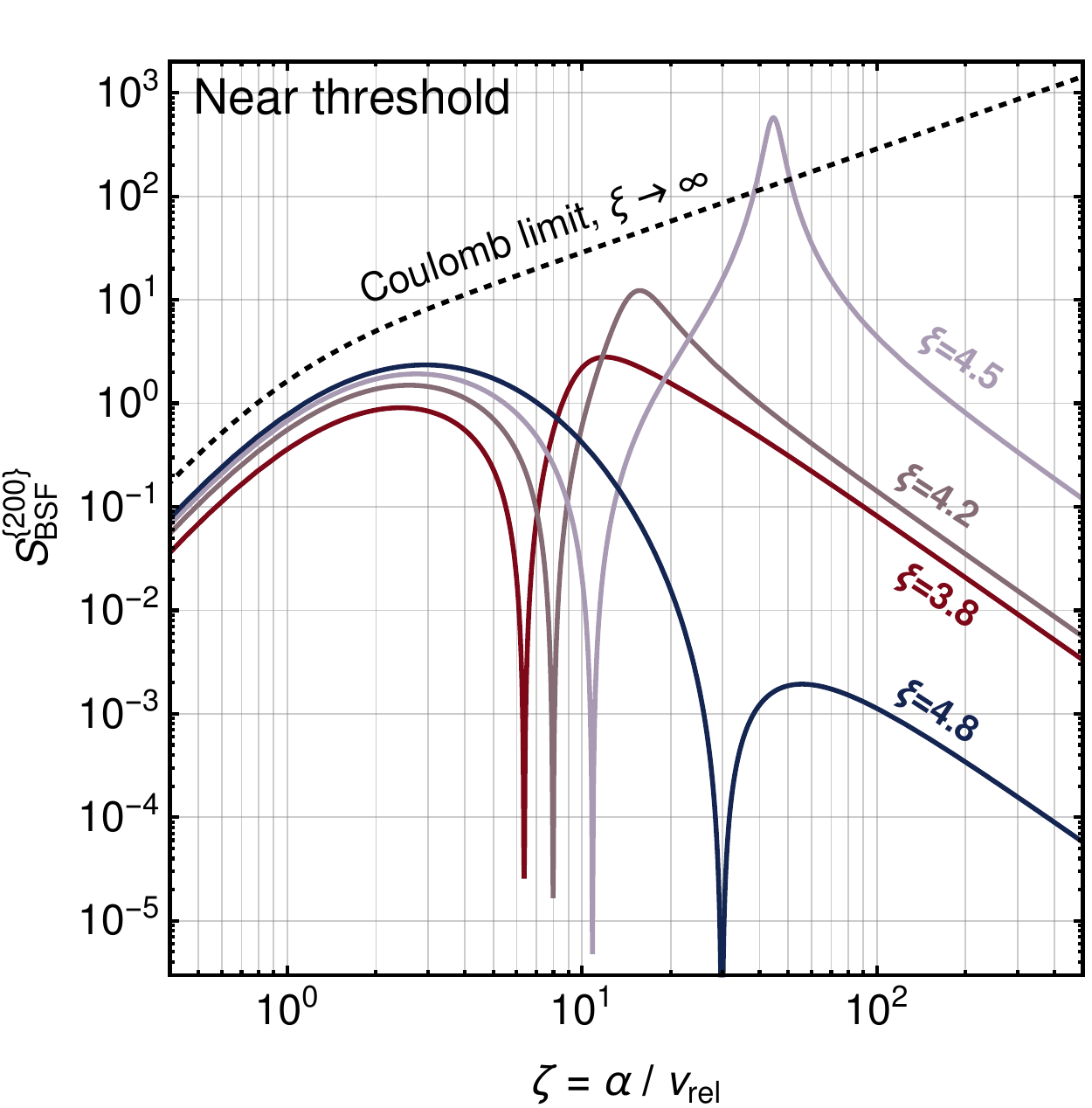}

\includegraphics[height=7cm]{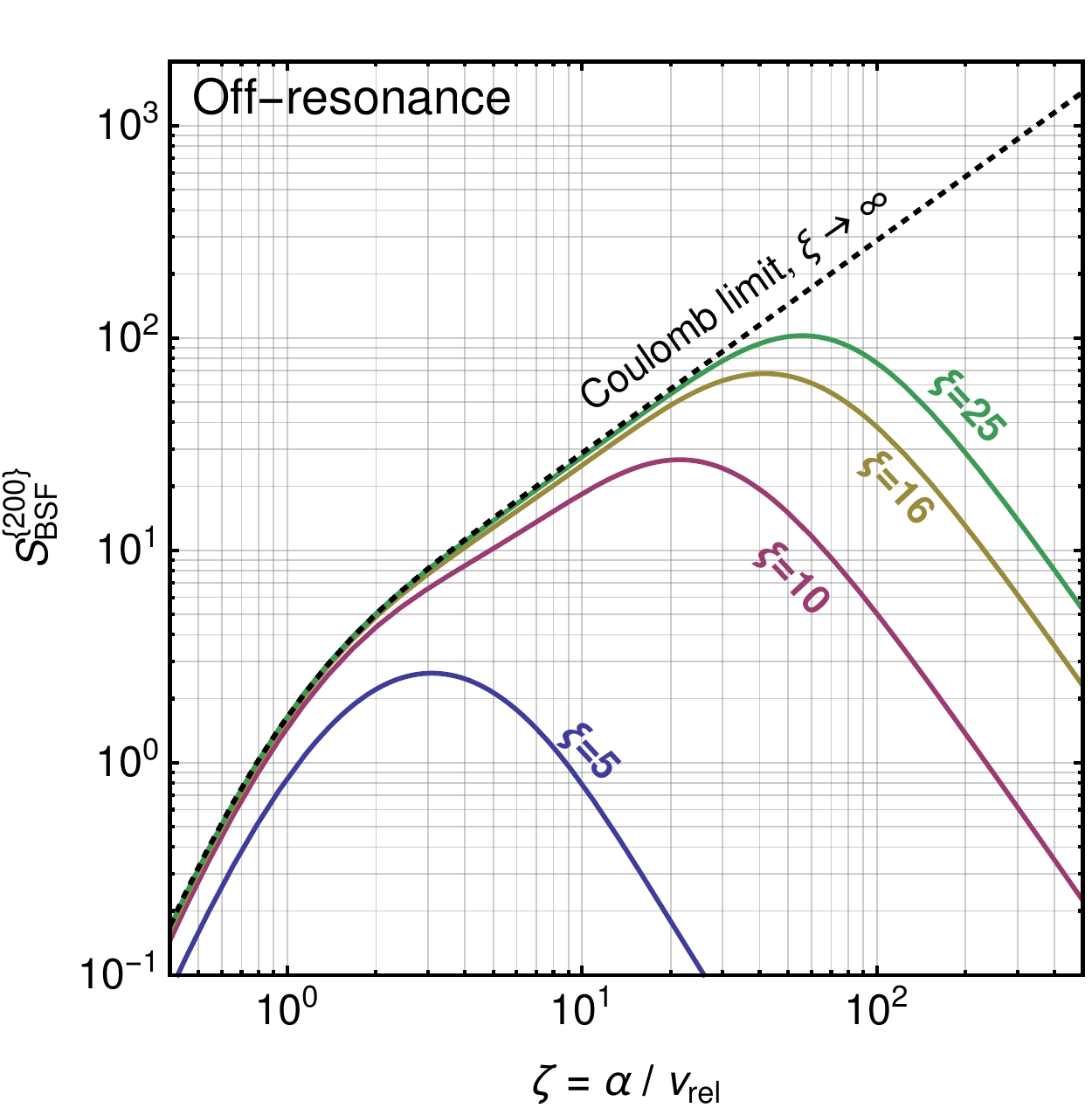}
~~~~~
\includegraphics[height=7cm]{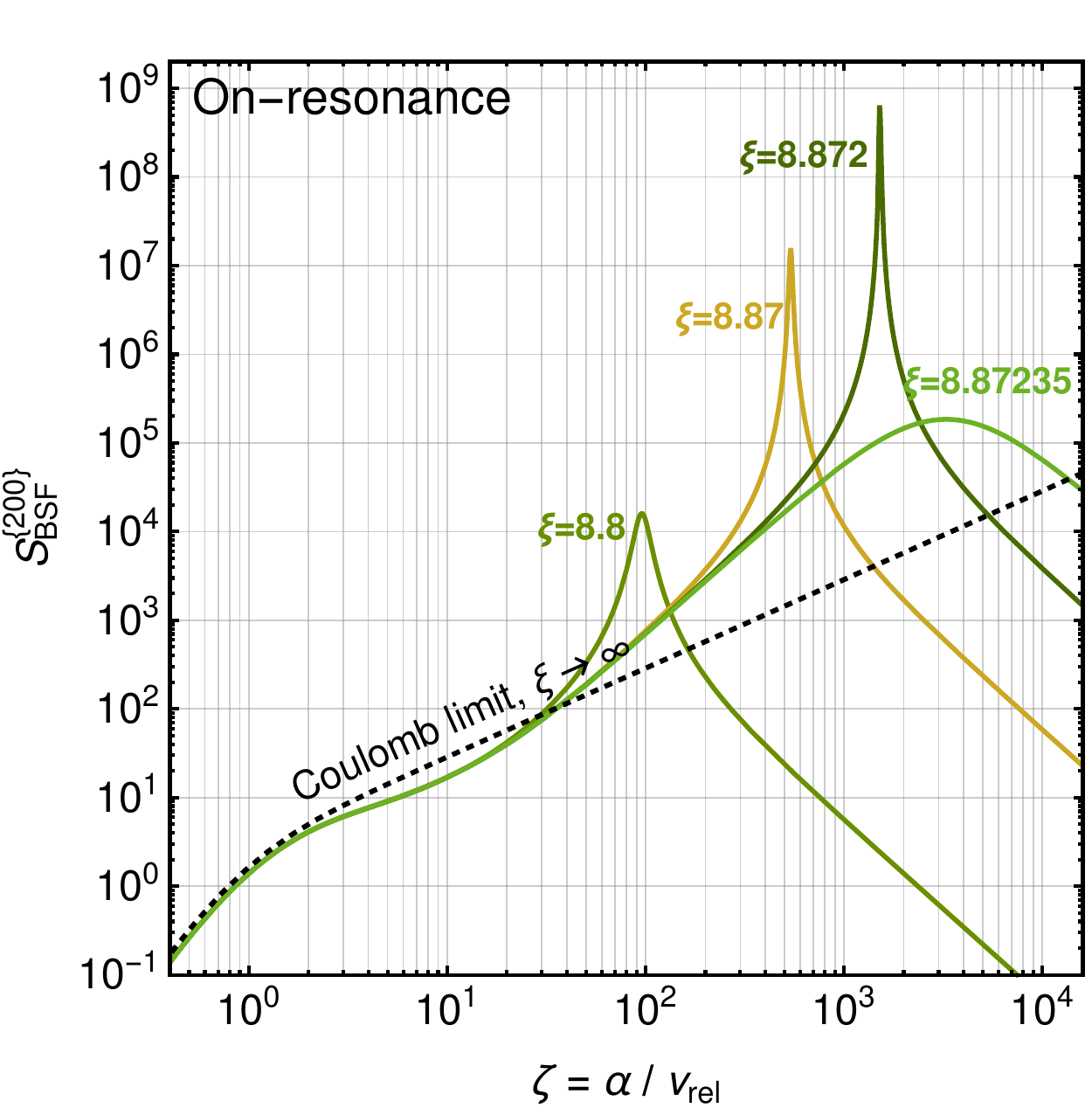}

\caption[]{\label{fig:VecMed_n=2_l=0}
Resonance structure and velocity dependence of the $S^{\{200\}}_{_{\rm BSF}}$ factor of the cross-section for the radiative formation of the first excited state $\{200\}$, with emission of a vector force mediator [cf.~eqs.~\eqref{eqs:VecMed_BSF_n00}]. In the bottom right panel, we show the velocity dependence near the $n=3$,~$\ell=1$ resonance. Besides the resonances, which we have already seen in the cross-section for capture to the ground state, the formation of excited states may feature anti-resonances.} 
\end{figure}

\begin{figure}[h!]
\centering

{\boldmath \bf Vector mediator: Capture into states 
with non-zero angular momentum $n=2, \ell=1$}

\includegraphics[height=7cm]{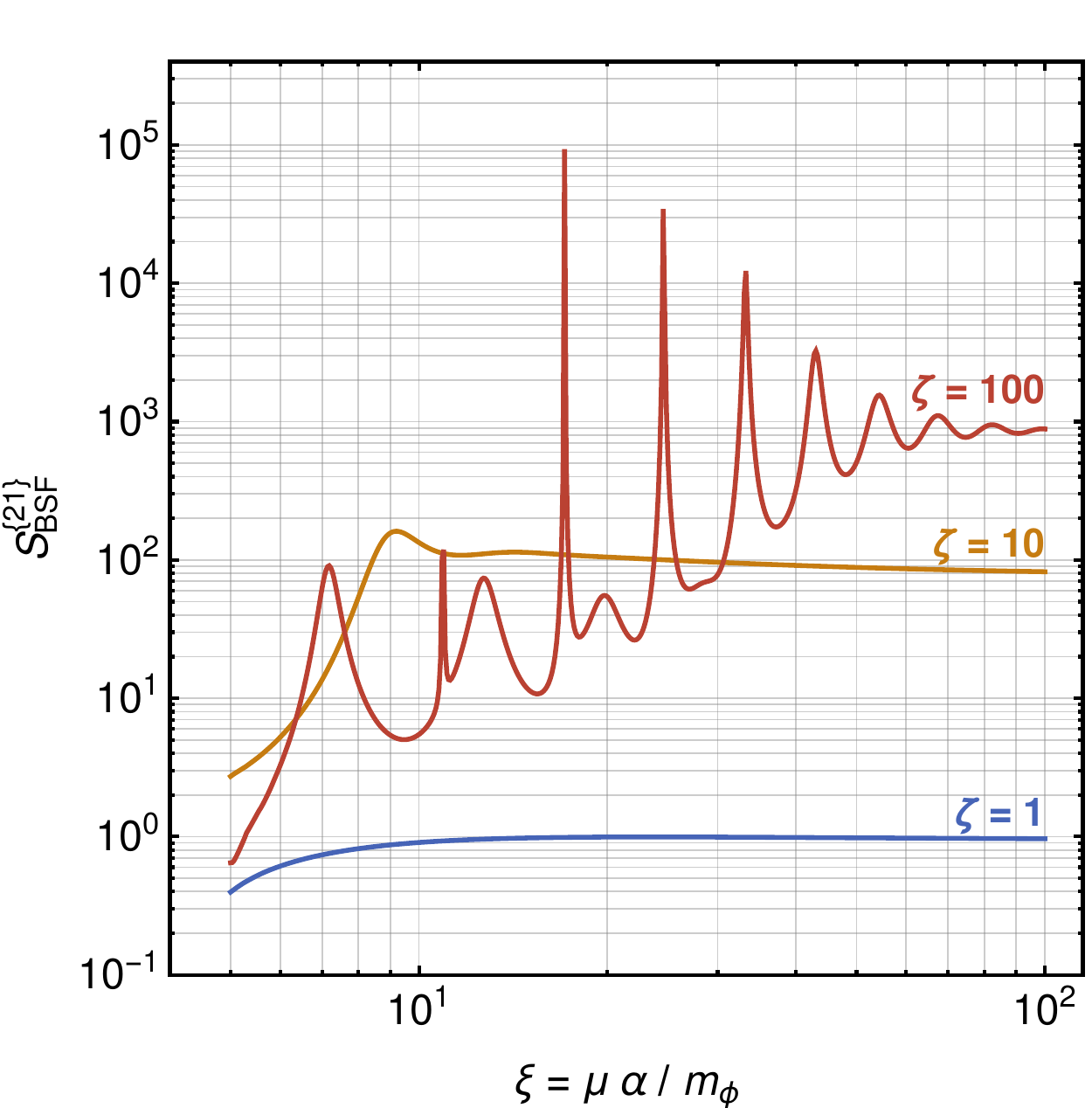}
~~~~
\includegraphics[height=7cm]{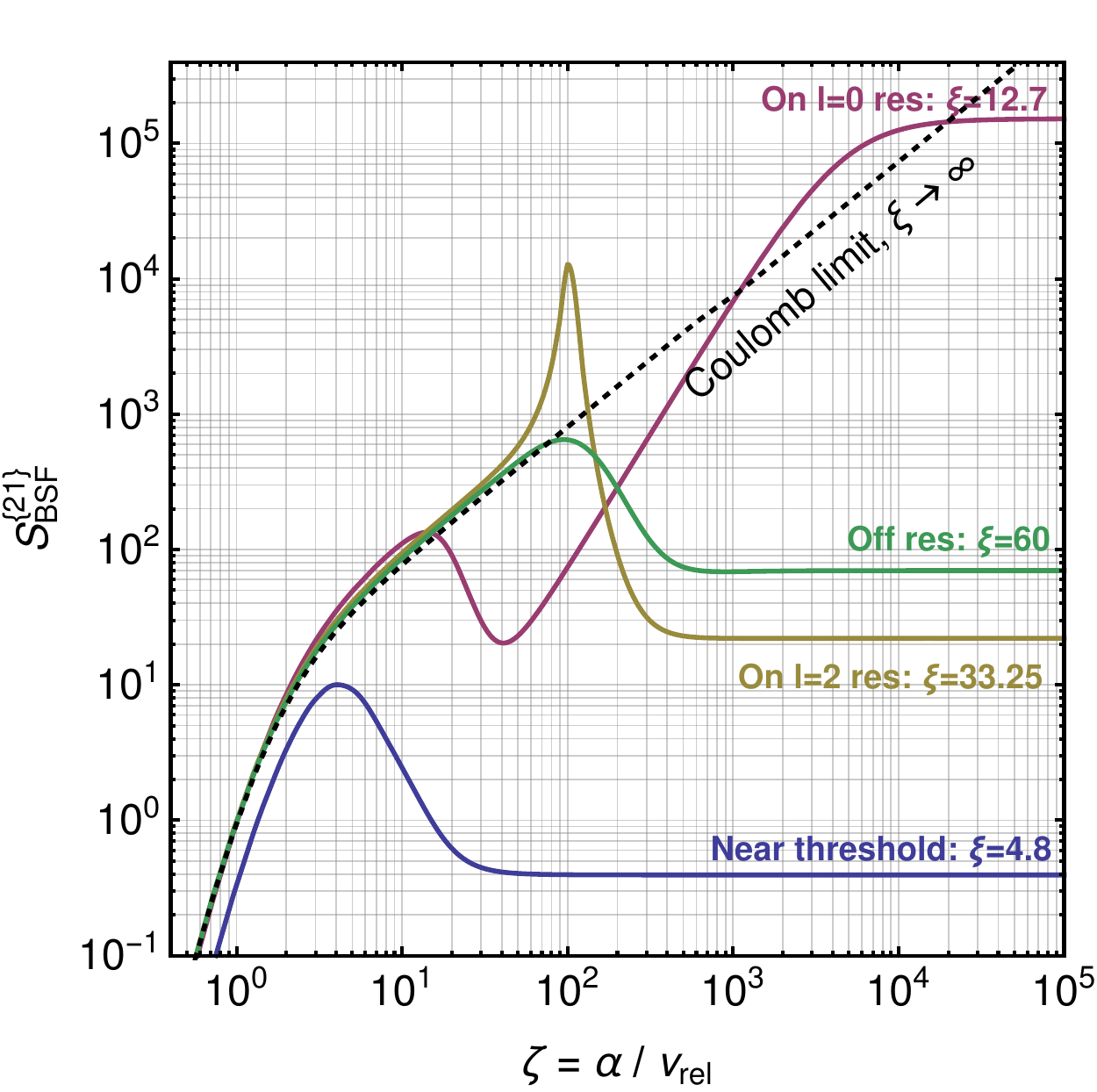}

\caption[]{\label{fig:VecMed_n=2_l=1}
Resonance structure and velocity dependence of the $S^{\{21\}}_{_{\rm BSF}}$ factor of the total cross-section for the radiative capture into any $n=2, \ell=1$ state, with emission of a vector force mediator [cf.~eqs.~\eqref{eqs:VecMed_BSF_n1}].

\smallskip

\emph{Left:} $S^{\{21\}}_{_{\rm BSF}}$ receives contributions from the $\ell = 0$ and the $\ell=2$ modes of the scattering state wavefunction. The $\ell=0$ mode dominates for small $\xi$, while the $\ell=2$ mode dominates for larger $\xi$ values (including the Coulomb regime) and gives rise to sharper resonances.

\smallskip

\emph{Right:}  At large $\zeta$, the contribution from the $\ell = 2$ mode decreases with decreasing velocity as $v_{\rm rel}^4$, while the contribution from the $\ell=0$ mode saturates to a constant value and dominates.

\smallskip

The superposition of two $\ell$ modes -- which have different velocity dependence and resonances -- gives rise to the various features of the total cross-section.
} 
\end{figure}

The BSF cross-sections exhibit a rich resonance structure, as seen in \cref{fig:VecMed_n=1_Xi,fig:VecMed_n=2_l=0,fig:VecMed_n=2_l=1}. The resonances are features of the scattering-state wavefunction, which determines the strength of the branch-cut singularity of the $X_1 - X_2$ 4-point function (see e.g.~\cite{Petraki:2015hla}). The resonances arise at the points of the parameter space where a pole of the 4-point function lives at zero energy, ${\cal E}_{n\ell} \to 0$, and thus overlaps with the branch-cut, which lives at ${\cal E}_{\vec k} \geqslant 0$ (see e.g.~Ref.~\cite[section~7.7]{Sakurai_QMbook}). The locations of resonances therefore denote the thresholds for the existence of genuine bound states with ${\cal E}_{n\ell} < 0$. These thresholds signify that if the mediator is massive, the potential has to be sufficiently strong for bound states to exist. For every bound-state energy level, this implies a minimum value for $\ks \equiv \mu \a / m_\vf$. In contrast, for a massless mediator, bound states exist independently of the strength of the coupling, as is the case with QED.

Unlike the Coulomb potential, the Yukawa potential does not imply the conservation of the Laplace-Runge-Lenz vector. The energy eigenvalues of the discrete spectrum depend on both the $n$ and $\ell$ quantum numbers (cf. \cref{app:BoundState WF num}). As a result, the resonances that appear in the scattering-state wavefunction depend on the $\ell$ mode.

Angular momentum conservation implies that the $\ell$ modes of the scattering-state wavefunction participating in a capture process depend on the angular momentum of the bound state formed, the spin of the emitted particle, and the orbital angular momentum of the final state (cf.~\cref{app:ConvIntegr_angular}). 
The formation of $\ell=0$ bound states with emission of a vector boson is dominated by the $\ell = 1$ mode of the scattering-state wavefunction [cf.~\cref{eq:VecMed_Sn00}]. On the other hand, the formation of $\ell=1$ bound states with vector emission is dominated by the $\ell = 0$~and~$\ell=2$ modes of the scattering-state wavefunction [cf.~\cref{eq:VecMed_Sn1,eq:VecMed_A0&A2}]. This explains the resonant patterns observed in \cref{fig:VecMed_n=1_Xi,fig:VecMed_n=2_l=0,fig:VecMed_n=2_l=1}.

In \cref{fig:VecMed_Sregularized_Vs_Xi}, we present the ratio of the factors $S\BSFgr$ and $S\BSF^{\{200\}}$ to $S\ann^{(1)}$, the Sommerfeld enhancement factor of $p$-wave annihilation processes,
\beq
S\ann^{(1)} \equiv \abs{
\frac{3\zeta}{2\kappa} \frac{d}{dr} \int_{-1}^1 d(\cos\theta_{\vec r})
P_1(\cos\theta_{\vec r}) \, \phi_{\vec k}(\vec{r})
}_{\vec r = 0}^2 = 
9\zeta^2 \abs{\frac{\chi_{|\vec{k}|,1}^{}(x)}{x^2}}_{x\to 0}^2 \, .
\label{eq:Sann1}
\eeq
As seen in \cref{eq:Sann1}, $S\ann^{(1)}$ depends only on the $\ell=1$ mode of the scattering-state wavefunction; in contrast $S\BSFgr$ and $S\BSF^{\{200\}}$ depend on the overlap of the wavefunctions of the scattering state, the bound state and the emitted vector boson, as seen in \cref{eqs:ConvIntegrals_Def}. The smoothness of the curves in \cref{fig:VecMed_Sregularized_Vs_Xi} attests that the resonances in the BSF cross-sections emanate from the scattering state. However, the capture to the $\{200\}$ bound state exhibits also anti-resonances (cf.~\cref{fig:VecMed_n=2_l=0}), whose origin is clearly not the scattering state alone. Instead, the anti-resonances arise from the convolution of the scattering-state wavefunction with the $\{200\}$ bound-state wavefunction that contains a node.

\subsubsection*{Velocity dependence away from the Coulomb regime}

At low velocities, $\vrel < m_\vf/\mu$, the inelastic cross-sections depart from the Coulombic scaling  $\s_{\rm inel} \vrel \propto 1/\vrel$. Their velocity dependence in this regime, is determined by the $\ell$ modes of the scattering-state wavefunction that participate in the corresponding processes.

For $\ks$ values away from resonances, the contribution of an $\ell$ mode of the scattering-state wavefunction to $\s_{\rm inel} \vrel$ scales as $\vrel^{2\ell}$ at $\vrel < m_\vf/\mu$. That is, as the velocity decreases, a contribution from the $\ell=0$ mode saturates to its Coulomb value at $\vrel \approx m_\vf/\mu$, while contributions from $\ell > 0$ modes drop below their Coulomb value at $\vrel \approx m_\vf/\mu$. The radiative capture into $\{n00\}$ bound states is dominated by the $\ell=1$ mode of the scattering-state wavefunction, and scales as $\vrel^2$ at $\vrel < m_\vf/\mu$, as can been seen in \cref{fig:VecMed_n=1_Zeta_NonRes,fig:VecMed_n=2_l=0} (bottom left panel).

For $\ks$ values near or on resonances, $\s_{\rm inel}\vrel$ grows faster than $1/\vrel$ at $\vrel < m_\vf/\mu$, and raises above its Coulomb value for a given velocity. This growth is sustained for a range of velocities that depends on how close $\ks$ is to a resonance value. Then, at sufficiently low velocities, the resonant growth stops, and the $\vrel^{2\ell}$ scaling ensues.\footnote{For $\ell=0$ and $\ks$ values exactly on resonance, $\s_{\rm inel} \vrel$ grows as $1/\vrel^2$ at $\vrel \lesssim m_\vf/\mu$, indefinitely. This is unphysical behaviour that needs to be regulated (see comments on unitarity below).} 
This behaviour can be observed in \cref{fig:VecMed_n=1_Zeta_Res,fig:VecMed_n=2_l=0} (bottom right panel).

If more than one $\ell$ modes of the scattering-state wavefunction participate in a process (at the same order in the coupling), then more complex patterns arise. This is the case with the radiative capture to $\ell=1$ bound states, which receives contributions from the $\ell=0$ and $\ell=2$ modes of the scattering-state wavefunction (cf.~\cref{fig:VecMed_n=2_l=1}).

Because of the different location of the resonances, as well as the different velocity dependence on- and off-resonance that the various BSF cross-sections exhibit, the relative strength of these processes at low velocities can be very different than in the Coulomb regime. In fact, at sufficiently low velocities, the capture to $\ell=1$ bound states, if kinematically allowed, should always dominate, since these are the only mono-photon transitions to which the $\ell=0$ mode of the scattering-state wavefunction participates.

\subsubsection*{Near-threshold behaviour}

As mentioned above, $\ks$ values near the thresholds for the existence of bound states imply resonances in the scattering-state wavefunction. These resonances then appear in the cross-sections of the processes in which this scattering state participates. However, the radiative formation of a bound state for $\ks$ values close to the threshold for the existence of the same bound state, is suppressed [cf.~\cref{fig:VecMed_n=1_Zeta_Threshold,fig:VecMed_n=2_l=0} (top right panel)]. In this limit, the bound state wavefunction becomes very spatially extended, and approaches zero. 

Note that this suppression is independent of the phase-space suppression due to the emission of a massive mediator. In fact, radiative BSF near threshold is kinematically allowed only if it occurs with emission of a nearly massless particle that therefore cannot be the force mediator itself.

\subsubsection*{Comparison of BSF and annihilation, for particle-antiparticle pairs}

At $\z \ll 1$, the BSF processes are rather suppressed. However, at $\z \gtrsim 1$, i.e.~in the regime where the Sommerfeld effect is important, BSF can be comparable to and even more significant than annihilation~\cite{vonHarling:2014kha}.

\begin{figure}[hp]
\centering

{\bf {}~~~~~~~ Vector mediator: Coulomb limit}

\includegraphics[height=9cm]{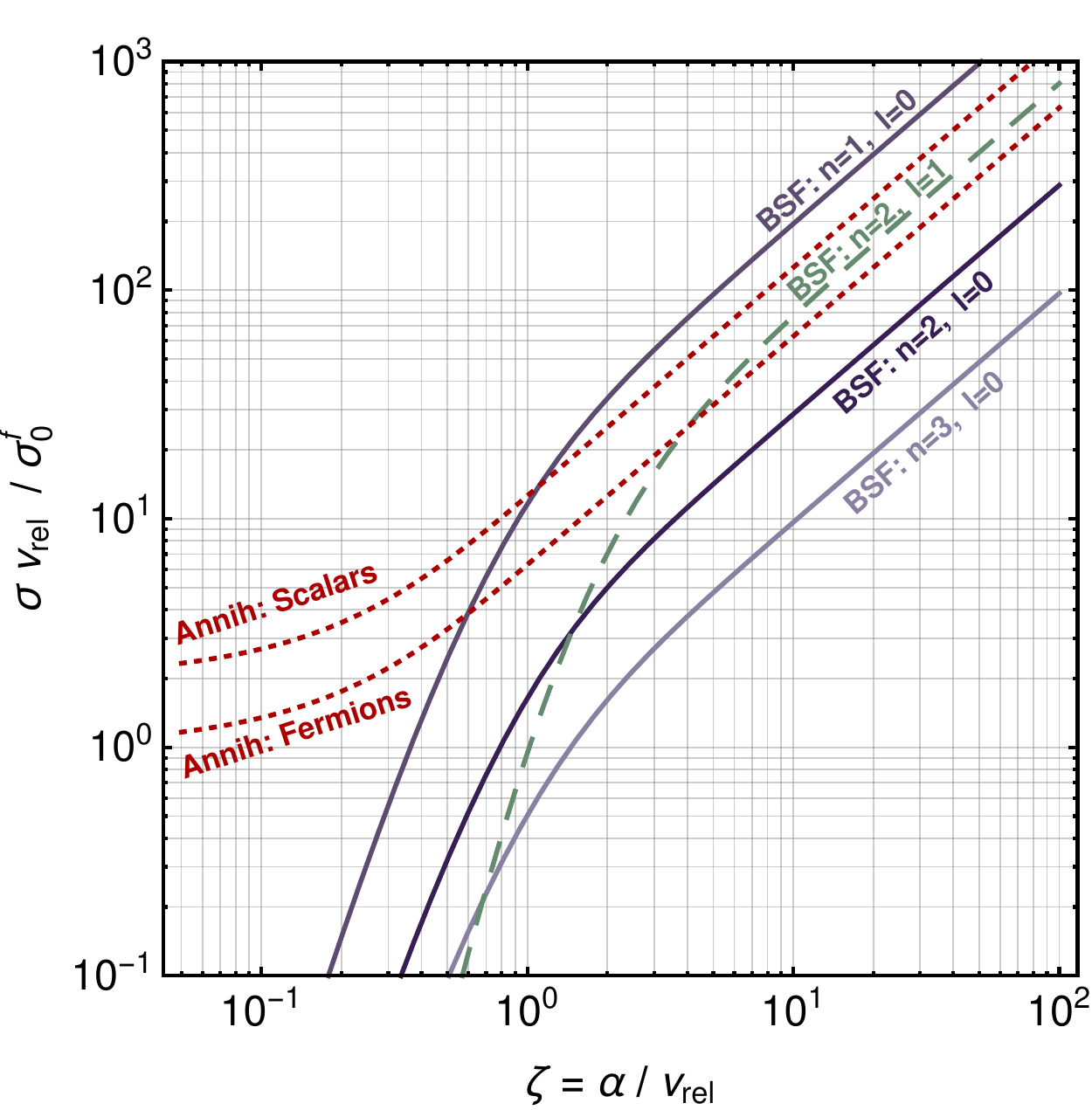}
\caption[]{
Velocity dependence of the cross-sections for the radiative formation of bound states, by a pair of particles charged under an unbroken dark U(1) force. The solid lines correspond to capture into zero angular momentum bound states, with $n$ denoting the principal quantum number. The dashed line is the total cross-section for capture into any $n=2, \, \ell=1$ state (i.e.~summed over all possible projections of the bound-state angular momentum on the $z$ axis). We also show the cross-sections for the annihilation of a particle-antiparticle pair of fermions  and scalars (red dashed lines).  All cross-section have been normalised to $\sigma_0^f \equiv \pi \alpha^2/(4\mu^2)$, where $\mu$ is the reduced mass of the interacting particles. $\sigma_0^f$ is the spin-averaged perturbative annihilation cross-section times relative velocity of a Dirac fermion-antifermion pair; for a complex scalar, this quantity is  $\sigma_0^s = 2\sigma_0^f$. 
The bound-state formation cross-sections do not depend on the spin of the incoming particles, which is conserved in the non-relativistic regime (at leading order), as particles get captured in a bound state. 
For particle-antiparticle pairs, at $\zeta = \alpha/v_{\rm rel} \gg 1$, the radiative capture into the ground state is the dominant inelastic process, $\sigma_{_{\rm BSF}}^{\{100\}} / \sigma_{\rm ann}^f 
\simeq 2\sigma_{_{\rm BSF}}^{\{100\}} /\sigma_{\rm ann}^s \simeq 3.13$ [cf.~eqs.~\eqref{eqs:VecMed_BSF_n00_Coul}, \eqref{eqs:VecMed_Ann} and table~\ref{tab:varrho n00 vector & scalar non-degen}].
\label{fig:VecMed_Coulomb}}
\end{figure}

In the Coulomb regime, we may easily compare the radiative capture to the ground state with the direct annihilation into force mediators, using \cref{eq:VecMed_BSF_n00_Coul,eq:VecMed_Ann}, 
\beq
  \frac{\s\BSFCgr}{\s\annC^f} 
= \frac{2\s\BSFCgr}{\s\annC^s} 
= \frac{2^9}{3^{}} \frac{\z^4}{(1+\z^2)^2} \, e^{-4\z {\rm arccot} \z} \, .
\label{eq:VecMed_BSFoverANN}
\eeq
At $\z \gtrsim 1$, this becomes $\s\BSFCgr / \s\annC^f = 2\s\BSFCgr / \s\annC^s \simeq 3.13$. In fact, the radiative formation of $n=2, \ell=1$ bound states is also faster than the annihilation of a fermion-antifermion pair, for $\z \gtrsim 4$. \Cref{fig:VecMed_Coulomb} compares the dominant BSF processes with annihilation, in the Coulomb limit. 
Since BSF is the dominant inelastic process for particle-antiparticle pairs, whose bound states are unstable and decay into radiation, the BSF via emission of a vector boson can significantly affect the relic density\footnote{
Reference~\cite{An:2016gad} argued that BSF cannot affect the DM relic density due to the rapid ionization of the bound states. However, ionisation was fully taken into account in a proper analysis in Ref.~\cite{vonHarling:2014kha}, which employed a set of coupled Boltzmann equations that incorporate bound-state formation, ionisation and decay processes. This analysis showed that, in a dark QED scenario, BSF reduces the DM density by a factor greater than 2 if the DM mass is $m_{_{\rm DM}} \gtrsim 15$~TeV (and up to factor of 4 for $m_{_{\rm DM}} \gtrsim 100$~TeV). As pointed out in~\cite{vonHarling:2014kha}, BSF depletes efficiently the DM density only after the ionisation rate drops below the decay rate of the bound states. The detailed timeline shows that this occurs around or before freeze-out for $m_{_{\rm DM}} \gtrsim 20$~TeV. Note that Sommerfeld-enhanced processes -- either annihilations or BSF -- remain important even after the DM freeze-out (conventionally defined as the time of departure of the DM density from its equilibrium value). This explains the significant effect of BSF on the relic density even for $m_{_{\rm DM}} \lesssim 20$~TeV.}~\cite{vonHarling:2014kha}
and enhance the indirect detection signals~\cite{Pospelov:2008jd, An:2016gad, Kouvaris:2016ltf, Cirelli:2016rnw} of symmetric DM.

Away from the Coulomb regime, the comparison of annihilation and BSF becomes more complex. \Cref{figs:VecMed_n=1} illustrate the main features, which are related to the discussion offered above, and which we now summarise: 
\bit
\item
The formation of zero-angular-momentum bound states and the direct annihilation into radiation exhibit resonances at different locations ($\ks$ values), as clearly seen in \cref{fig:VecMed_n=1_Xi}. This is due to the different $\ell$ modes of the scattering-state wavefunction that contribute to each process: $\ell=1$ for the former and $\ell=0$ for the latter. In fact, the locations of the $\ell=1$ resonances exhibit a mild velocity dependence, in contrast to the $\ell=0$ resonances. 

\item
The $\ell$ modes of the scattering-state wavefunction also determine the velocity dependence away from the Coulomb limit. At low velocities, $\s\ann \vrel$ saturates to a constant value, while $\s\BSFgr \vrel$ scales as $\vrel^2$. This scaling, together with the relative strength of the two processes in the Coulomb regime that we discussed above, imply that the radiative capture to the ground state dominates over annihilation within a finite range of velocities, as can be seen in \cref{fig:VecMed_n=1_Zeta_NonRes,fig:VecMed_n=1_Zeta_Res}. This range is roughly $1 \lesssim \z \lesssim \ks$ (or equivalently $m_\vf \lesssim \m \vrel \lesssim \m \a$) for non-resonant $\ks$ values, but it may extend to much lower velocities (by orders of magnitude) for $\ks$ values near an $\ell=1$ resonance.

\item
Near the threshold for the existence of the ground state, i.e.~for $\ks \approx 1$, the radiative BSF is always suppressed with respect to its Coulomb value, while annihilation is on resonance. This behaviour can be seen in \cref{fig:VecMed_n=1_Zeta_Threshold}. (We repeat that in this regime, BSF is kinematically allowed to occur only via emission a nearly massless particle, which cannot therefore be the force mediator itself.)

\eit

\subsubsection*{Partial-wave unitarity}

Partial-wave unitarity implies an upper bound on the inelastic cross-sections, which in the non-relativistic regime is~\cite{Griest:1989wd},
\beq 
(\s_{{\rm uni}})_J^{} \, \vrel = \frac{(2J+1)\p}{\mu^2 \vrel} \, , 
\label{eq:SigmaUni} 
\eeq 
where $J$ is the partial wave. 

In the Coulomb limit, inelastic cross-sections have the same velocity scaling as \cref{eq:SigmaUni}; setting $\s_{\rm inel} \leqslant \s_{\rm uni}$ then implies an upper bound on $\a$ that does not depend on any other physical parameter, and is roughly $\a \lesssim 0.85$~\cite{vonHarling:2014kha,Baldes:2017gzw}. Around this upper bound on $\a$, higher order corrections (of perturbative or non-perturbative origin) need to be considered. Notably, $\a\sim 0.85$ is well below the naive perturbativity limit, $\a \sim 4\p$.

However, the resonances that appear away from the Coulomb limit, imply that the leading-order computations presented here, may violate the unitarity bound even at much lower values of $\a$ that now also depend on $\z$ and $\ks$. Evidently, the peaks of the resonances can be unphysical.

The resonances can be regulated by taking into account the short-range elastic scattering of the interacting particles (see e.g.~\cite{Kong:1998sx,Kong:1999tw}). For particle-antiparticle pairs, the short-range inelastic scattering -- in particular the annihilation processes -- may also contribute to taming the unphysical behaviour~\cite{Blum:2016nrz}.


\bigskip\bigskip\bigskip\bigskip

\begin{subfigures}

\label{fig:VecMed_n=1}
\label[pluralfigure]{figs:VecMed_n=1}

\begin{figure}[h!]
\centering

{\bf Vector mediator: Resonances}

\includegraphics[width=0.46\linewidth]{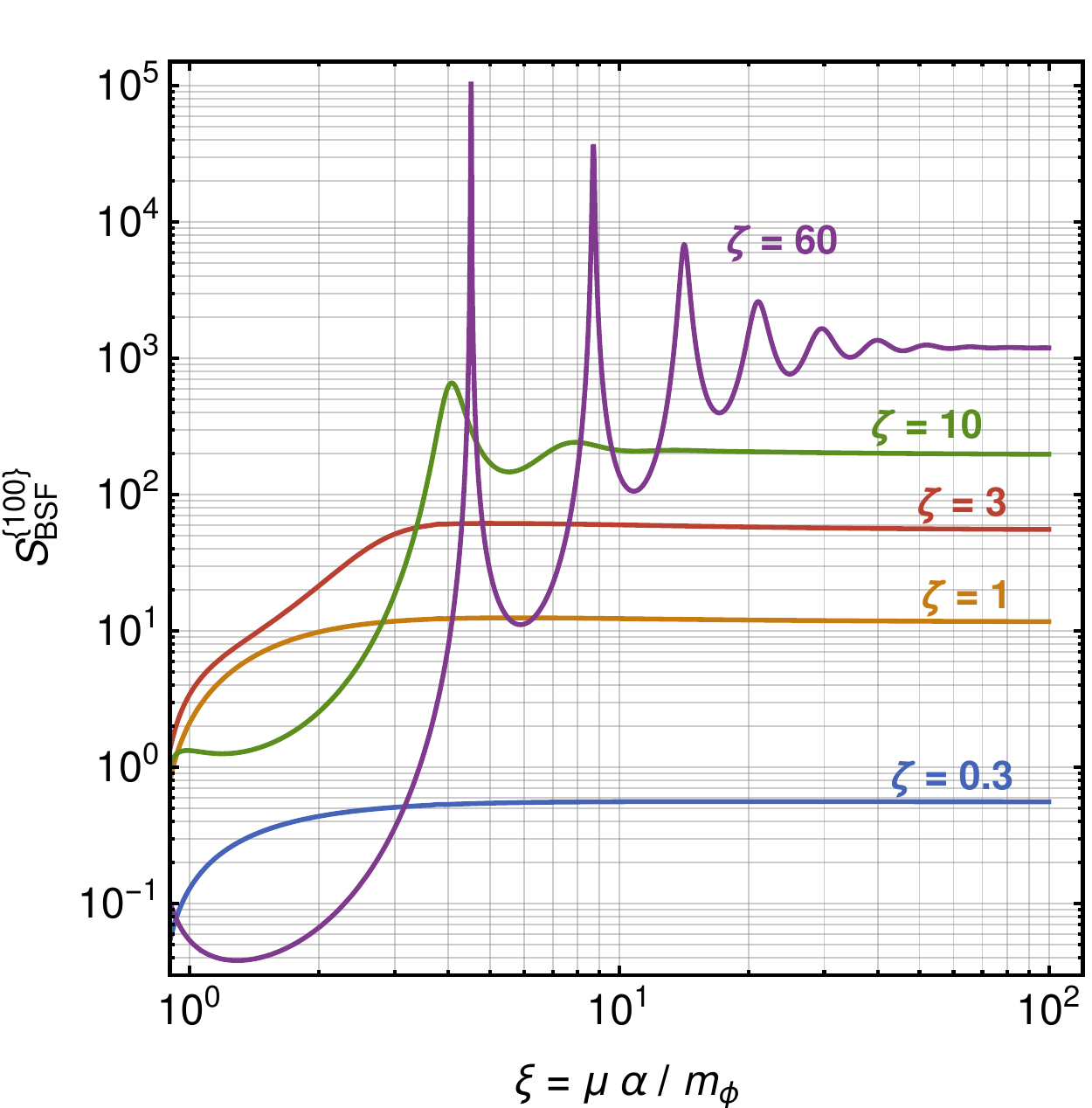}~~~~~
\includegraphics[width=0.46\linewidth]{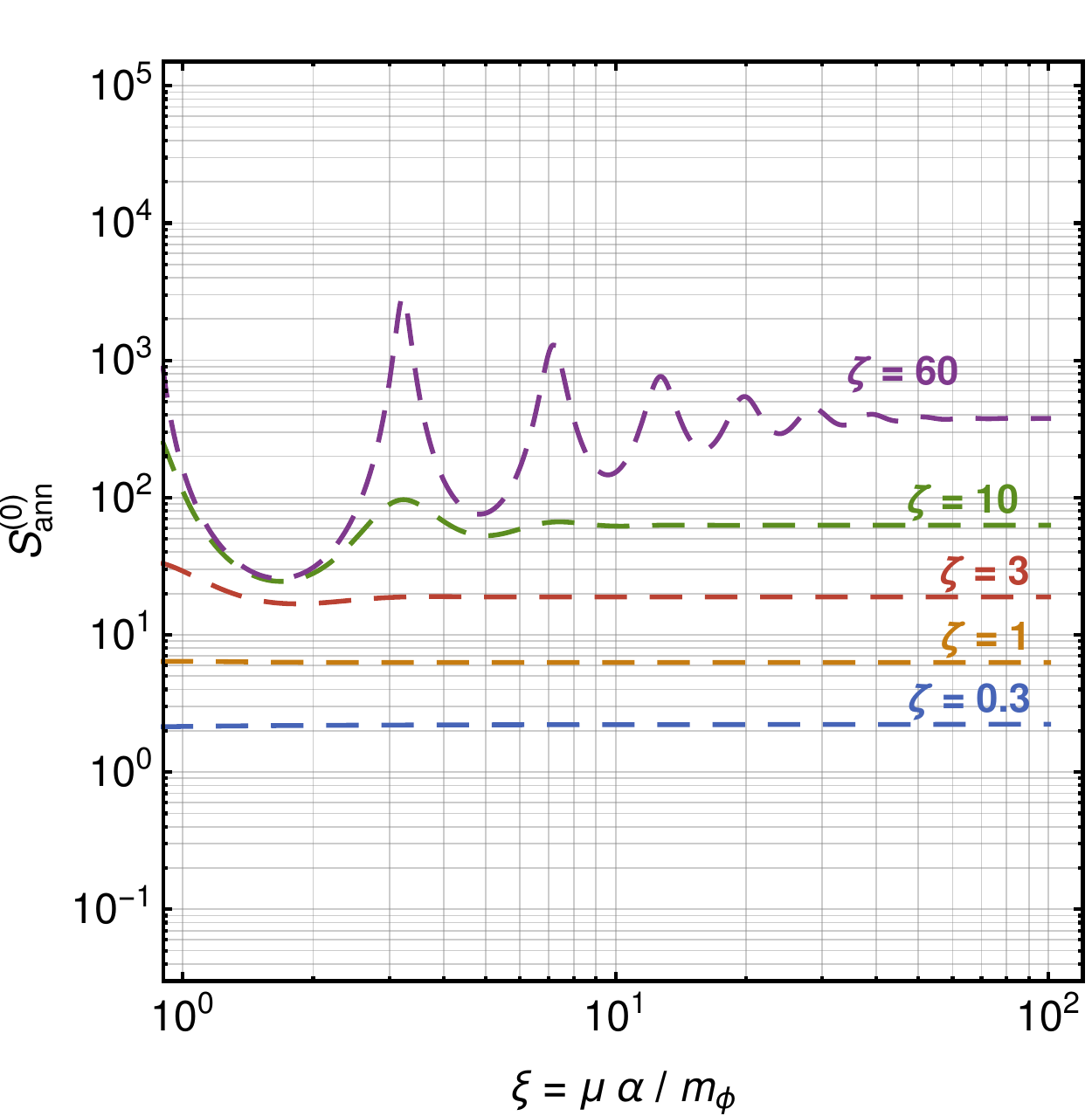}

\parbox[c]{0.46\linewidth}{
\includegraphics[width=\linewidth]{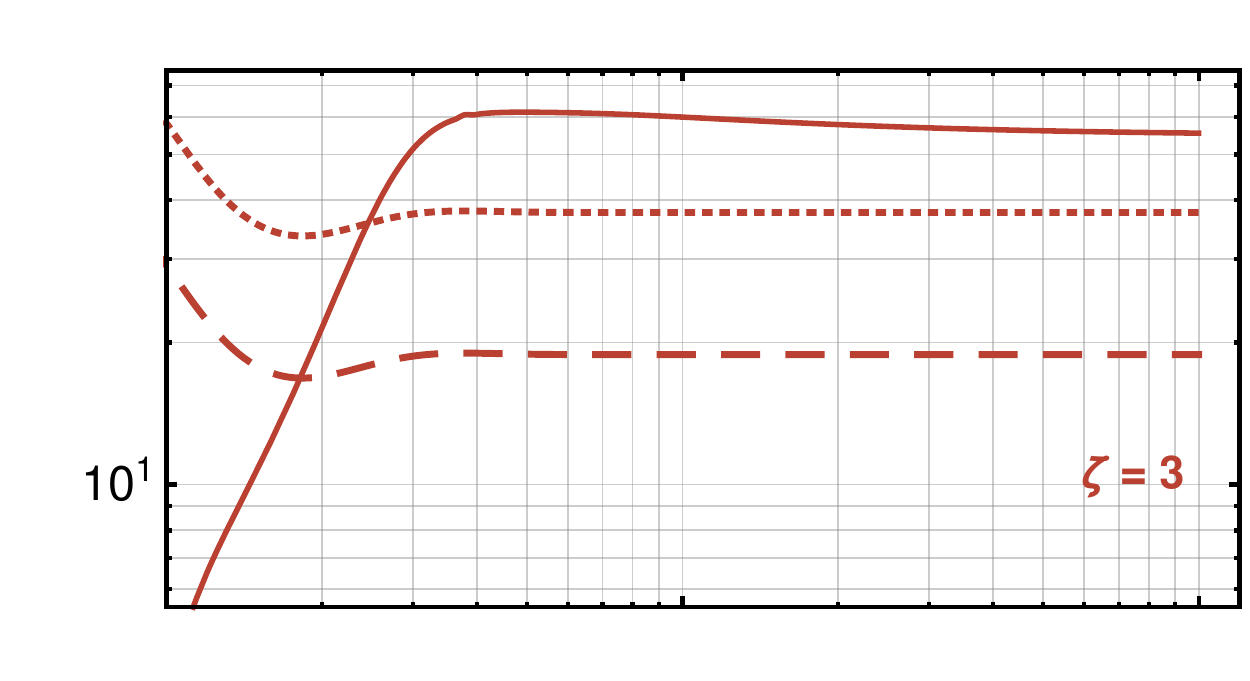}

\vspace{-0.7cm}
\includegraphics[width=\linewidth]{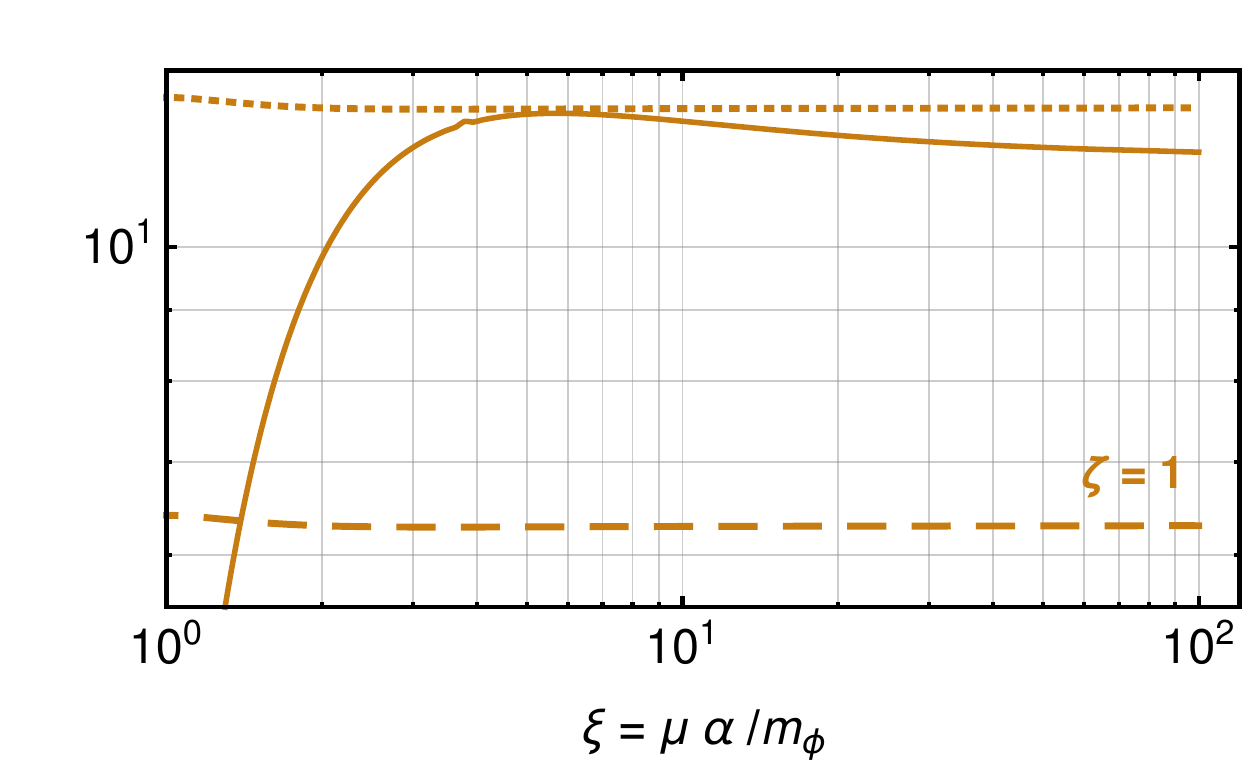}
}
~~~
\parbox[c]{0.46\linewidth}{
\includegraphics[width=\linewidth]{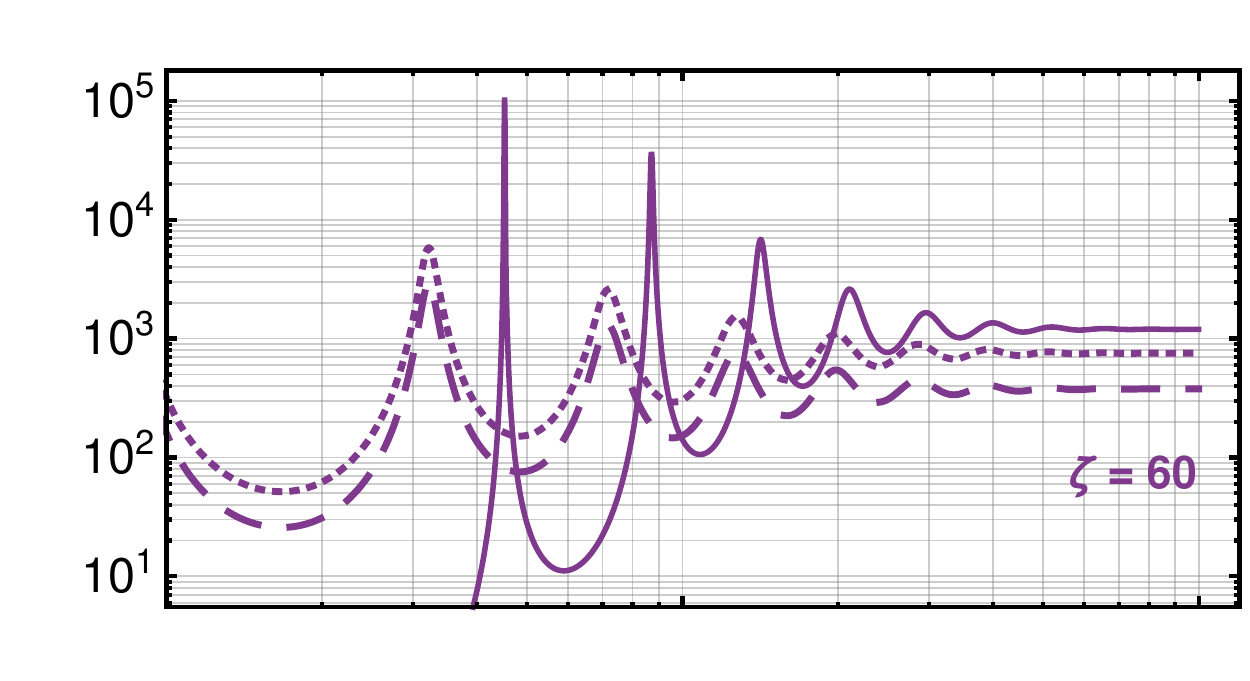}

\vspace{-0.7cm}
\includegraphics[width=\linewidth]{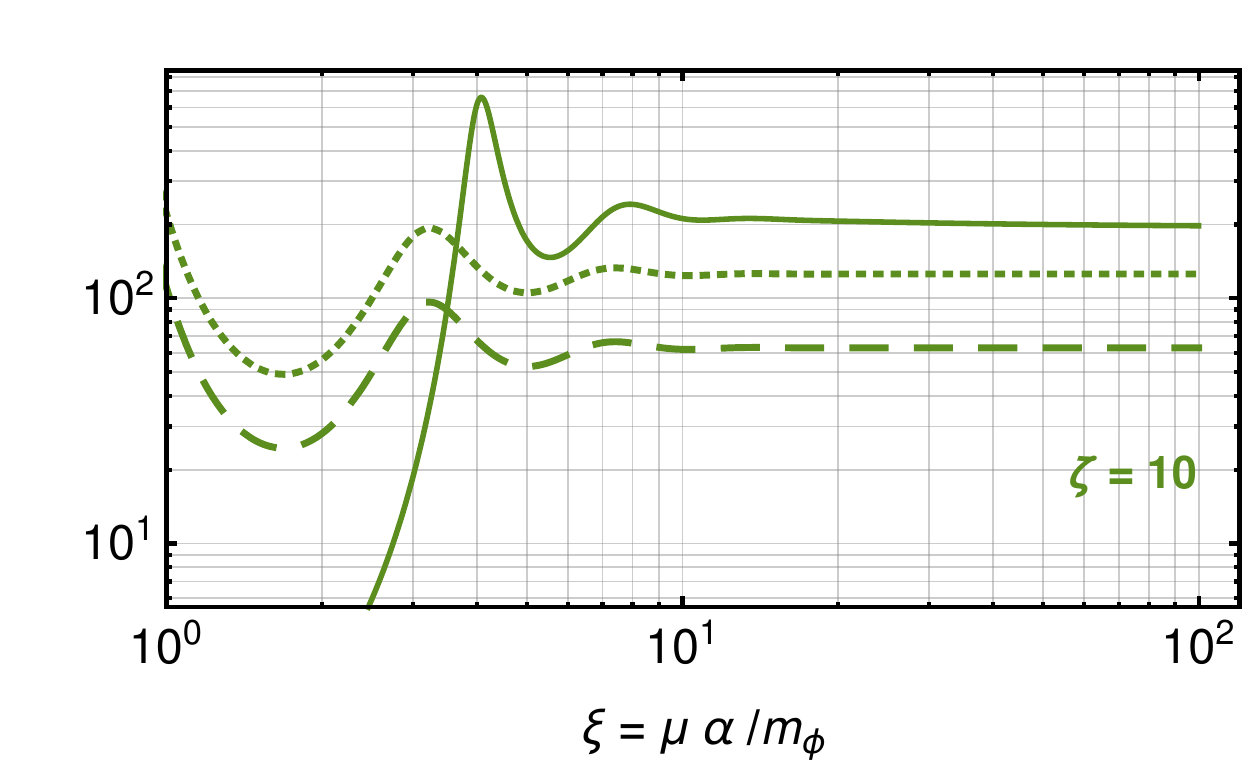}
}
\caption[]{\label{fig:VecMed_n=1_Xi}
Resonances due to bound states at threshold (zero binding energy) appear at discrete values of $\xi = \mu \alpha / m_\varphi$, with a magnitude that increases at low energies (large $\zeta = \alpha/v_{\rm rel}$). They arise from the wavefunction of the (initial) two-particle scattering state, and depend on the $\ell$ modes that contribute to the process of interest.

\smallskip

\emph{Top left:}
The factor $S_{_{\rm BSF}}^{\{100\}}(\zeta,\xi)$, appearing in the cross-section for capture into the ground state, with emission of a vector force mediator. The $\ell = 1$ mode of the scattering state wavefunction yields the dominant contribution to this process, and results in sharp resonances whose precise location has a mild $\zeta$ (velocity) dependence [cf.~eqs.~\eqref{eqs:VecMed_BSF_n00}, \eqref{eqs:VecMed_BSF_n00_Coul}].

\smallskip

\emph{Top right:} The Sommerfeld enhancement factor $S_{\rm ann}^{(0)}$ of the annihilation of a particle-antiparticle pair into two dark photons. The $\ell = 0$ mode of the scattering state wavefunction yields the dominant contribution to this process. The resonances are less sharp than those of the $\ell= 1$ mode, and their location does not depend on $\zeta$ [cf.~eqs.~\eqref{eqs:VecMed_Ann}].

\smallskip

\emph{Bottom:} Comparison of $S_{_{\rm BSF}}^{\{100\}}$ (solid lines), $S_{\rm ann}^{(0)}$ (dashed lines), $2S_{\rm ann}^{(0)}$ (dotted lines), for values of $\zeta$ shown in the top panels. 
For a particle-antiparticle pair, this is a comparison between the strength of the radiative capture into the ground state (ignoring the phase-space suppression), and the spin-averaged annihilation of a pair of fermions, or the annihilation of a pair of scalars, respectively. The location and the magnitude of the $S_{_{\rm BSF}}^{\{100\}}$ and $S_{\rm ann}^{(0)}$ resonances is clearly seen to be different.
}
\end{figure}

\begin{figure}
\centering

{\boldmath \bf Vector mediator: 
$\xi$ values away from $\ell = 0$ and $\ell = 1$ resonances}

\includegraphics[width=0.46\linewidth]{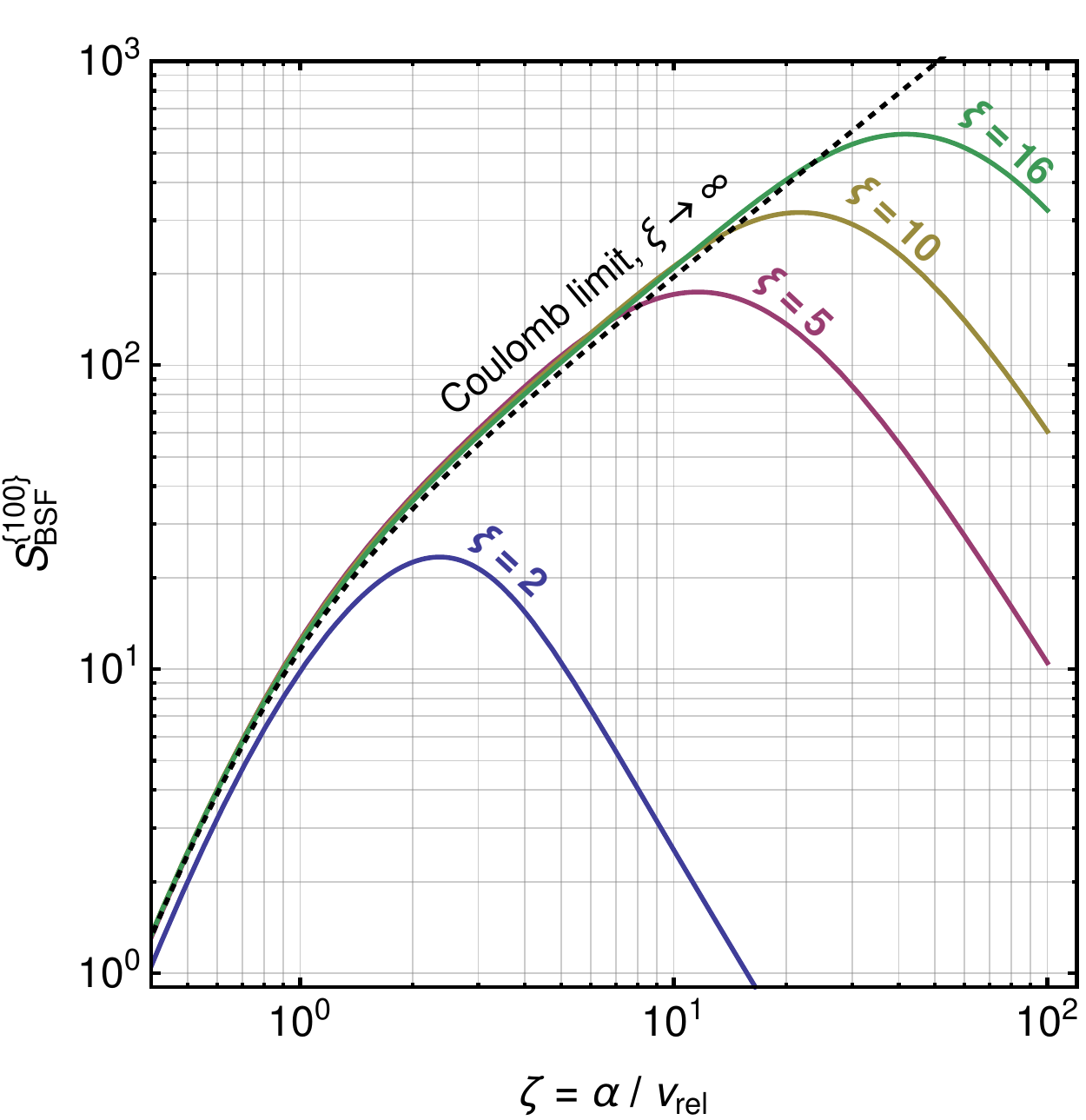}~~~~~
\includegraphics[width=0.46\linewidth]{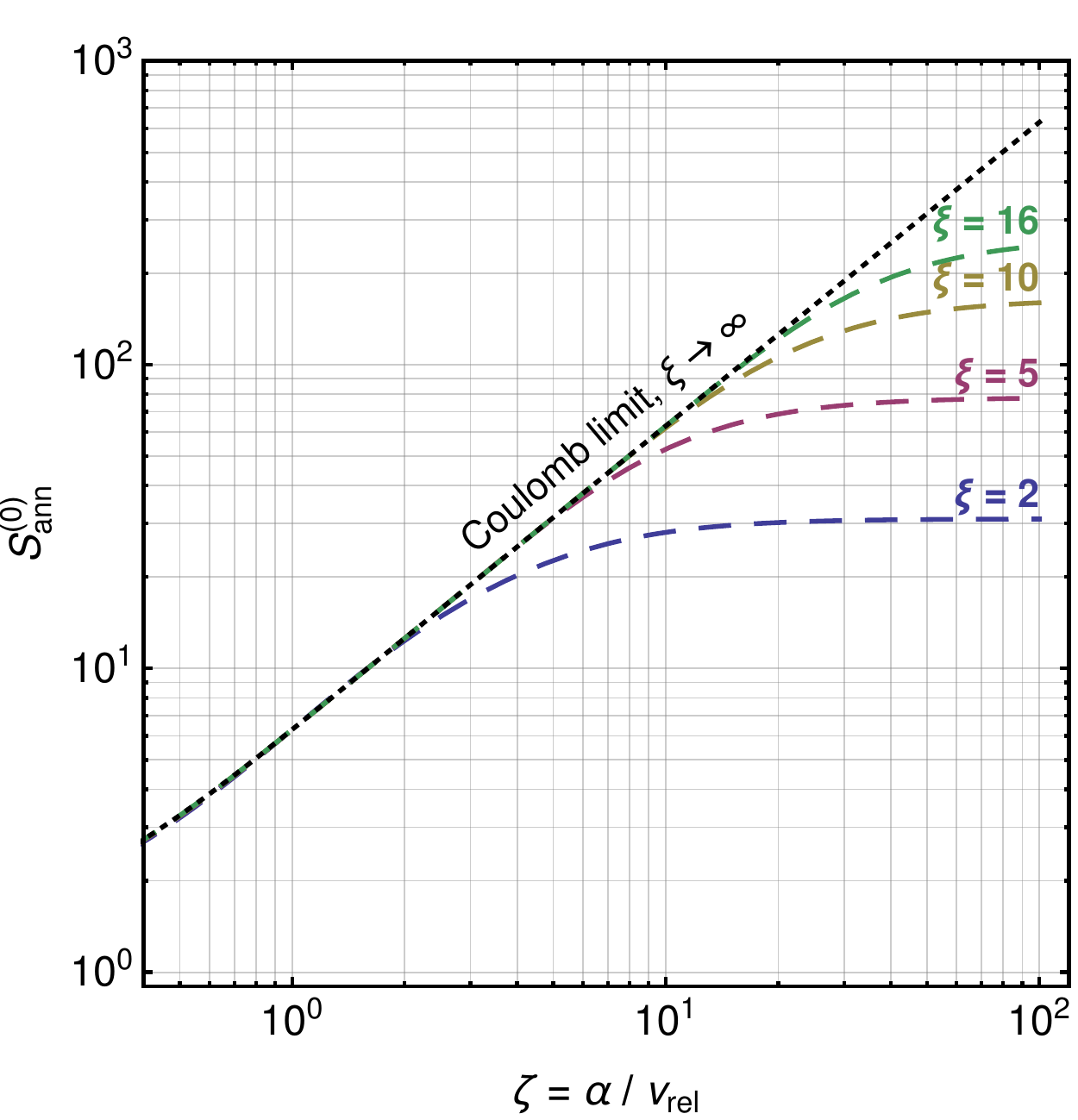}

\smallskip

\parbox[c]{0.46\linewidth}{
\includegraphics[width=\linewidth]{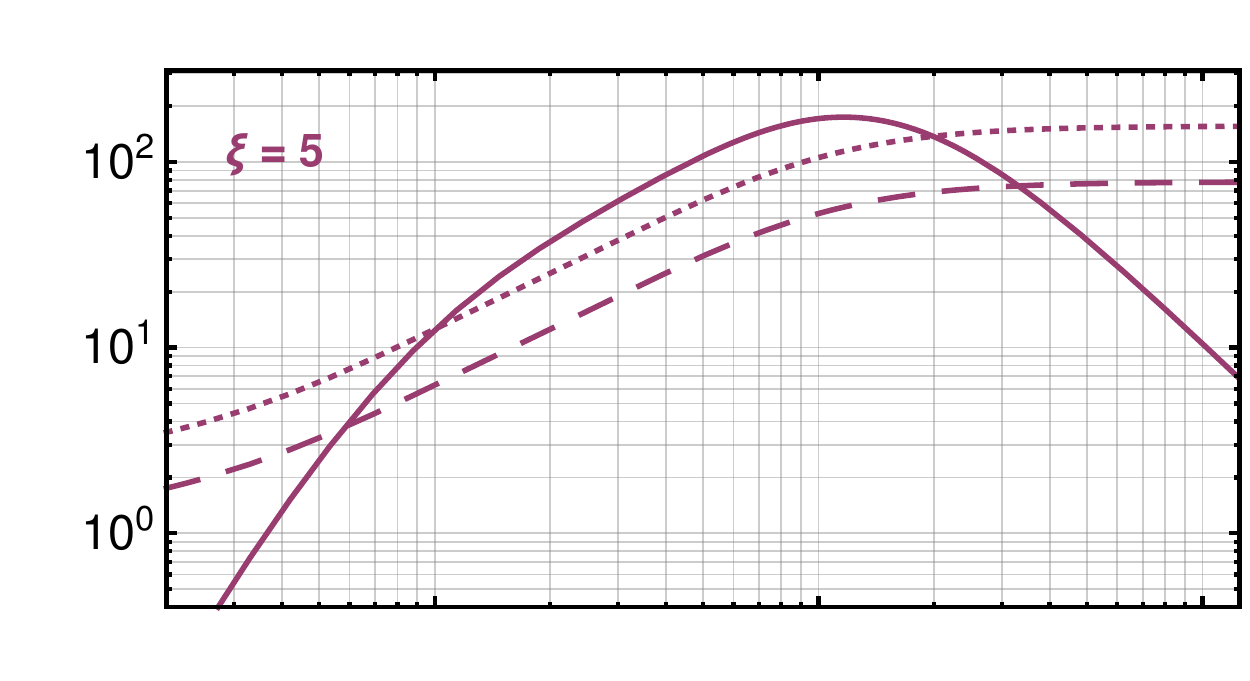}

\vspace{-0.7cm}
\includegraphics[width=\linewidth]{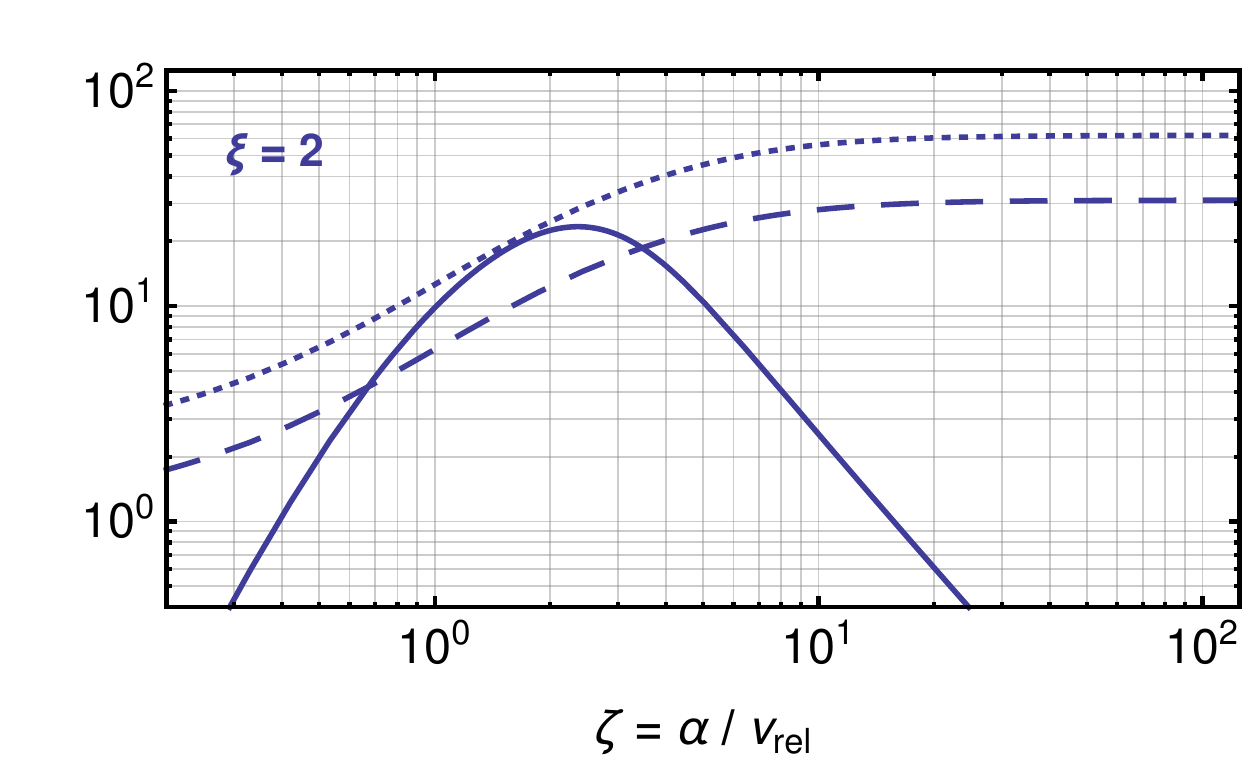}
}
~~~~
\parbox[c]{0.46\linewidth}{
\includegraphics[width=\linewidth]{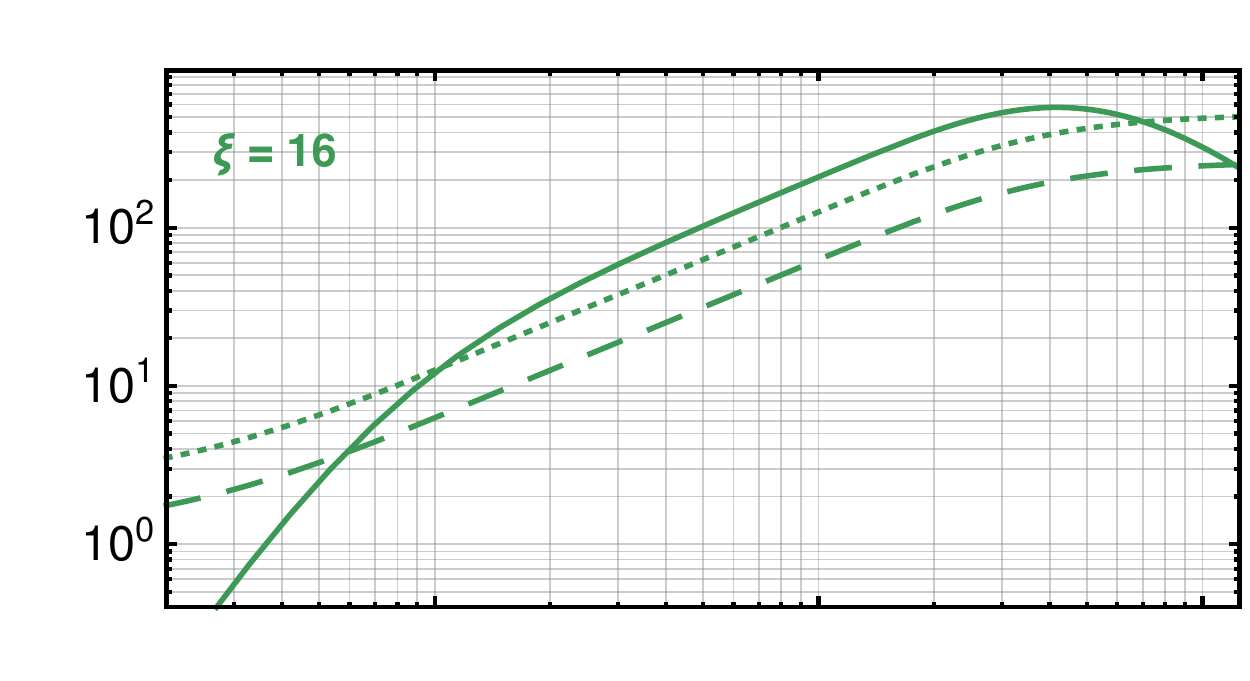}

\vspace{-0.7cm}
\includegraphics[width=\linewidth]{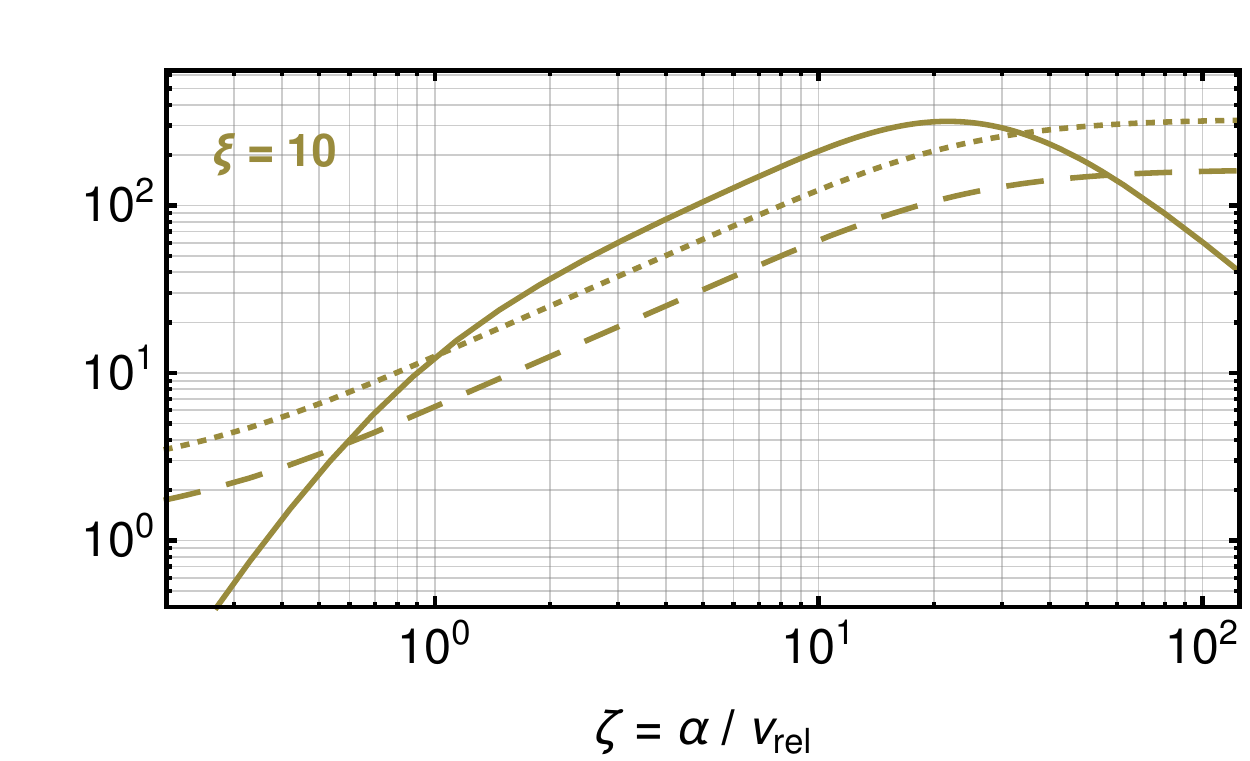}
}
\caption[]{\label{fig:VecMed_n=1_Zeta_NonRes}
Velocity dependence of $S_{_{\rm BSF}}^{\{100\}}$ and $S_{\rm ann}^{(0)}$, for non-resonant values of $\xi = \mu \alpha/m_\varphi$. When the momentum transfer between the interacting particles drops below the mediator mass, $\mu v_{\rm rel} \lesssim m_\vf$ (equivalently, $\zeta \gtrsim \xi$), the velocity dependence of the cross-sections departs from the Coulombic behaviour. 

\smallskip
\emph{Top left:} The cross-section for radiative capture into the ground state with emission of a vector mediator drops approximately as $\zeta^{-2} \propto v_{\rm rel}^2$ (see also fig.~\ref{fig:VecMed_Sregularized_Vs_Zeta}).

\smallskip 
\emph{Top right:} For $s$-wave annihilation processes, the Sommerfeld enhancement saturates to a constant value.

\smallskip
\emph{Bottom:} 
Comparison of the velocity dependence of $S_{_{\rm BSF}}^{\{100\}}$ (solid lines), $S_{\rm ann}^{(0)}$ (dashed lines), $2S_{\rm ann}^{(0)}$ (dotted lines), for various values of $\xi$. For a non-zero mediator mass, and for a particle-antiparticle pair, the bound-state formation cross-section is larger than the annihilation cross-section within a finite range of velocities, roughly
$m_\vf \lesssim \mu v_{\rm rel} \lesssim \mu \alpha$, or equivalently $1 \lesssim \zeta \lesssim \xi$. (This comparison ignores the phase-space suppression of the bound-state formation process.)
}
\end{figure}

\begin{figure}
\centering
{\bf Vector mediator}
\\
{\boldmath \bf $\xi$ values near $n = 2$, $\ell = 1$ resonance}~~~~~~
{\boldmath \bf $\xi$ values near $n = 2$, $\ell = 0$ resonance}

\includegraphics[width=0.45\textwidth]{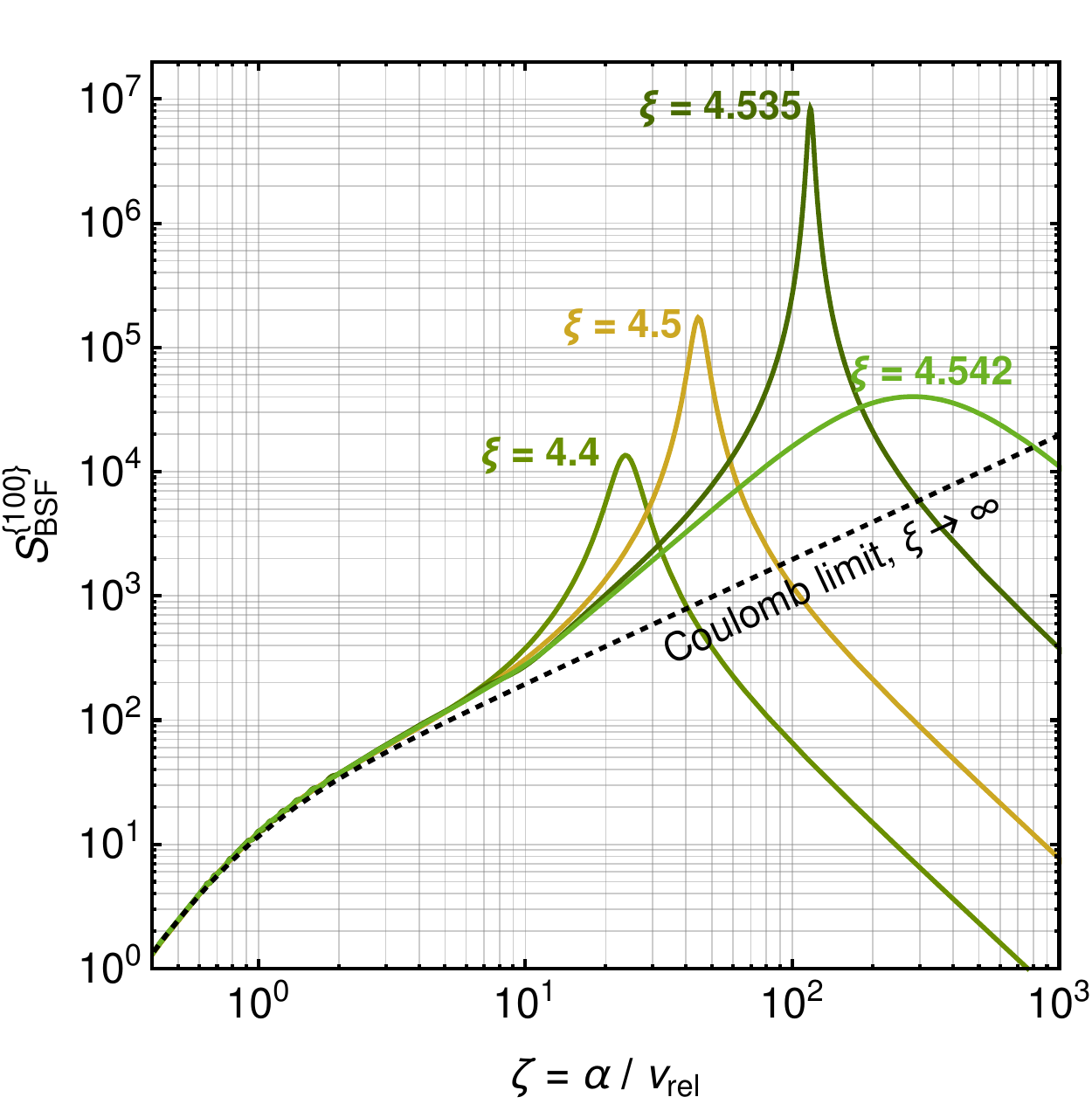}~~~~~
\includegraphics[width=0.45\textwidth]{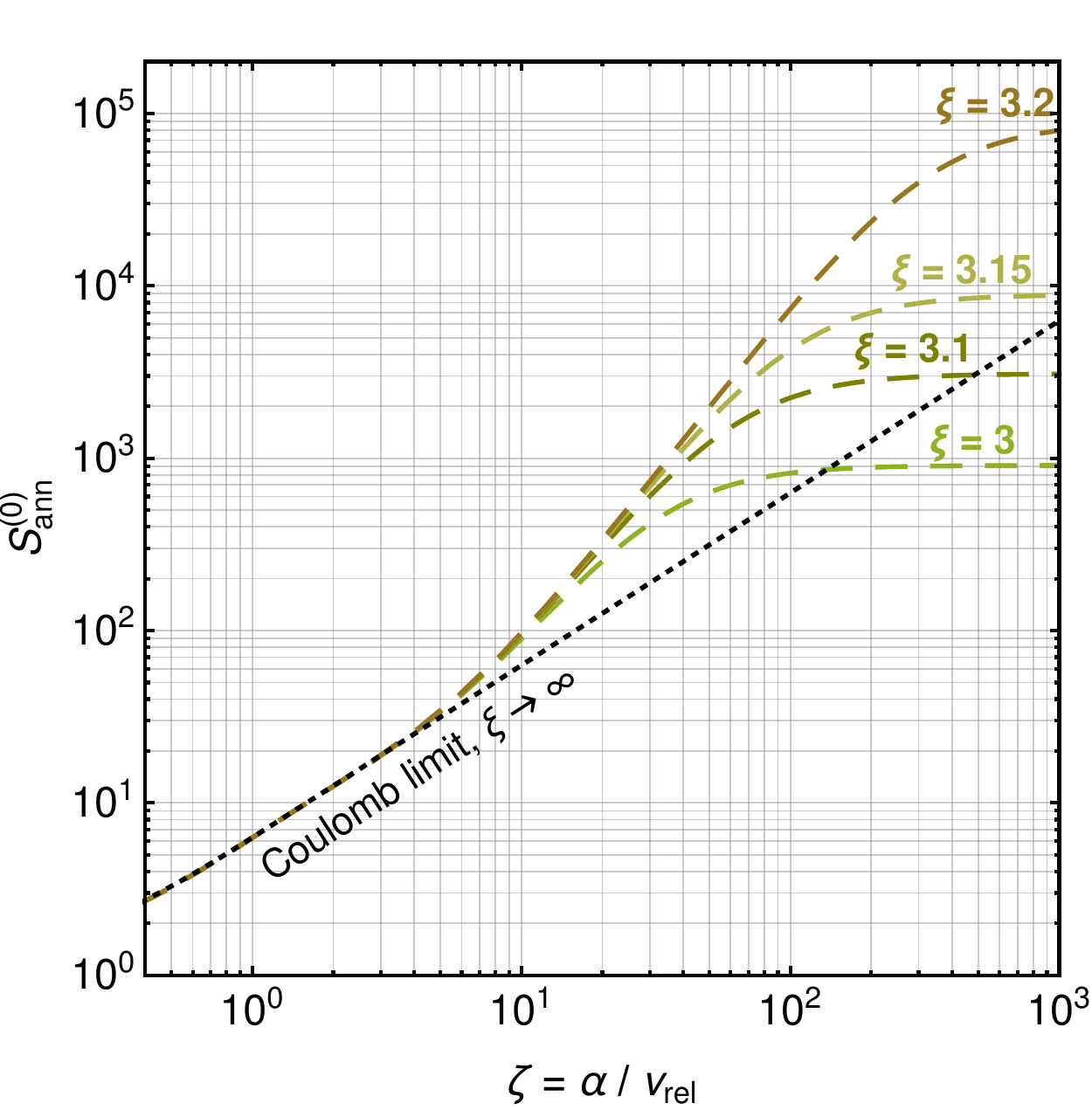}

\smallskip

\parbox[c]{0.45\linewidth}{
\includegraphics[width=\linewidth]{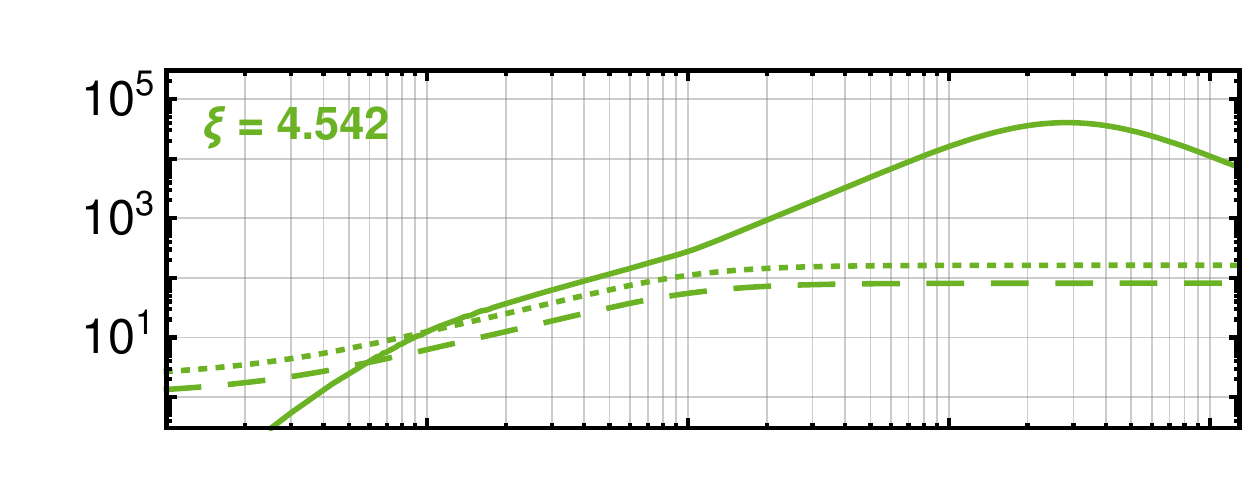}

\vspace{-0.7cm}
\includegraphics[width=\linewidth]{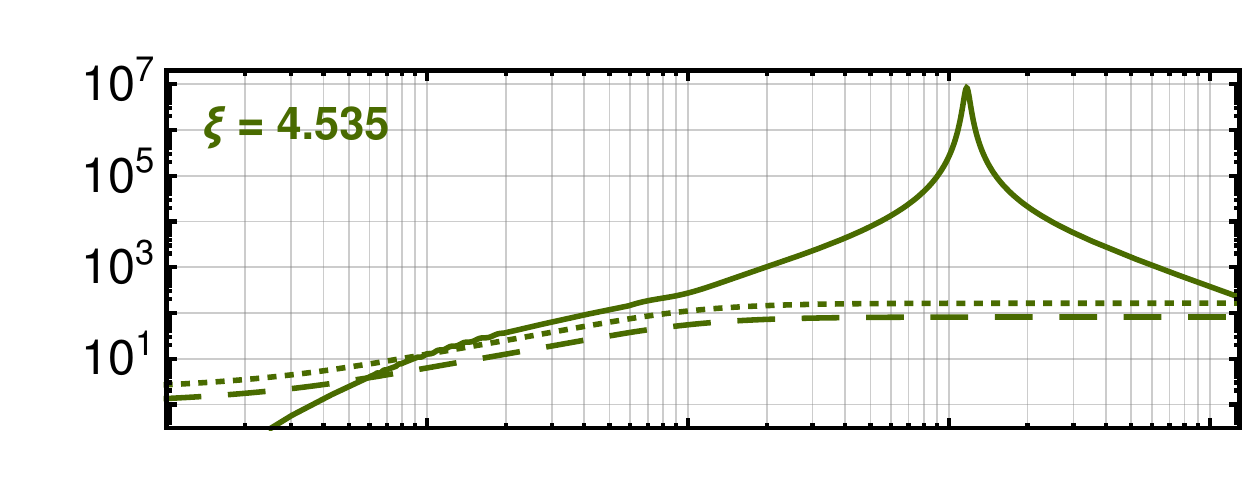}

\vspace{-0.7cm}
\includegraphics[width=\linewidth]{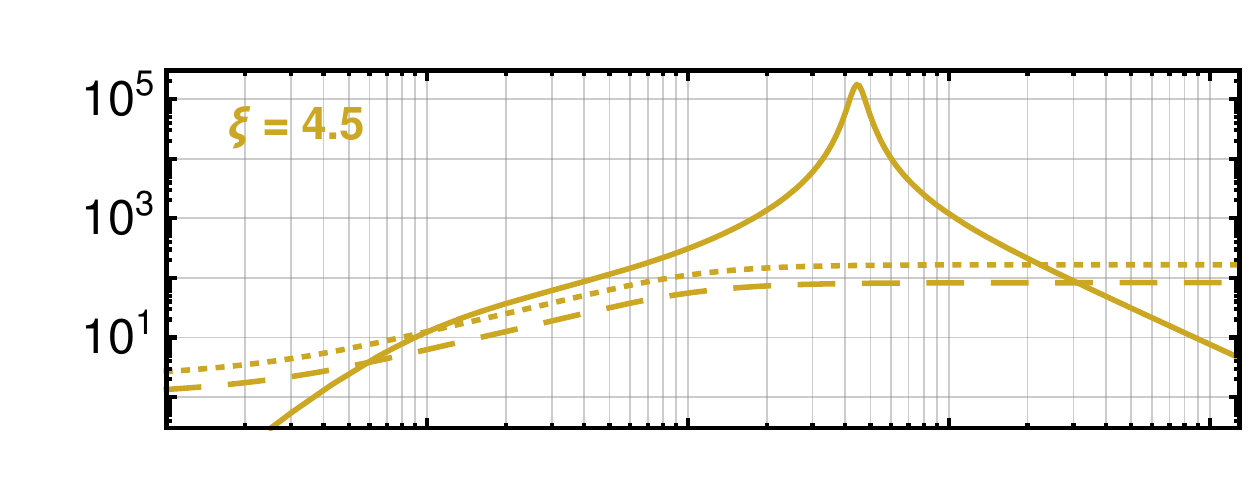}

\vspace{-0.7cm}
\includegraphics[width=\linewidth]{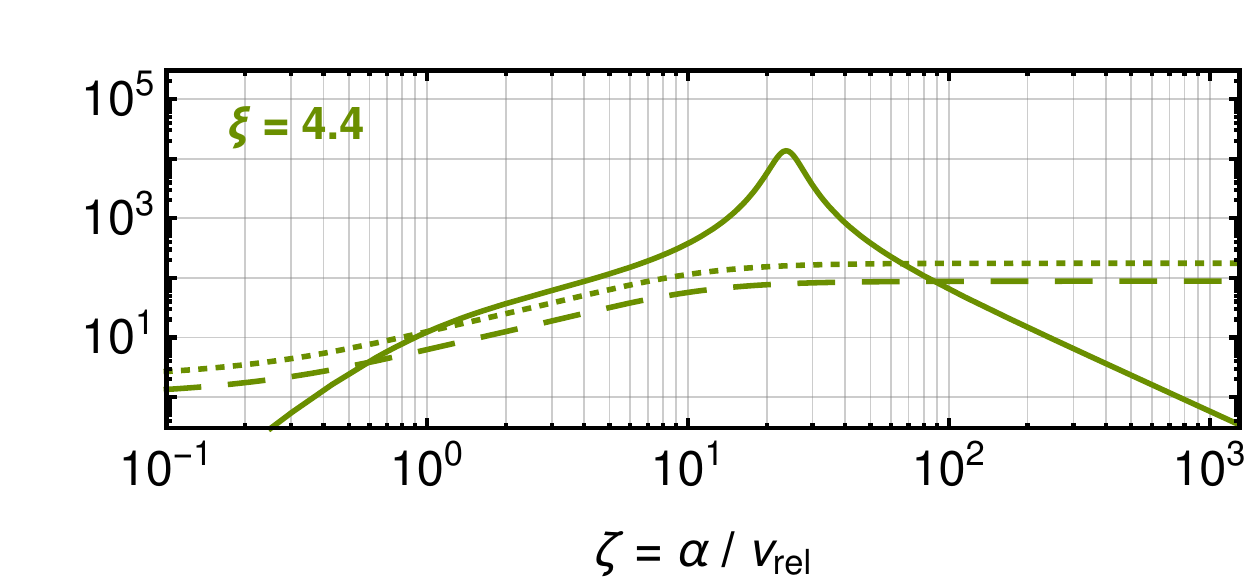}
}
~~~
\parbox[c]{0.46\linewidth}{
\includegraphics[width=\linewidth]{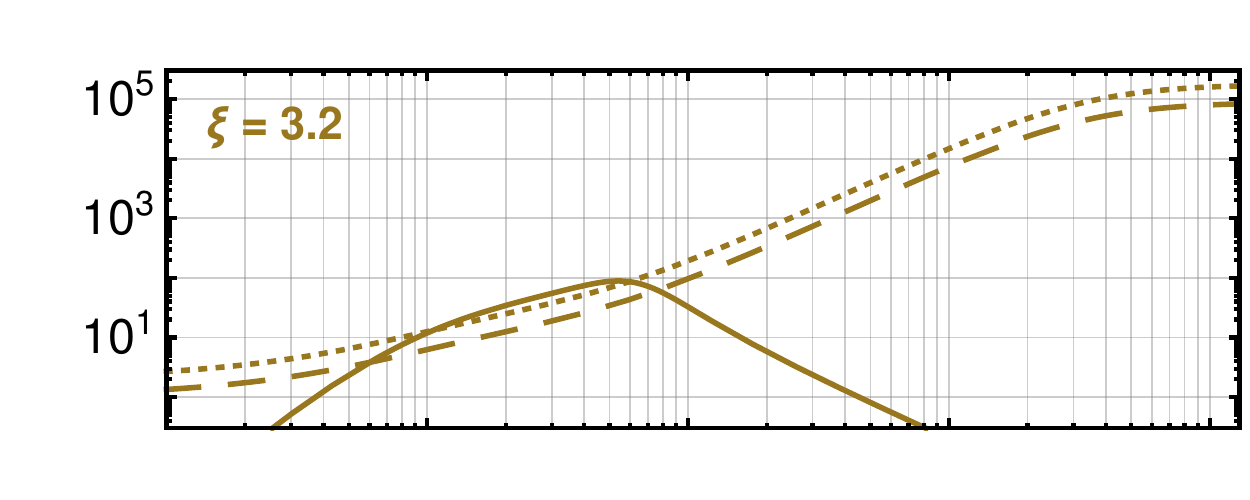}

\vspace{-0.7cm}
\includegraphics[width=\linewidth]{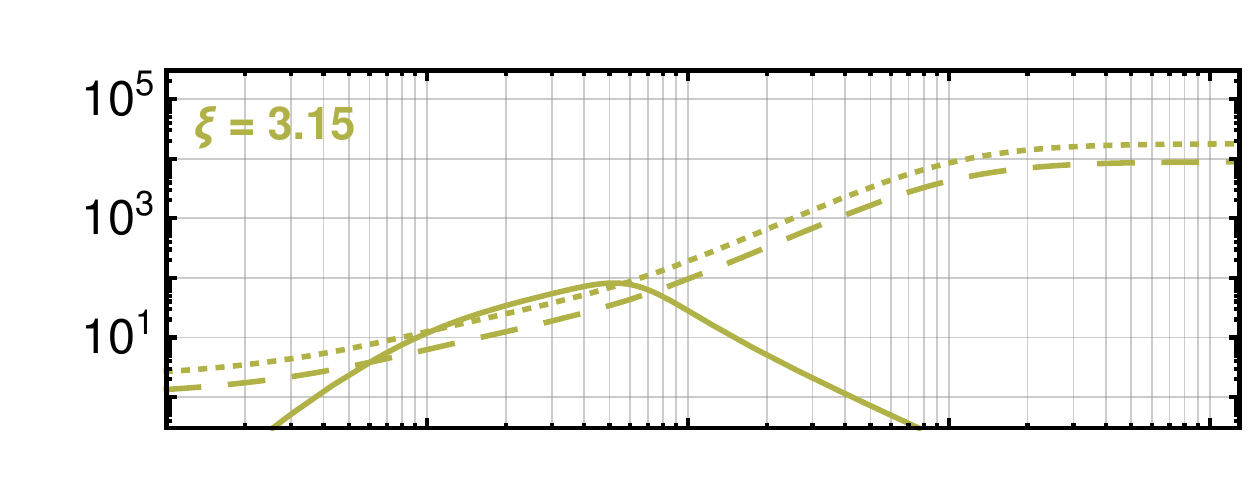}

\vspace{-0.7cm}
\includegraphics[width=\linewidth]{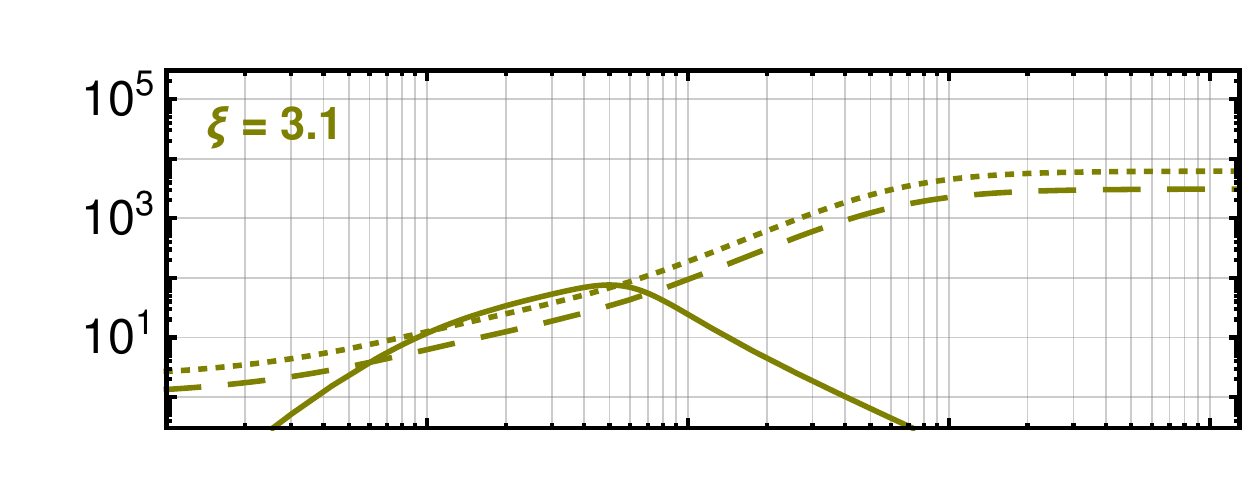}

\vspace{-0.7cm}
\includegraphics[width=\linewidth]{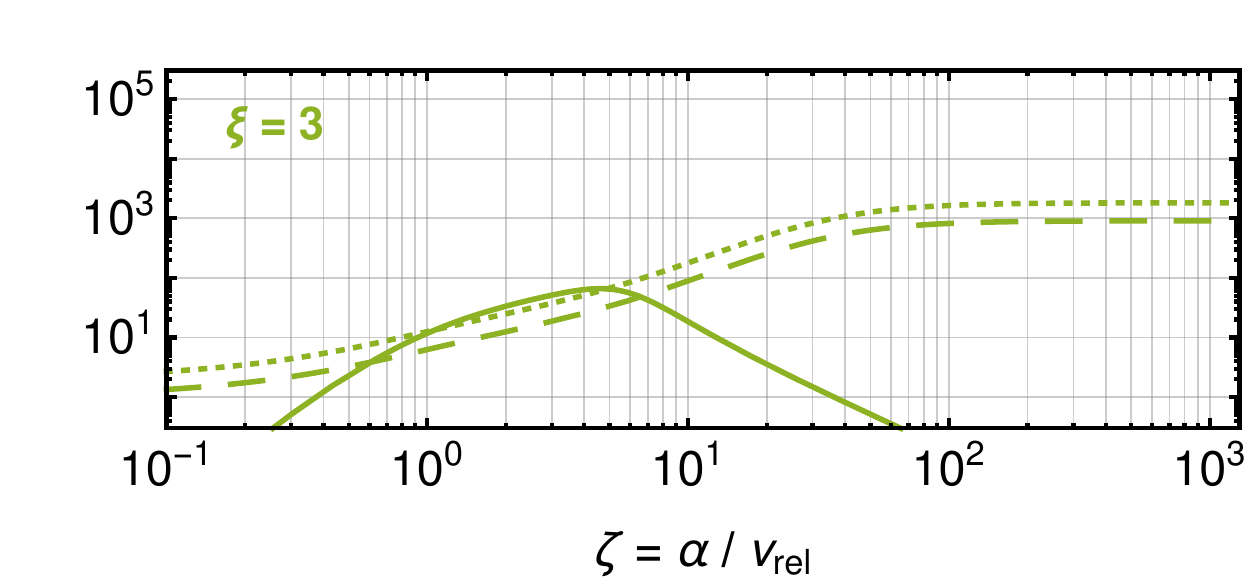}
}

\caption[]{\label{fig:VecMed_n=1_Zeta_Res}
Velocity dependence of $S_{_{\rm BSF}}^{\{100\}}$ and $S_{\rm ann}^{(0)}$, for values of $\xi = \mu \alpha/m_\varphi$ near the $n=2$ resonances. As in the non-resonant case, the velocity dependence of the cross-sections departs from the Coulombic behaviour when the momentum transfer between the interacting particles drops below the mediator mass, $\mu v_{\rm rel} \lesssim m_\vf$ ($\zeta \gtrsim \xi$). In contrast to non-resonant $\xi$ values though, for $\xi$ near a resonance and at $\zeta \gtrsim \xi$, the interaction cross-sections become larger than their Coulomb values at the same velocity, within a finite range of velocities.

\emph{Top left:} The cross-section for radiative capture into the ground state with emission of a vector mediator rises above its Coulomb limit, and then drops approximately as $\zeta^{-2} \propto v_{\rm rel}^2$  at $\zeta \gg \xi$ (see also fig.~\ref{fig:VecMed_Sregularized_Vs_Zeta}).

\emph{Top right:} For $s$-wave annihilation processes, the Sommerfeld enhancement rises monotonically with $\zeta$;  at $\zeta \gg \xi$, it asymptotes to a value that can be much larger than its value at $\zeta \approx \xi$.

\emph{Bottom:} 
Comparison of the velocity dependence of $S_{_{\rm BSF}}^{\{100\}}$ (solid lines), $S_{\rm ann}^{(0)}$ (dashed lines), $2S_{\rm ann}^{(0)}$ (dotted lines), for various values of $\xi$ near resonances arising in the $\ell = 1$ (left) and $\ell = 0$ (right) modes of the scattering state wavefunction. 
}
\end{figure}

\begin{figure}
\centering

{\boldmath \bf Vector mediator: 
$\xi$ values near the $n = 1$, $\ell = 0$ threshold/resonance}

\includegraphics[width=0.48\textwidth]{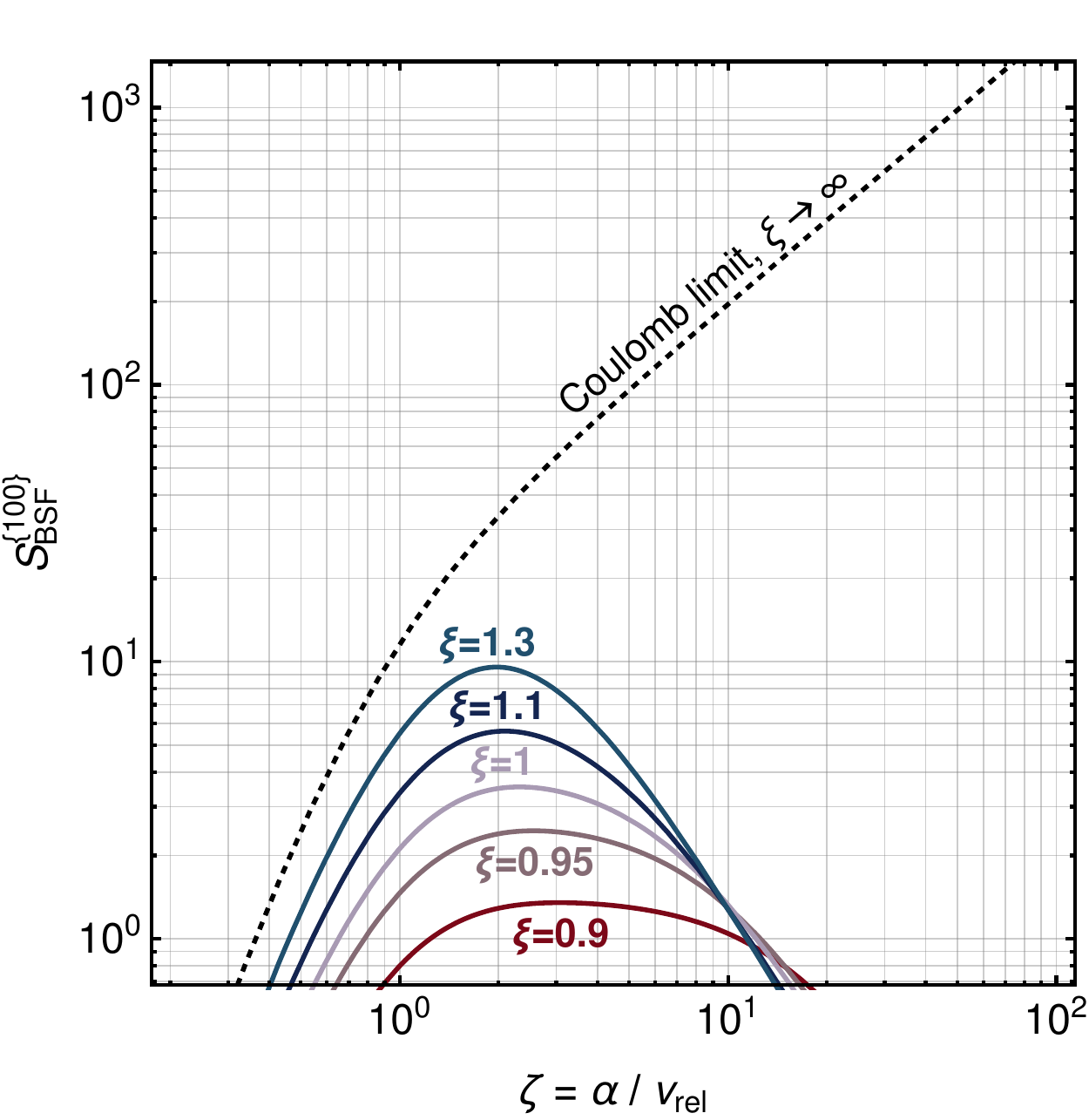}~~~~~
\includegraphics[width=0.48\textwidth]{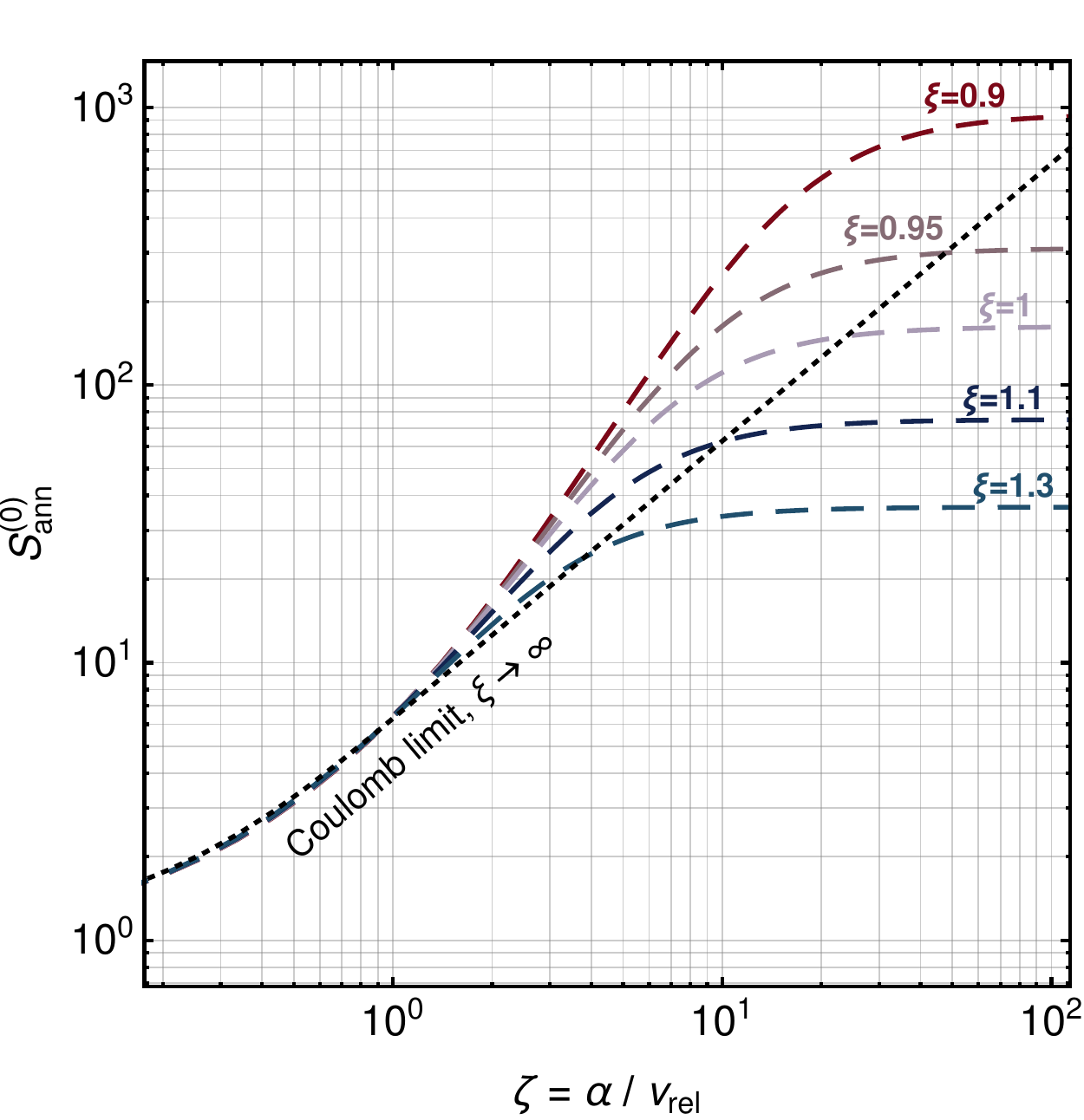}

\caption{\label{fig:VecMed_n=1_Zeta_Threshold}
\emph{Left:} For $\xi$ values close to the threshold for the existence of a bound state, the cross-section for the radiative formation of this bound state is suppressed with respect to its Coulomb value, for all velocities. This suppression arises from the bound-state wavefunction, which becomes very extended for $\xi$ values near threshold, and is independent of the phase-space suppression due to the emission of a massive force mediator.
\emph{Right:} The same values of $\xi$ yield a resonance in the annihilation processes.
}

\end{figure}
\end{subfigures}

\begin{subfigures}

\label{fig:VecMed_Sregularized}
\label[pluralfigure]{figs:VecMed_Sregularized}

\begin{figure}
\centering

{\boldmath \bf Vector mediator: \\ 
Formation of zero angular momentum bound states vs. $p$-wave annihilation}

\includegraphics[height=7cm]{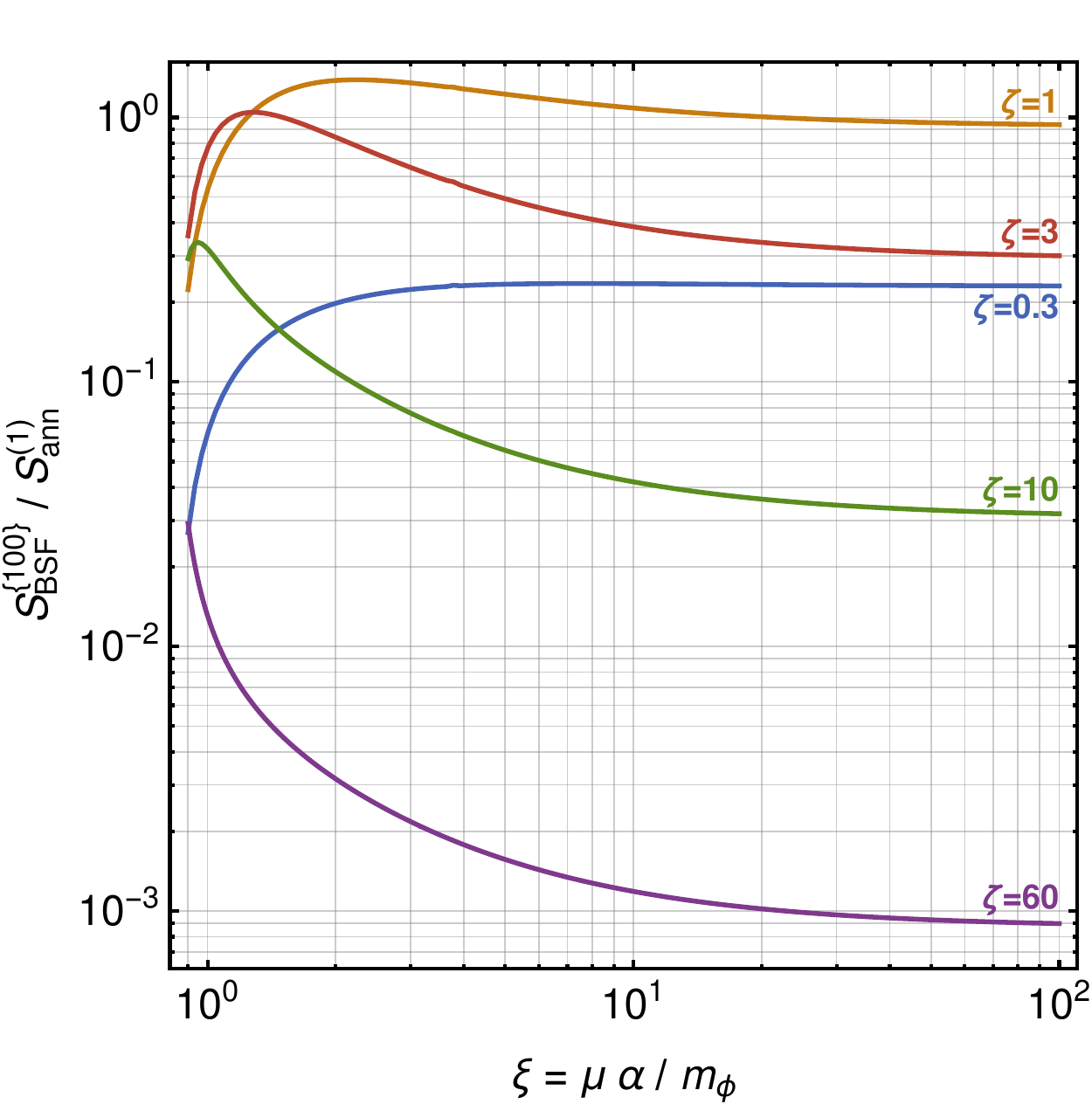}~~~~~
\includegraphics[height=7cm]{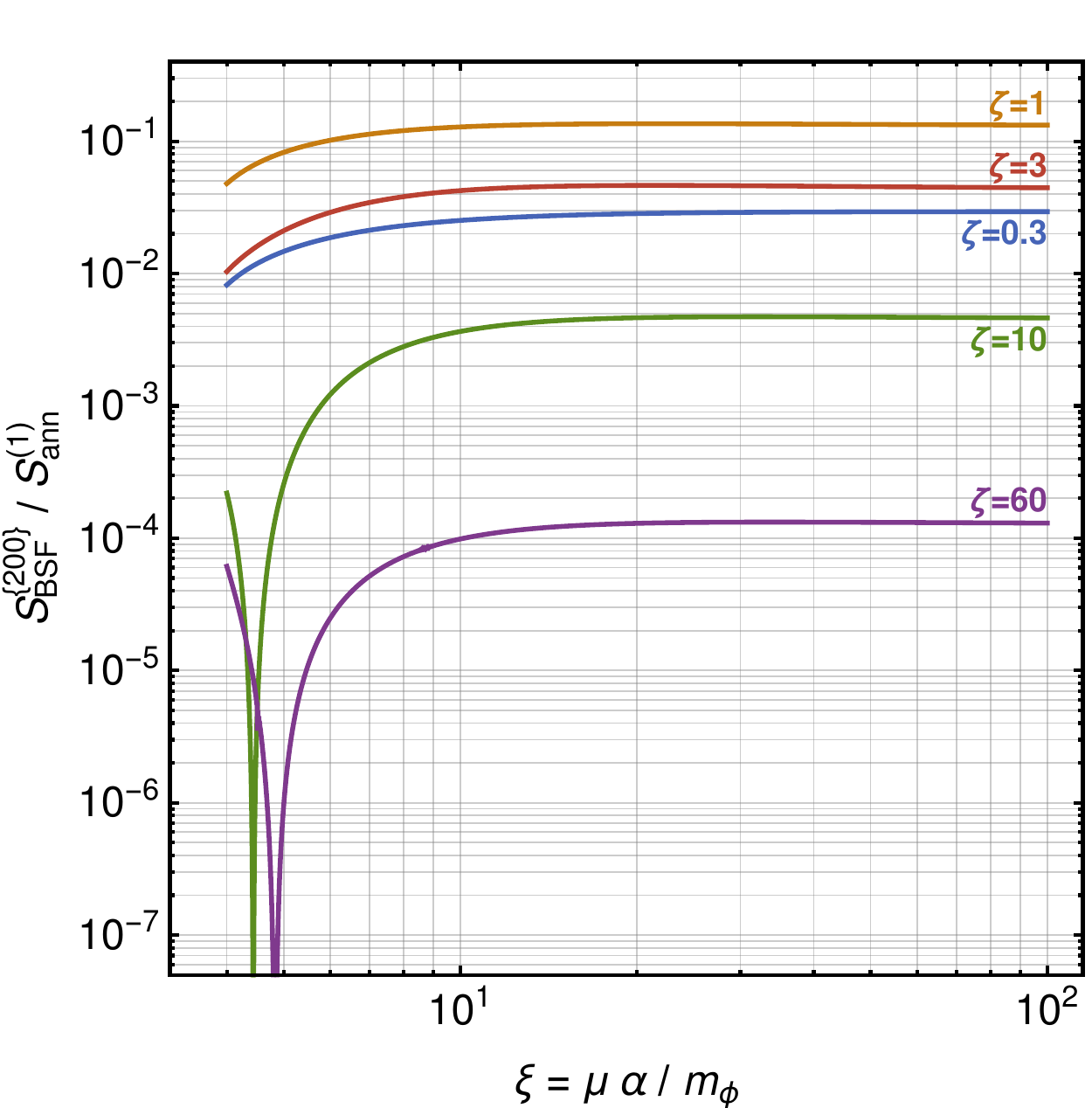}

\caption[]{\label{fig:VecMed_Sregularized_Vs_Xi}
The ratios 
$S_{_{\rm BSF}}^{\{100\}} / S_{\rm ann}^{(1)}$ (left) and 
$S_{_{\rm BSF}}^{\{200\}} / S_{\rm ann}^{(1)}$ (right), 
where $S_{\rm ann}^{(1)} (\zeta,\xi)$ is the Sommerfeld enhancement factor of $p$-wave annihilation processes, which depends on the $\ell=1$ component of the scattering state wavefunction only [cf.~eq.~\eqref{eq:Sann1}]. The absence of any resonances in these ratios implies that the resonances in $S_{_{\rm BSF}}^{\{100\}}$ and $S_{_{\rm BSF}}^{\{200\}}$ arise solely from the scattering-state wavefunction. On the other hand, the anti-resonances arise from the interference of the scattering-state and bound-state wavefunctions.
}

\includegraphics[height=7cm]{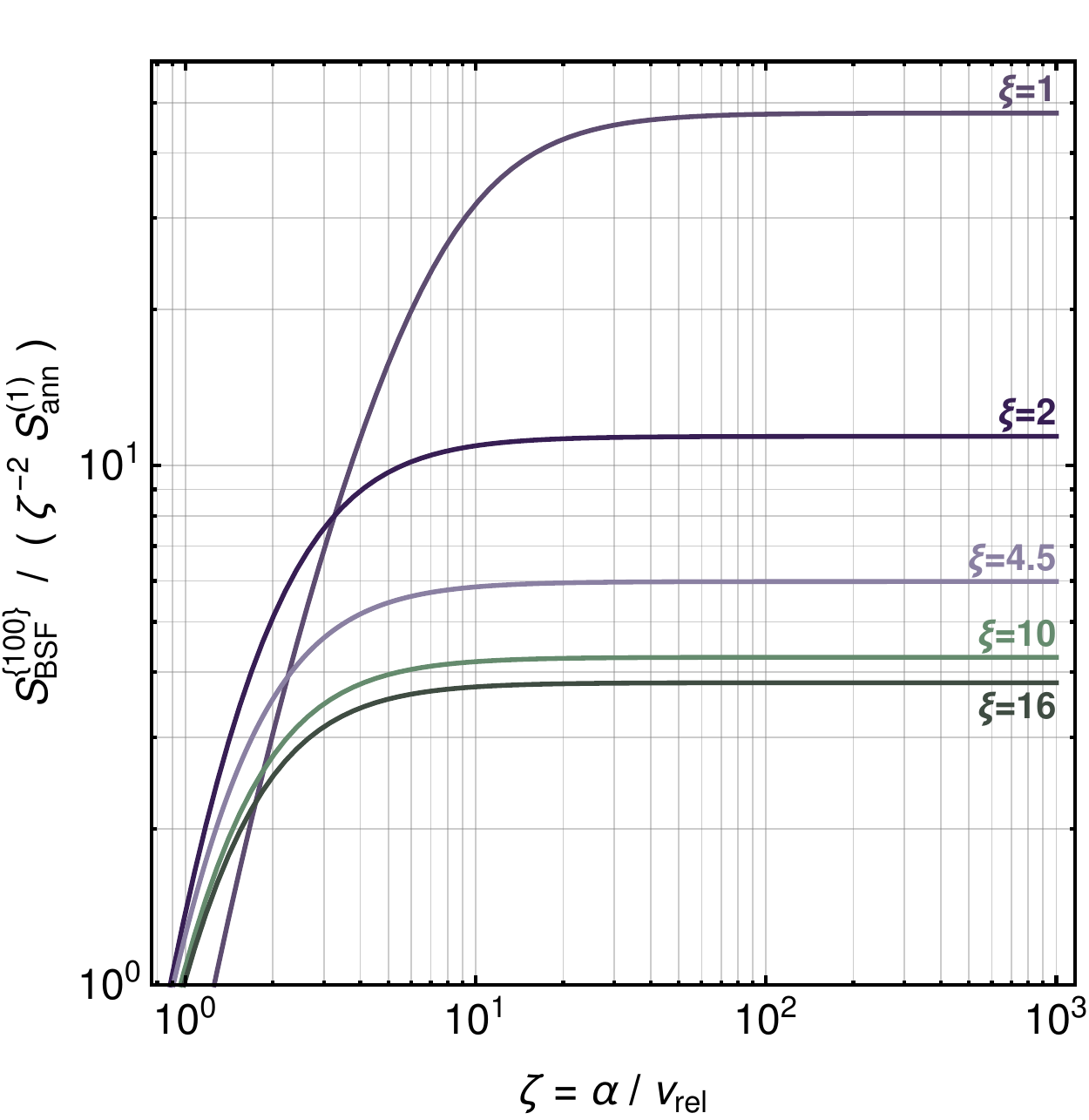}~~~~~
\includegraphics[height=7cm]{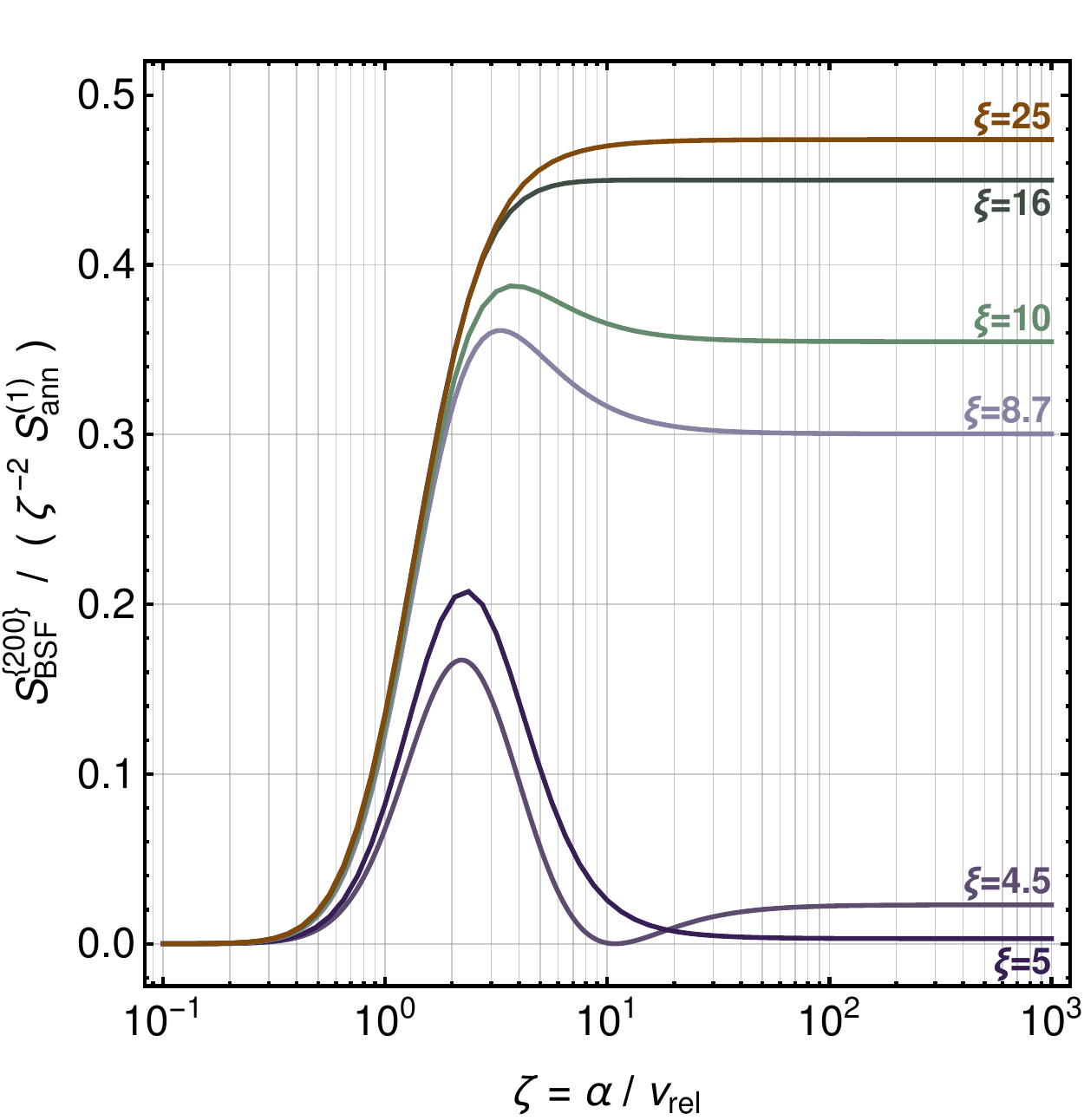}
\caption[]{\label{fig:VecMed_Sregularized_Vs_Zeta}
Comparison of the velocity dependence of radiative capture processes with emission of a vector force mediator, to  $p$-wave annihilation processes.
The factor $\zeta^{-2} S_{\rm ann}^{(1)}$ captures the entire velocity dependence of the latter. At large $\zeta$, the ratio $S^{\{n00\}}_{_{\rm BSF}} / [\zeta^{-2} S_{\rm ann}^{(1)}]$ tends to a $\xi$-dependent constant, which itself saturates to a fixed value (its Coulomb limit) at large $\xi$. For the ground state and first excited state, 
$S^{\{100\}}_{_{\rm BSF}} / [\zeta^{-2} S_{\rm ann}^{(1)}] \simeq 2^9/(3e^4) \simeq 3.13$ 
and 
$S^{\{200\}}_{_{\rm BSF}} / [\zeta^{-2} S_{\rm ann}^{(1)}] \simeq 2^{12}/(3e^8) \simeq 0.46$ 
[cf.~eq.~\eqref{eq:VecMed_Sn00_Coul} and table~\ref{tab:varrho n00 vector & scalar non-degen}].
}

\end{figure}
\end{subfigures}

\clearpage
\section{Scalar force mediator \label{Sec:ScalMed}}

We now consider the interaction Lagrangians
\begin{subequations} 
\label{eq:ScalMed_L}
\label[pluralequation]{eqs:ScalMed_L}
\begin{align}
{\cal L} 
&= \frac{1}{2} \partial_\m X_1 \, \partial^\m X_1
 + \frac{1}{2} \partial_\m X_2 \, \partial^\m X_2
 + \frac{1}{2} \partial_\m \vf  \, \partial^\m \vf
 - \frac{1}{2} m_1^2 X_1^2 - \frac{1}{2} m_2^2 X_2^2 - \frac{1}{2} m_\vf^2 \vf^2
\nn \\
&- \frac{1}{2} g_1 m_1 \vf X_1^2 - \frac{1}{2} g_2 m_2 \vf X_2^2
\, ,
\label{eq:ScalMed_L_Real}
\\
{\cal L} 
&= \partial_\m X_1^\dagger \, \partial^\m X_1
 + \partial_\m X_2^\dagger \, \partial^\m X_2
 + \frac{1}{2}\partial_\m \vf \, \partial^\m \vf
 - m_1^2 |X_1|^2 - m_2^2 |X_2|^2 - \frac{1}{2} m_\vf^2 \vf^2
\nn \\
&- g_1 m_1 \vf |X_1|^2 - g_2 m_2 \vf |X_2|^2
\, ,
\label{eq:ScalMed_L_Complex}
\\
{\cal L} 
&= \bar{X}_1 i\slashed{D} X_1  +\bar{X}_2 i\slashed{D} X_2
 + \frac{1}{2}\partial_\m \vf \, \partial^\m \vf
 - m_1 \bar{X}_1 X_1 - m_2  \bar{X}_2 X_2 - \frac{1}{2} m_\vf^2 \vf^2
\nn \\
&- g_1 \vf \bar{X}_1 X_1 - g_2 \vf \bar{X}_2 X_2
\, .
\label{eq:ScalMed_L_Fermions}
\end{align}
\end{subequations}
In eqs.~\eqref{eq:ScalMed_L_Real}, \eqref{eq:ScalMed_L_Complex}, and \eqref{eq:ScalMed_L_Fermions}, 
$X_1$ and $X_2$ are real scalar fields, complex scalar fields, and Dirac fermions, respectively. $\vf$ is a real scalar boson, and $g_1, g_2$ are dimensionless couplings. The interaction between $X_1, X_2$ via $\vf$ exchange is described by the Yukawa potential of \cref{eq:Yukawa}, with $\a = \a_{sc}$ or $\a = \a_f$ depending on whether the interacting particles are scalars or fermions respectively, where~\cite{Petraki:2015hla}
\beq
\a_{sc} = \frac{g_1g_2}{16\p} 
\qquad \text{and} \qquad
\a_f = \frac{g_1g_2}{4\p} \, .
\label{eq:ScalMed_alpha}
\eeq
The interaction is attractive if $g_1 g_2 >0$. In the following, the parameters $\z\equiv \a/\vrel$ and $\ks = \a\mu /m_\vf$ are always defined using the appropriate $\a$.

\subsection{Pairs of non-degenerate particles \label{sec:ScalarMed_NonDegen} }

The BSF amplitude is~\cite{Petraki:2015hla}
\beq
\M_{\vec k \to n\ell m} \simeq - M \sqrt{2\mu} \: 
\[ g_1 {\cal I}_{\vec k, n\ell m} (\h_2 \vec P_\vf) + g_2 {\cal I}_{\vec k, n\ell m} (-\h_1 \vec P_\vf)  \]  \, .
\label{eq:ScalMed_NonDeg_M}
\eeq
The dominant contributions to the amplitudes $\M_{\vec k \to n00}$ and $\M_{\vec k \to n1m}$ that we will consider below, are of order $|\vec{P}_\vf|/\k$.

\subsubsection[Capture into $\ell=0$ bound states]{Capture into $\boldsymbol{\ell=0}$ bound states}

Using \cref{eq:ScalMed_NonDeg_M,eq:In00 expansion}, we find
\begin{multline}
|\M_{\vec{k}\to n00}|^2 \simeq 
\frac{2^5 \pi^2 M^2}{\mu^2} 
\[\frac{(g_1 \h_2-g_2\h_1)^2}{16\p\a}\]
pss_{n,0}^{} \ \times \
\\ \times
\(\frac{1+\z^2\g_{n,0}^2}{\z^2}\)^2
\ \left|\int_0^\infty dx \ x \ \chi_{n,0}^*(x) \, \chi_{|\vec k|,1}^{}(x) \right|^2  
\ \cos^2\theta \, .
\label{eq:ScalMed_NonDegen_M_n00}
\end{multline}
In the limit $g_1=g_2, \ \h_1 \gg \h_2$, the factor in the square brackets is $(g_1 \h_2-g_2\h_1)^2/(16\p\a)~=$~1~or~1/4, for a bosonic and fermionic pair respectively. From \cref{eq:DiffSigma,eq:ScalMed_NonDegen_M_n00}, we find the corresponding cross-section to be
\begin{subequations} 
\label{eq:ScalMed_NonDegen_BSF}
\label[pluralequation]{eqs:ScalMed_NonDegen_BSF}
\beq
\s\BSF^{\{n00\}} \vrel \simeq 
\frac{\p\a^2}{\mu^2} \: \[\frac{(g_1 \h_2-g_2\h_1)^2}{16\p\a}\] 
\ pss_{n,0}^{3/2} \times S\BSF^{\{n00\}} (\z,\ks) \, , 
\label{eq:ScalMed_NonDegen_sigma_n00}
\eeq
where
\beq
S\BSF^{\{n00\}} (\z,\ks) =  \frac{1}{3} \(\frac{1+\z^2\gamma_{n,0}^2(\ks)}{\z^2}\)^3 
\ \left|\int_0^\infty dx \ x \ \chi_{n,0}^*(x) \, \chi_{|\vec k|,1}^{}(x) \right|^2  \, .
\label{eq:ScalMed_NonDegen_Sn00}
\eeq
\end{subequations}

\subsubsection*{Coulomb limit}  

\begin{figure}[p]
\centering
{}~~~~~~~~{\bf Scalar mediator, non-degenerate particles} 

\includegraphics[height=6.8cm]{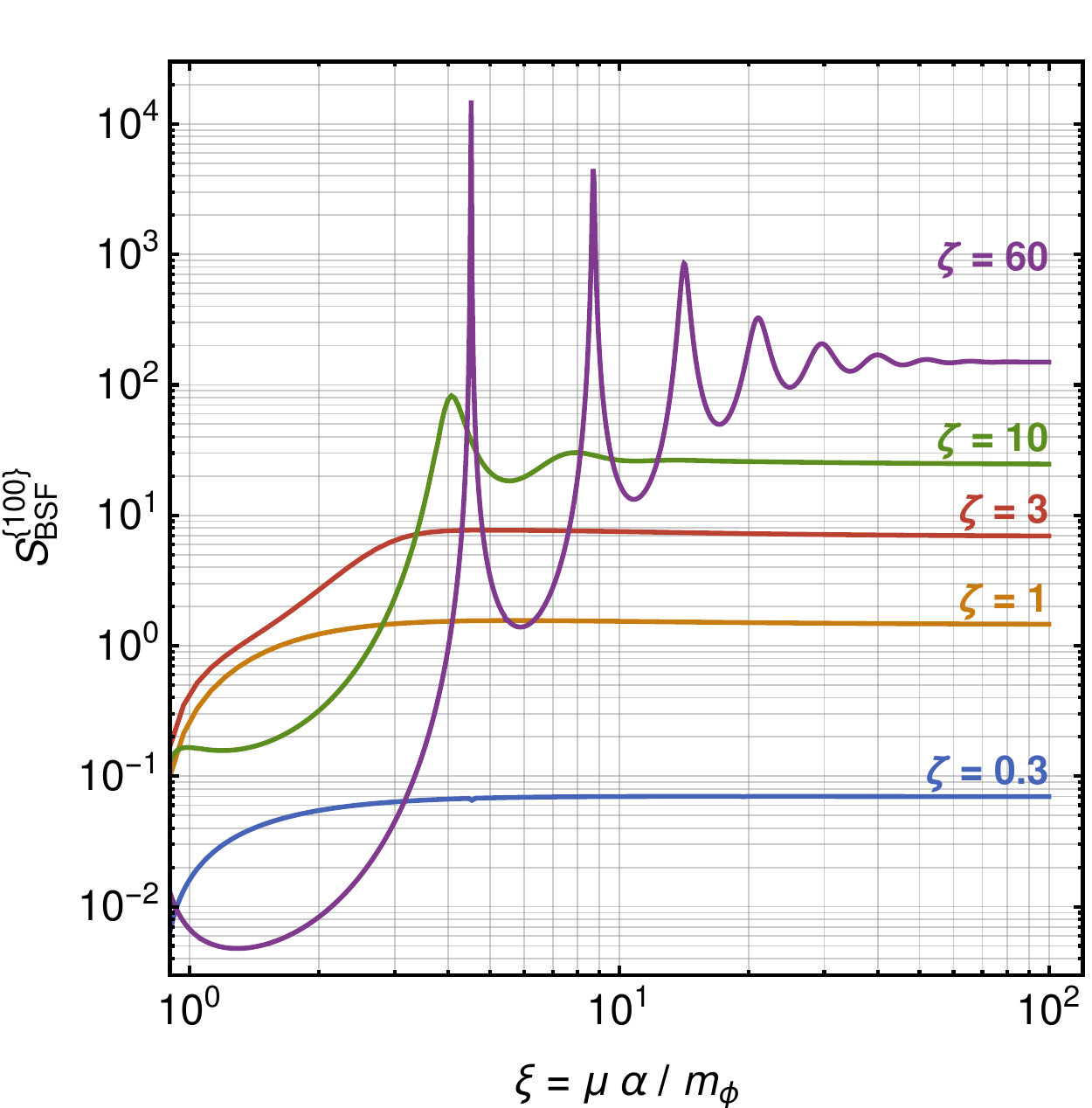}~~
\includegraphics[height=6.8cm]{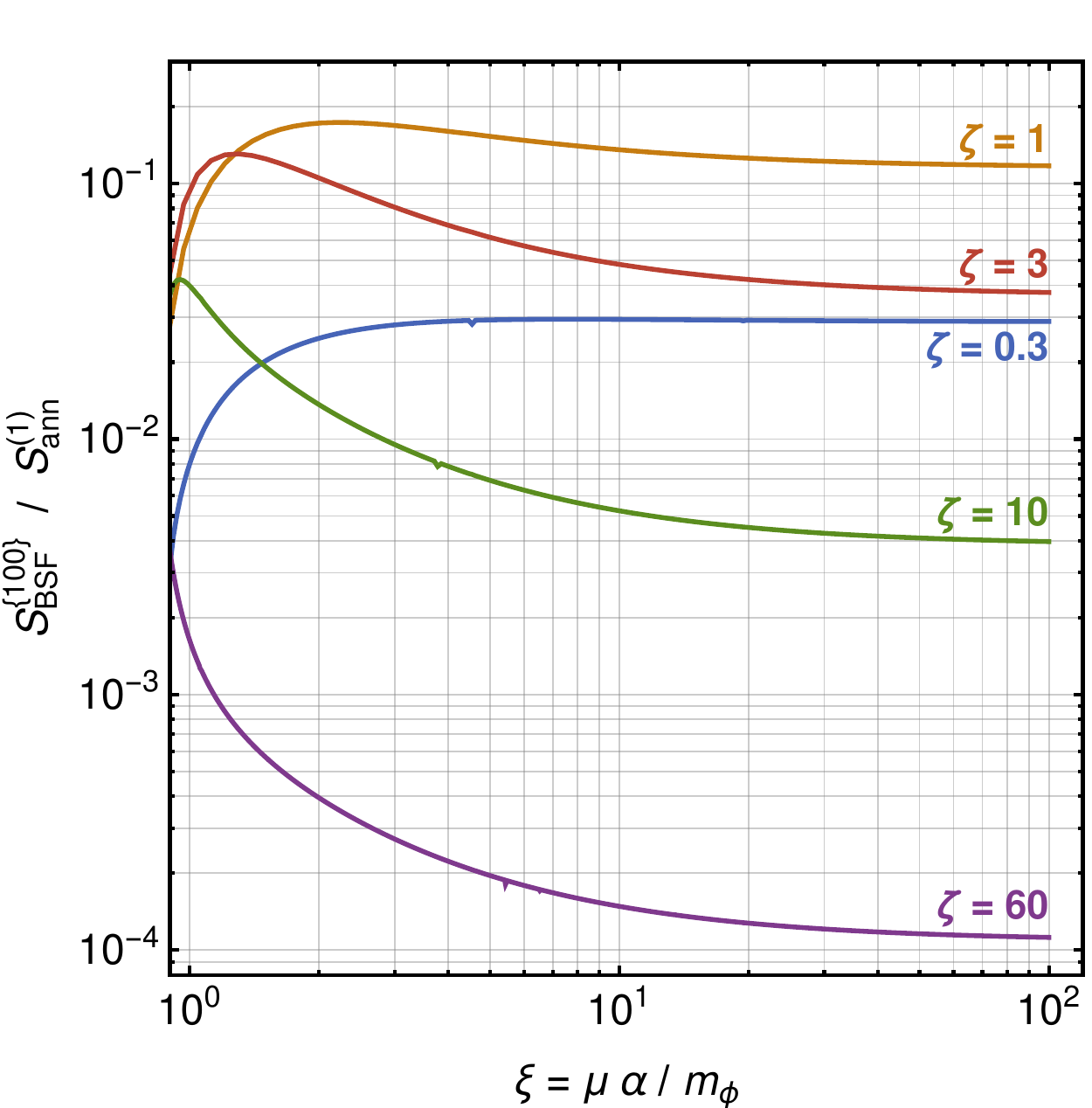}

\medskip

\includegraphics[height=6.8cm]{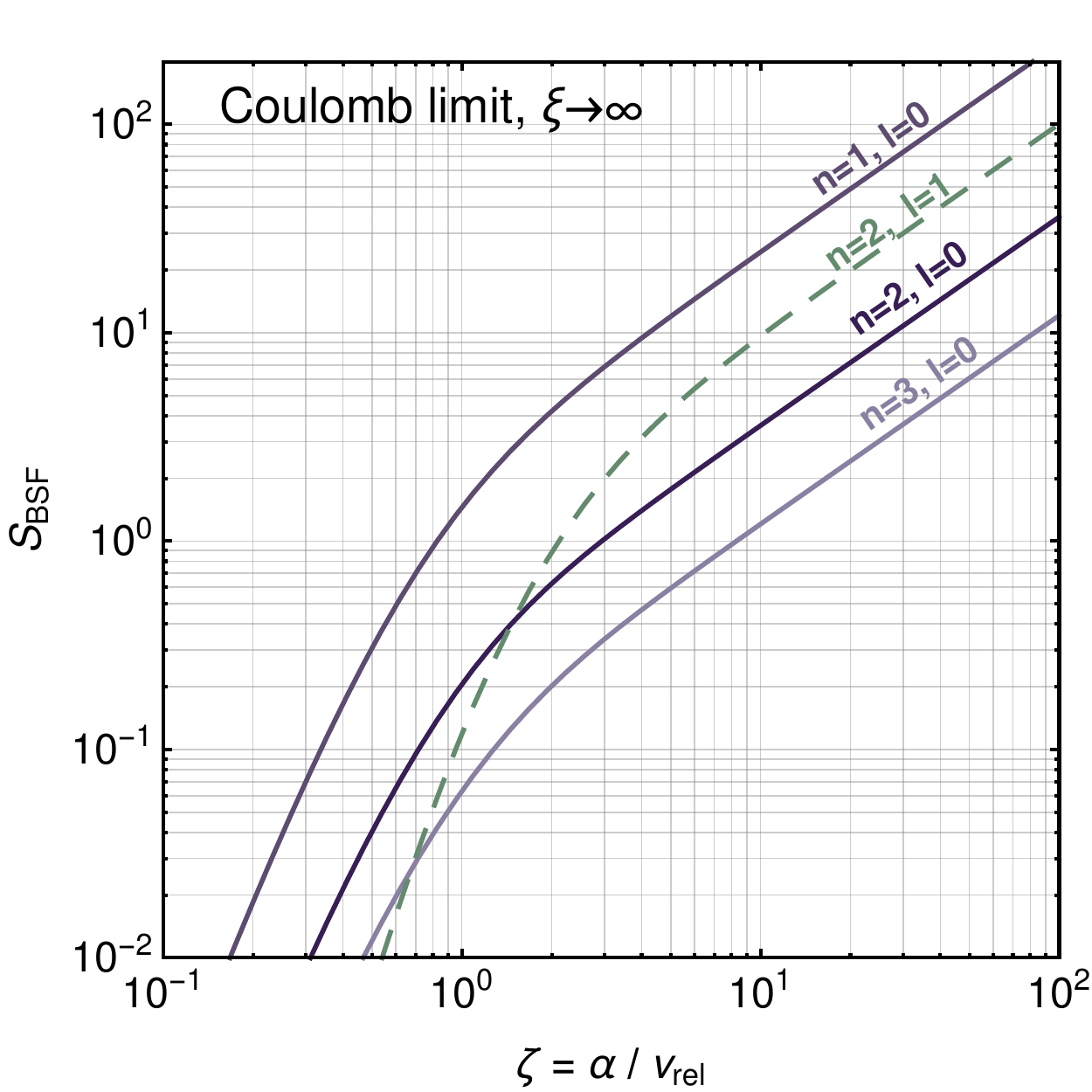}
~~
\includegraphics[height=6.8cm]{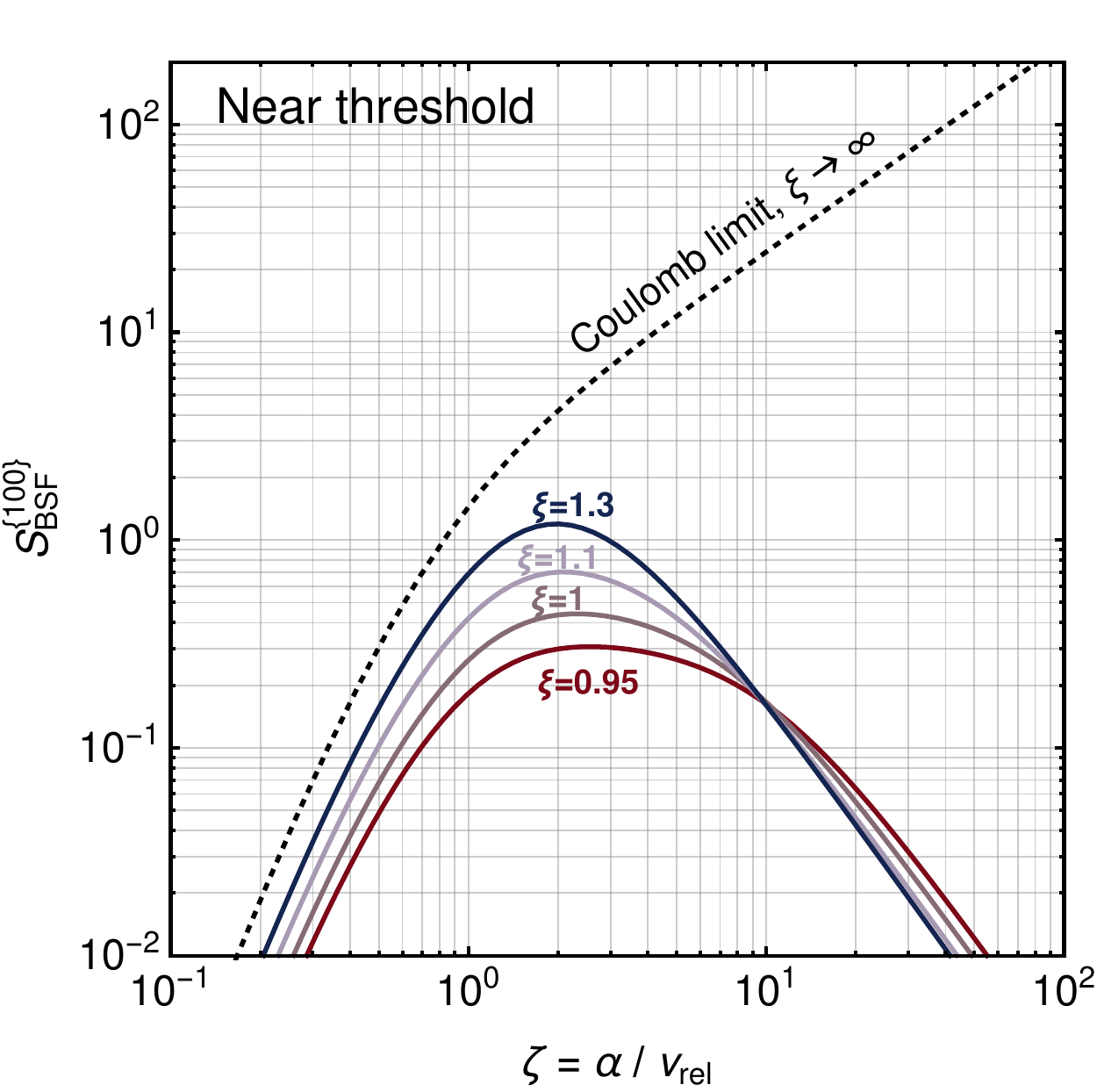}

\includegraphics[height=6.8cm]{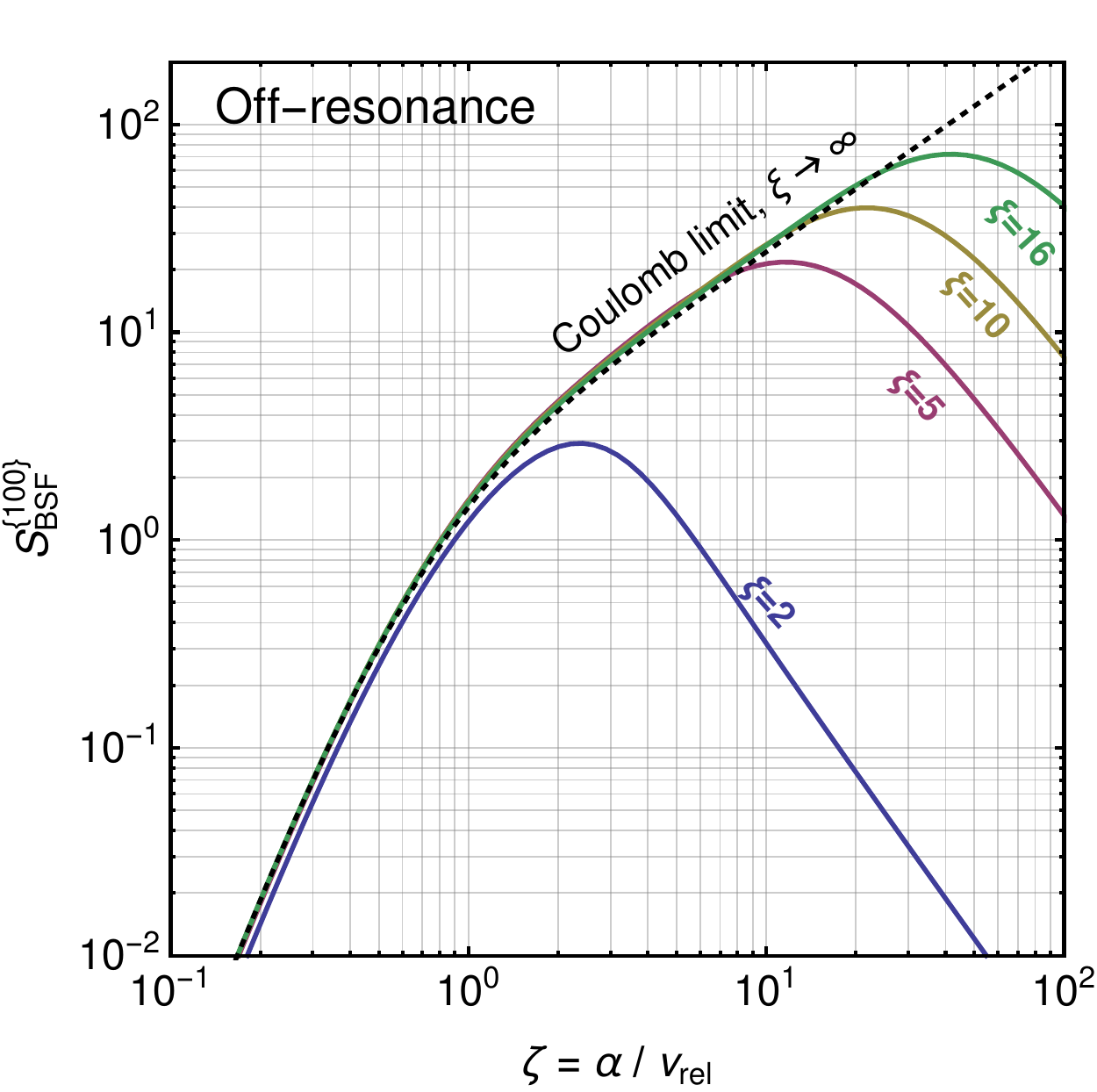}
~~
\includegraphics[height=6.8cm]{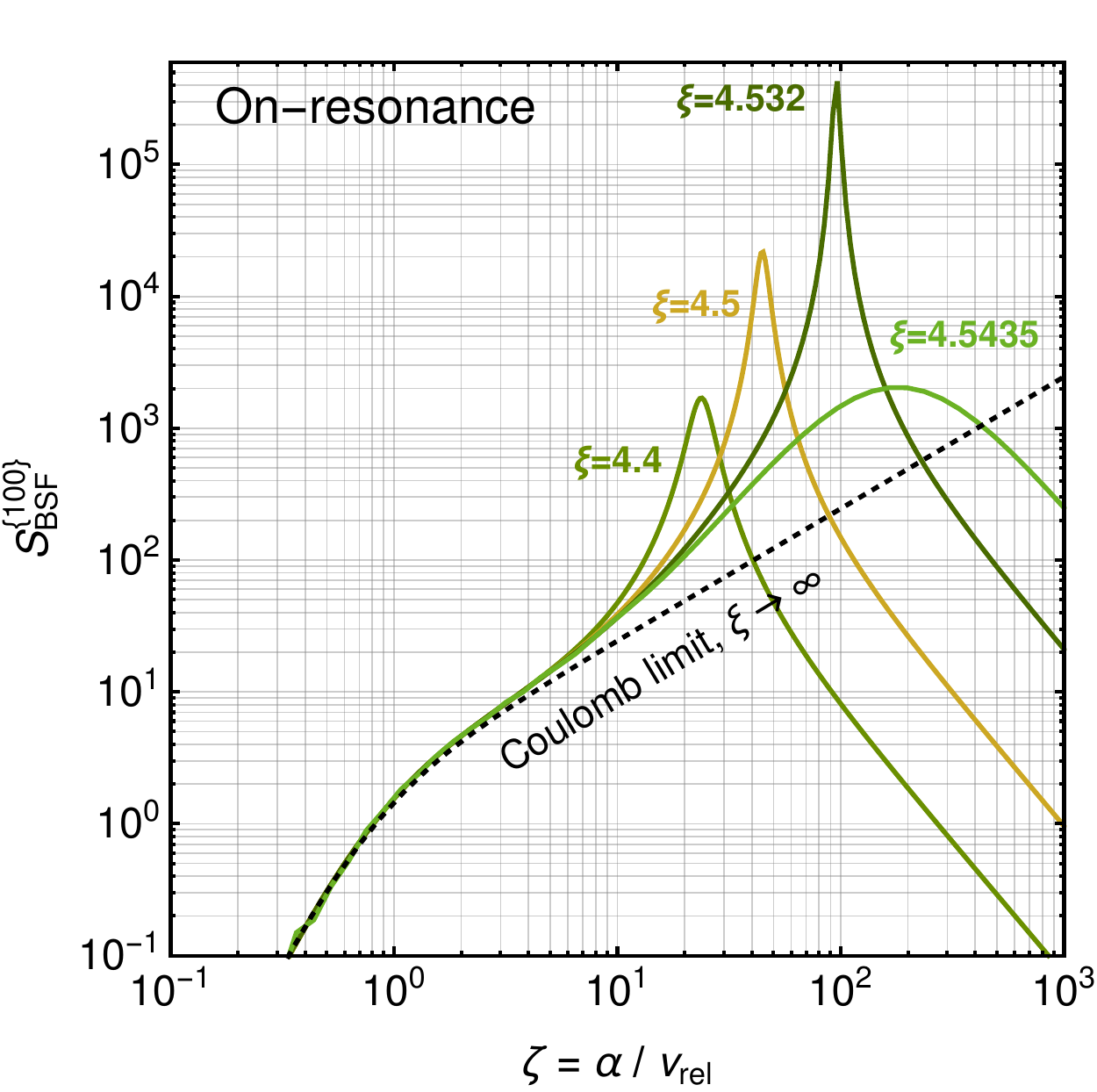}~~

\caption[]{\label{fig:ScalarMed_NonDegen_100}
Resonance structure (upper two panels) and velocity dependence (lower four panels) of the $S_{_{\rm BSF}}$ factors determining the cross-sections for the formation of bound states with emission of a scalar force mediator, by two particles with different masses and/or couplings [cf.~eqs.~\eqref{eqs:ScalMed_NonDegen_BSF} and \eqref{eqs:ScalMed_NonDegen_BSF_Coul}].}

\end{figure}

The cross-section \eqref{eqs:ScalMed_NonDegen_BSF} becomes
\begin{subequations} 
\label{eq:ScalMed_NonDegen_BSF_Coul}
\label[pluralequation]{eqs:ScalMed_NonDegen_BSF_Coul}
\beq
\lim_{\ks\to \infty} \s\BSF^{\{n00\}} \vrel 
\simeq  \frac{\p\a^2}{\mu^2} 
\: \[\frac{(g_1 \h_2-g_2\h_1)^2}{16\p\a}\]  
\ S\BSFC^{\{n00\}} (\z,\ks) \, ,
\label{eq:ScalMed_NonDegen_sigma_n00_Coul}
\eeq
where from \cref{eq:In00 Coul expansion}, and keeping terms of order $|\vec P_\vf|/\k$, we find
\begin{multline}
S\BSFC^{\{n00\}}(\z) 
= \(\frac{2\p\z}{1-e^{-2\p\z}}\) \, \frac{2^6}{3n^5} 
\: \z^2 \, (1+\z^2) \, (1+\z^2/n^2)^3 \, \times
\\
\times \left| \sum_{s=0}^{n-1} 
\frac{n! \, 2^s \, \z_n^s}{(n-s-1)! \, (s+1)! \, s!} \
\frac{d^s}{d\z_n^s} \[ \frac{(\z-2\z_n) \: e^{-2\z\, {\rm arccot}\, \z_n}}{(1+\z_n^2)^3} \]
\right|^2_{\z_n = \z/n}  \, .
\label{eq:ScalMed_NonDegen_Sn00_Coul_sum}
\end{multline}
Expanding the sum, \cref{eq:ScalMed_NonDegen_Sn00_Coul_sum} becomes
\beq
S\BSFC^{\{n00\}}(\z) 
= \(\frac{2\p\z}{1-e^{-2\p\z}}\) \, \frac{2^6}{3n^3} 
\ \frac{\z^4 (1+\z^2)}{(1+\z^2/n^2)^{2n-1}}
\ [\varrho_n(\z)]^2
\ e^{-4\z\, {\rm arccot}\, (\z/n)}
\, ,
\label{eq:ScalMed_NonDegen_Sn00_Coul}
\eeq
where $\varrho_n(\z)$ is a rational function of $\z^2$ with $\lim_{\z\to 0}\varrho_n(\z) = 1$. The explicit expressions of $\varrho_n(\z)$ for $1\leqslant n \leqslant 5$ are given in \cref{tab:varrho n00 vector & scalar non-degen}. For $n=1$,
\beq
S\BSFC^{\{100\}}(\z) = 
\(\frac{2\p\z}{1-e^{-2\p\z}}\) \times 
\frac{2^6}{3} \(\frac{\z^2}{1+\z^2}\)^2 \ e^{-4\z {\rm arc cot}\,\z} \, ,
\label{eq:ScalMed_NonDegen_S100_Coul}
\eeq
\end{subequations}
We inspect the features of $S\BSF^{\{100\}} (\z,\ks)$ and $S\BSFC^{\{100\}} (\z)$ in \cref{fig:ScalarMed_NonDegen_100}.

\subsubsection[Capture into $\ell=1$ bound states]{Capture into $\boldsymbol{\ell=1}$ bound states}

\begin{figure}[p]
\centering

{\boldmath \bf Scalar mediator, non-degenerate particles: 

Capture into $n=2, \ell=1$ bound states}

\includegraphics[height=7cm]{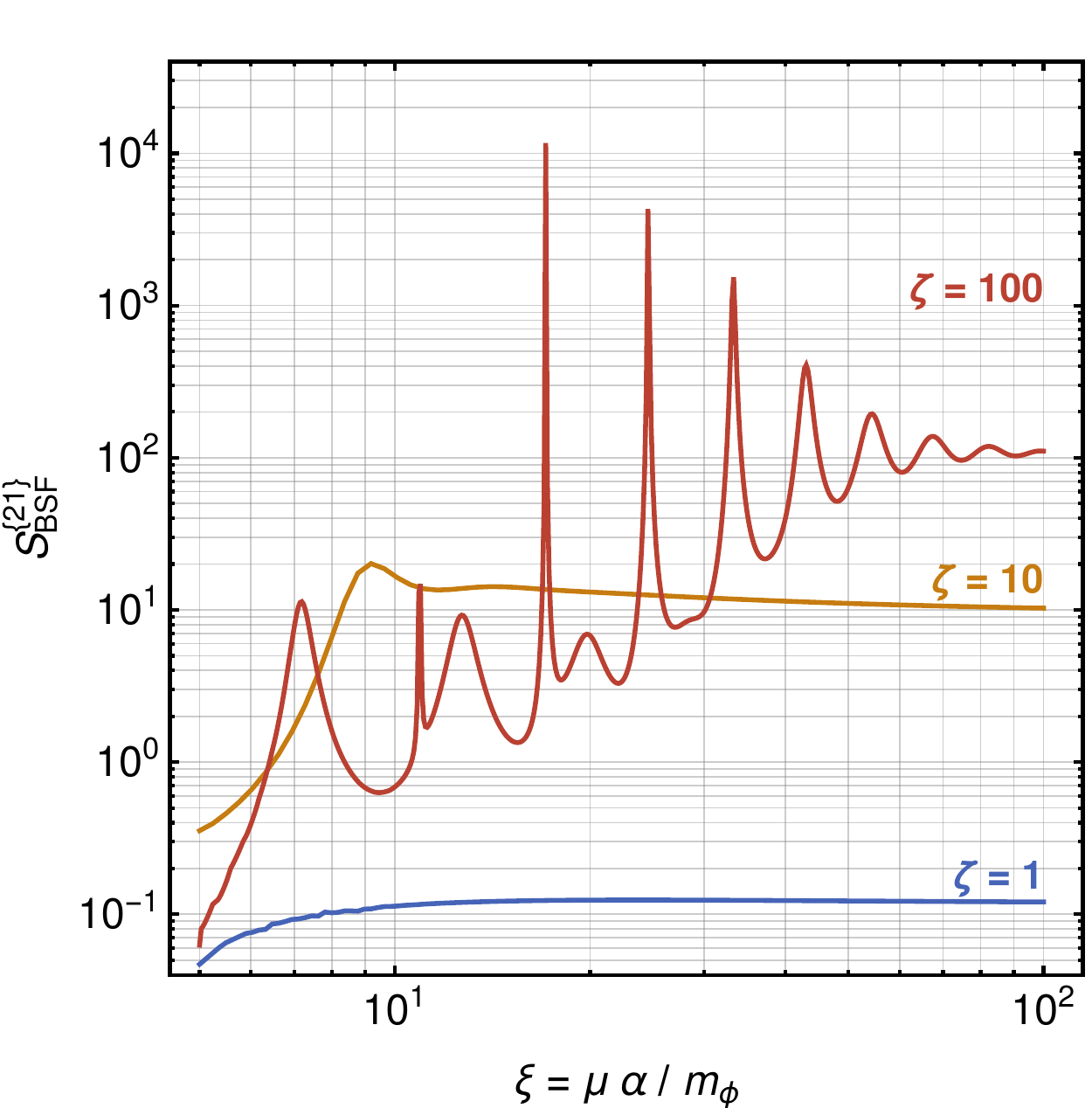}
~~~~
\includegraphics[height=7cm]{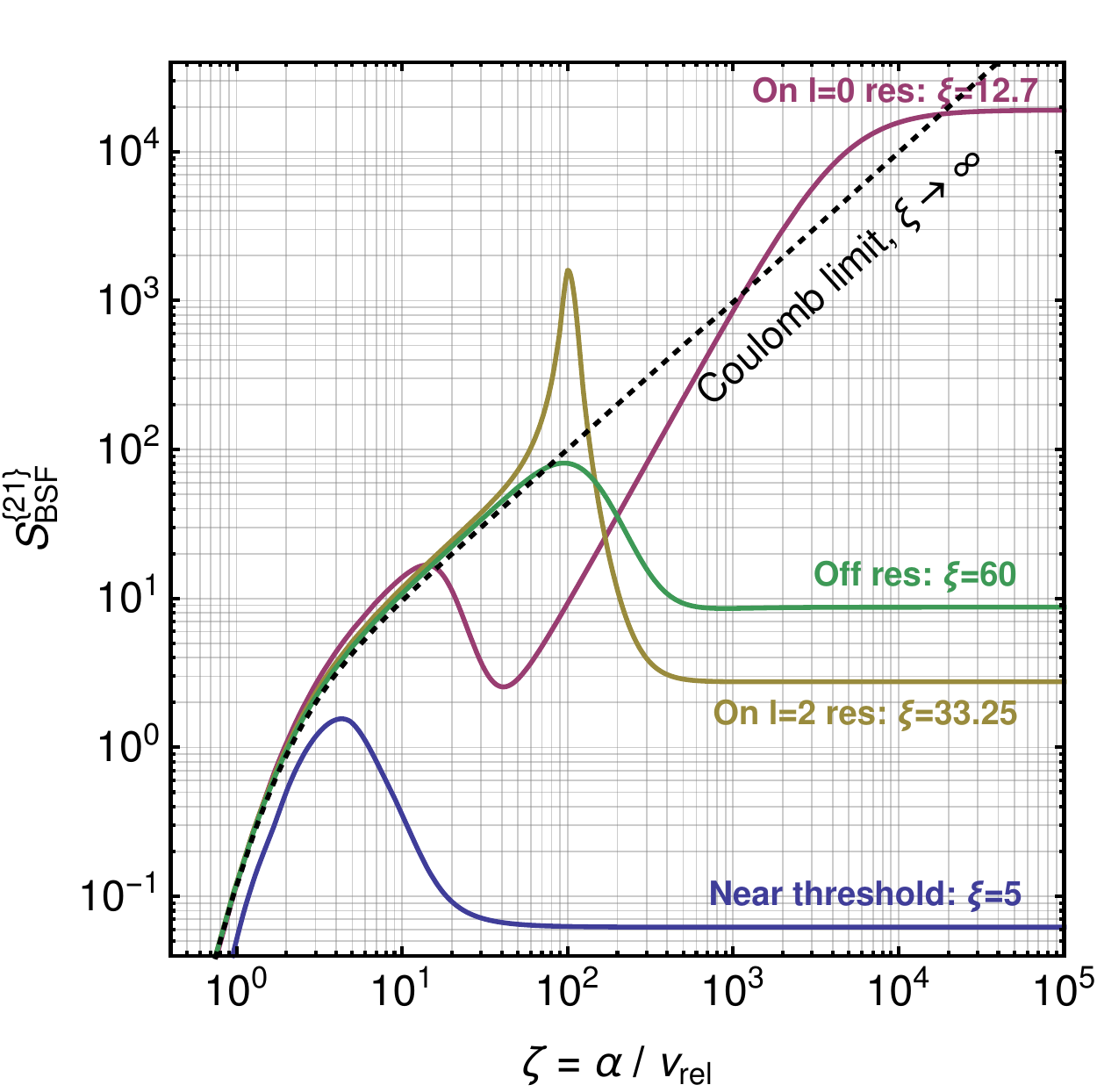}

\caption[]{\label{fig:ScalarMed_NonDeg_n=2_l=1}
Resonance structure and velocity dependence of the $S^{\{21\}}_{_{\rm BSF}}$ factor of the total cross-section for the radiative capture into any $n=2, \ell=1$ state of two particles with different masses and/or couplings, with emission of a scalar force mediator [cf.~eqs.~\eqref{eqs:ScalMed_NonDegen_BSF_l=1}].

\smallskip

\emph{Left:} $S^{\{21\}}_{_{\rm BSF}}$ receives contributions from the $\ell = 0$ and the $\ell=2$ modes of the scattering-state wavefunction. The $\ell=0$ mode dominates for small $\xi$, while the $\ell=2$ mode dominates for larger $\xi$ values (including the Coulomb regime) and gives rise to sharper resonances.

\smallskip

\emph{Right:}  At large $\zeta$, the contribution from the $\ell = 2$ mode decreases with decreasing velocity as $v_{\rm rel}^4$, while the contribution from the $\ell=0$ mode saturates to a constant value and dominates.
} 
\end{figure}

Similarly to above, we combine \cref{eq:DiffSigma,eq:ScalMed_NonDeg_M,eq:Inlm general}, and find the BSF cross-section to be
\begin{subequations} 
\label{eq:ScalMed_NonDegen_BSF_l=1}
\label[pluralequation]{eqs:ScalMed_NonDegen_BSF_l=1}
\beq
\s\BSF^{\{n1\}} \vrel \equiv
\sum_{m=-1}^1 \s\BSF^{\{n1m\}} \vrel \simeq 
\frac{\p\a^2}{\mu^2} \: \[\frac{(g_1 \h_2-g_2\h_1)^2}{16\p\a}\] 
\ pss_{n,1}^{3/2} \times S\BSF^{\{n1\}} (\z,\ks) \, , 
\label{eq:ScalMed_NonDegen_sigma_n00 l=1}
\eeq
where
\begin{multline}
S\BSF^{\{n1\}} (\z,\ks) =  \frac{1}{3} \(\frac{1+\z^2\gamma_{n,1}^2(\ks)}{\z^2}\)^3 \times
\\ \times
\( 
  \left|\int_0^\infty dx \ x \ \chi_{n,1}^*(x) \, \chi_{|\vec k|,0}^{}(x) \right|^2  +  
2 \left|\int_0^\infty dx \ x \ \chi_{n,1}^*(x) \, \chi_{|\vec k|,2}^{}(x) \right|^2  
\)
\, .
\label{eq:ScalarMed_NonDegen_Sn1}
\end{multline}
\end{subequations}
We inspect the features of $S\BSF^{\{21\}} (\z,\ks)$ in \cref{fig:ScalarMed_NonDeg_n=2_l=1}.

\clearpage
\subsection{Bosonic particle-antiparticle pairs (non-identical) \label{sec:ScalarMed_Degen}}

We now consider particle-antiparticle pairs of a non-self-conjugate species $X$. In this case, $g_1=g_2=g$, $\h_1=\h_2=1/2$ and $\mu=M/4$. As seen in \cref{eq:ScalMed_NonDegen_BSF,eq:ScalMed_NonDegen_BSF_l=1}, the lowest-order terms cancel, and we are forced to consider the next order contributions. For this reason, the computations that follow, which are based on the formalism of Ref.~\cite{Petraki:2015hla} that was developed for bosonic species, will be valid for bosonic particle-antiparticle pairs only. A computation of BSF cross-sections for fermion-antifermion pairs with emission of a scalar force mediator, can be found in Ref.~\cite{An:2016kie}.

In the radiative BSF amplitude, we shall now include higher-order contributions from (i) the relativistic normalisation of states, and (ii) the off-shellness of the incoming and outgoing fields in the perturbative part of the amplitude that includes the radiative vertex (i.e.~the part of the diagrams of \cref{fig:BSF} that remains when the incoming and outgoing ladders are amputated).\footnote{We believe that the difference between the bosonic case considered here, and the fermionic case considered in Ref.~\cite{An:2016kie}, is the corrections due to (ii), which are not present in the latter case.} 
Then, the BSF amplitude for a particle-antiparticle pair of non-self-conjugate bosons is~\cite{Petraki:2015hla}
\begin{multline}
\M_{\vec k \to n\ell m} \simeq - M \sqrt{2\mu} \: 
\[ g_1 {\cal I}_{\vec k, n\ell m} (\h_2 \vec P_\vf) + g_2 {\cal I}_{\vec k, n\ell m} (-\h_1 \vec P_\vf)  +
\right.  \\  \left.
+ \frac{g_1 {\cal K}_{\vec k, n\ell m} (\h_2 \vec P_\vf)+ g_2 {\cal K}_{\vec k, n\ell m} (-\h_1 \vec P_\vf)}{2M\mu}   
\]  \, .
\label{eq:ScalMed_M_extended}
\end{multline}

\subsubsection[Capture into $\ell=0$ bound states]{Capture into $\boldsymbol{\ell=0}$ bound states  \label{sec:ScalarMed_Degen_n00}}

Keeping the terms of order $(|\vec P_\vf| / \k)^2$ from the ${\cal I}_{\vec k, n00}$ integral [cf.~\cref{eq:In00 expansion}] and the zeroth-order terms from the ${\cal K}_{\vec k, n00}$ integral [cf.~\cref{eq:Kn00 expansion}], the amplitude \eqref{eq:ScalMed_M_extended} becomes
\begin{multline}  
|\M_{\vec{k}\to n00}|^2  \simeq  2^7 \pi^2 \a^2 \ \times
\\ \times
\left| 
-\int_0^\infty dx \(\frac{2e^{-x/\ks}}{x}\)\chi_{n,0}^*(x) \, \chi_{|\vec k|,0}^{}(x)
+
\frac{pss_{n,0}^{}}{12}
\(\frac{1+\z^2 \g_{n,0}^2(\ks)}{\z^2}\)^2
\int_0^\infty  dx \: x^2 \, \chi_{n,0}^*(x) \, \chi_{|\vec k|,0}^{}(x)  
\right.  \\  \left.  + \ 
P_2(\hat{\vec k} \cdot \hat{\vec P}_\vf) \ \frac{pss_{n,0}^{}}{6}
\(\frac{1+\z^2 \g_{n,0}^2(\ks)}{\z^2}\)^2 
\int_0^\infty dx \: x^2 \, \chi_{n,0}^*(x) \, \chi_{|\vec k|,2}^{}(x)
\right|^2  \, .
\label{eq:ScalMed_Degen_M_n00}
\end{multline}
Note that here, $\M_{\vec{k}\to n00}$ receives its dominant contribution from the $\ell=0$ and $\ell=2$ components of the scattering state wavefunction. This is in contrast to the formation of zero angular momentum bound states via vector emission [cf.~\cref{eq:VecMed_Mn00_final}], or via scalar emission but by non-degenerate particles [cf.~\cref{eq:ScalMed_NonDegen_M_n00}], where the $\ell=1$ mode of the scattering-state wavefunction dominates.

Substituting \cref{eq:ScalMed_Degen_M_n00} into \cref{eq:DiffSigma}, we find the corresponding cross-section,
\begin{subequations}
\label{eq:ScalMed_Degen_BSF}
\label[pluralequation]{eqs:ScalMed_Degen_BSF}
\beq
\s\BSF^{\{n00\}} \vrel \simeq 
\( \frac{\p\a^2}{\mu^2} \)
\, \a^2 \, S\BSF^{\{n00\}} (\z,\ks;\a) 
\: pss_{n,0}^{1/2}
\, , 
\label{eq:ScalMed_Degen_sigma_n00}
\eeq
where $S\BSF^{\{n00\}}$ includes terms suppressed by different powers of $pss_{n,0}$,
\beq
S\BSF^{\{n00\}} (\z,\ks;\a) = 
\Sigma_0^{\{n00\}} (\z,\ks) 
+ pss_{n,0}    \times  \Sigma_1^{\{n00\}} (\z,\ks) 
+ pss_{n,0}^2  \times  \Sigma_2^{\{n00\}} (\z,\ks) 
\, ,
\label{eq:ScalMed_Degen_Sn00}
\eeq
with
\begin{align}
\Sigma_0^{\{n00\}} (\z,\ks) 
&=  \frac{1}{4} 
\(\frac{1+\z^2 \g_{n,0}^2(\ks)}{\z^2}\) \times
\left|\int_0^\infty dx \(\frac{2e^{-x/\ks}}{x}\) \chi_{n,0}^*(x) \ \x_{|\vec k|,0}^{}(x)
\right|^2
\, ,
\label{eq:Sigma0 n00 ScalarMed Degen}
\\
\Sigma_1^{\{n00\}} (\z,\ks) 
&= -\frac{1}{2^4 \cdot 3} 
\(\frac{1+\z^2 \g_{n,0}^2(\ks)}{\z^2}\)^3 \times 
\nn \\
&\times \(
\[\int_0^\infty dx \(\frac{2e^{-x/\ks}}{x}\)\chi_{n,0}^*(x) \ \x_{|\vec k|,0}^{}(x) \] 
\[\int_0^\infty dx \: x^2 \, \x_{n,0}^{}(x) \, \x_{|\vec k|,0}^*(x) \]
+ \cc \) \, ,
\label{eq:Sigma1 n00 ScalarMed Degen}
\\
\Sigma_2^{\{n00\}} (\z,\ks) 
&=  \frac{1}{2^4 \, 3^2} 
\(\frac{1+\z^2 \g_{n,0}^2(\ks)}{\z^2}\)^5 \ \times
\nn \\
&\times
\(
\frac{1}{4}\left| 
\int_0^\infty dx \: x^2 \, \x_{n,0}^*(x) \, \x_{|\vec k|,0}^{}(x)  
\right|^2  
+
\frac{1}{5} \left|
\int_0^\infty dx \: x^2 \, \x_{n,0}^*(x) \, \x_{|\vec k|,2}^{}(x)
\right|^2
\) \, .
\label{eq:Sigma2 n00 ScalarMed Degen}
\end{align}
\end{subequations}
Note that the dependence of $S\BSF^{\{n00\}}(\z,\ks; \a)$ on $\a$ (independently of $\z$ and $\ks$) arises from the phase-space suppression factor $pss_{n,0}$ in \cref{eq:ScalMed_Degen_Sn00}.

\subsubsection*{Coulomb limit}

The BSF cross-section \eqref{eqs:ScalMed_Degen_BSF} becomes
\begin{subequations} 
\label{eq:ScalMed_Degen_BSF_Coul}
\label[pluralequation]{eqs:ScalMed_Degen_BSF_Coul}
\beq
\lim_{\ks \to \infty} 
\sigma\BSF^{\{n00\}} \vrel 
\simeq \( \frac{\p\a^2}{\mu^2} \) \, \a^2 \, S\BSFC^{\{n00\}} (\z) 
\, , 
\label{eq:ScalMed_Degen_sigma_n00_Coul }
\eeq
where the analytic expression for $S\BSFC^{\{n00\}}$ can be found using 
\cref{eq:DiffSigma,eq:ScalMed_M_extended,eq:In00 Coul expansion,eq:Kn00 Coul expansion}. 
Since the general expression is rather lengthy, here we give the explicit form only for $n=1,2,3$,
\begin{align}
S\BSFCgr (\z) &= 
\(\frac{2\p\z}{1-e^{-2\p\z}}\) \!\times 
\frac{2^6 \, \z^2 (3+2\z^2)}{15 \, (1+\z^2)^2} 
\ e^{-4\z {\rm arccot}\,\z} \, ,
\label{eq:ScalMed_Degen_S100_Coul}
\\
S\BSFC^{\{200\}} (\z) &= 
\(\frac{2\p\z}{1-e^{-2\p\z}}\) \!\times 
\frac{2^5 \, \z^2 (192+272\z^2+100\z^4+15\z^6)}{15 \, (4+\z^2)^4} 
\ e^{-4\z {\rm arccot}\,(\z/2)} \, ,
\label{eq:ScalMed_Degen_S200_Coul}
\\
S\BSFC^{\{300\}} (\z) &= 
\(\frac{2\p\z}{1-e^{-2\p\z}}\) \!\times 
\frac{2^6 \, \z^2 (3^3+7\z^2) (3^7 + 2^2 3^6\z^2 + 915\z^4 + 122\z^6)}{3^3 5 \, (9+\z^2)^5} 
\ e^{-4\z {\rm arccot}\,(\z/3)} \, .
\label{eq:ScalMed_Degen_S300_Coul}
\end{align}
\end{subequations}
At $\z \gg 1$, we find $S\BSFCgr (\z) \simeq  0.16  \times 2\p\z$.

\begin{figure}[b!]
\centering

{\bf Scalar mediator, non-self-conjugate bosonic particle-antiparticle pairs} 
\\
{\bf ~~~Coulomb limit}

\includegraphics[height=7cm]{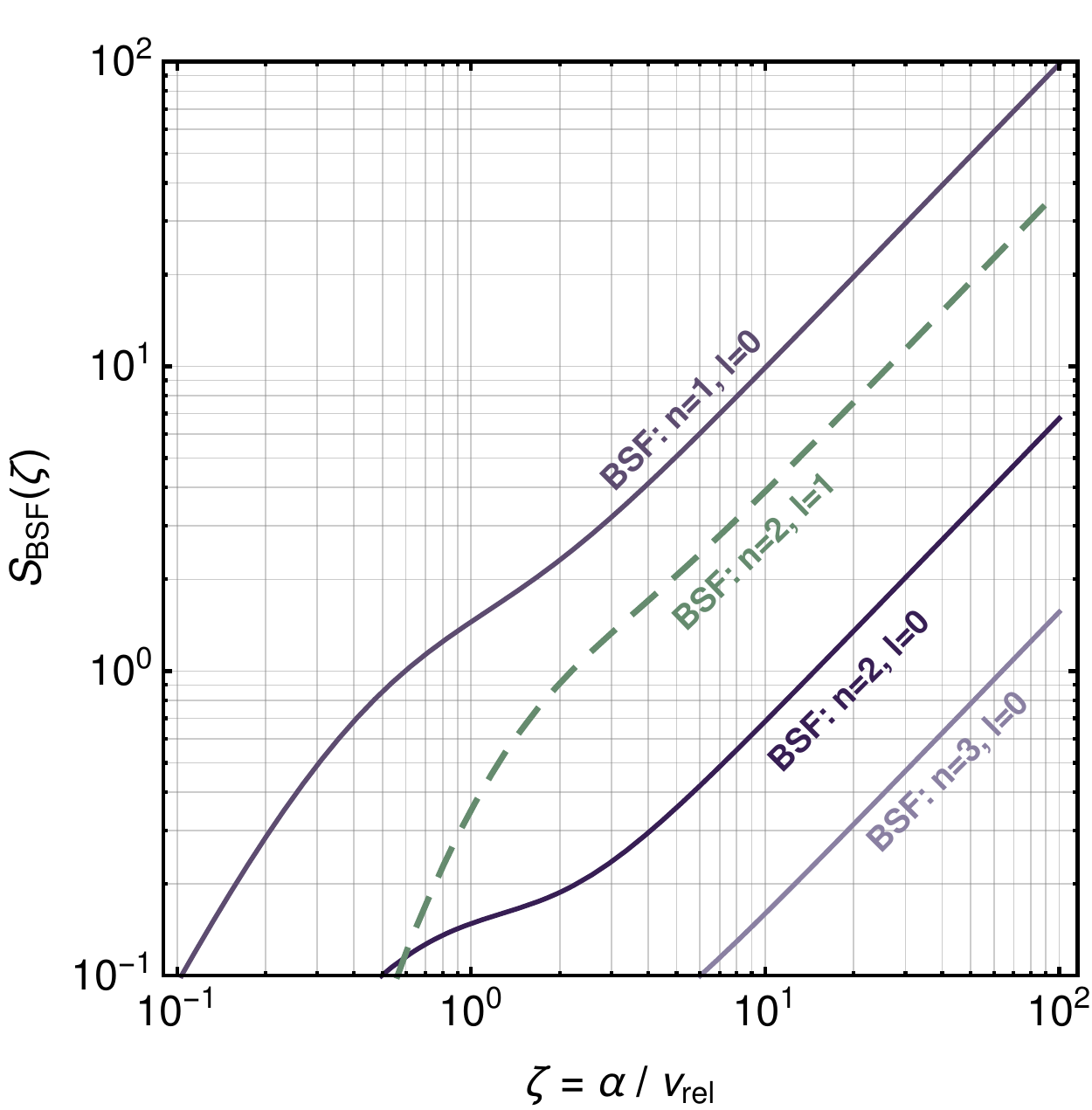}

\caption[]{\label{fig:ScalarMed_Degen_Coulomb}
Velocity dependence of the $S_{\rm BSF}$ factors for the radiative bound-state formation by a bosonic particle-antiparticle pair coupled to a nearly massless scalar. $n$ and $\ell$ denote the principal and the angular-momentum quantum number of the bound states formed. For capture into $\ell=1$ states, we have summed over all $m$ (the angular-momentum projections on one axis). Annihilation (not shown here) is the dominant inelastic process; in the Coulomb regime, 
$\sigma_{_{\rm BSF}}^{\{100\}} / \sigma_{\rm ann}^{sc} = \a^2 \, S_{_{\rm BSF}}^{\{100\}}(\zeta) / S_{\rm ann}^{(0)}(\zeta) \simeq 0.16\alpha^2$, at  $\zeta = \alpha/v_{\rm rel} \gg 1$  
[cf.~eqs.~\eqref{eqs:ScalMed_Degen_BSF_Coul}, \eqref{eq:ScalMed_Ann}, \eqref{eq:ScalMed_BSFoverANN}].
}
\end{figure}

We illustrate the features of $S\BSFgr (\z,\ks;\a)$ and $S\BSFCgr (\z)$ in \cref{fig:ScalarMed_Degen_Coulomb,fig:ScalarMed_Degen}.

\subsubsection[Capture into $\ell=1$ bound states]{Capture into $\boldsymbol{\ell=1}$ bound states}

\begin{figure}[h!]
\centering

{\bf Scalar mediator, non-self-conjugate bosonic particle-antiparticle pairs} 
\\
{\bf Capture into $\boldsymbol{n=2, \ \ell=1}$ states: Resonances}

\parbox[c]{0.42\linewidth}{
\centering
\includegraphics[width=\linewidth]{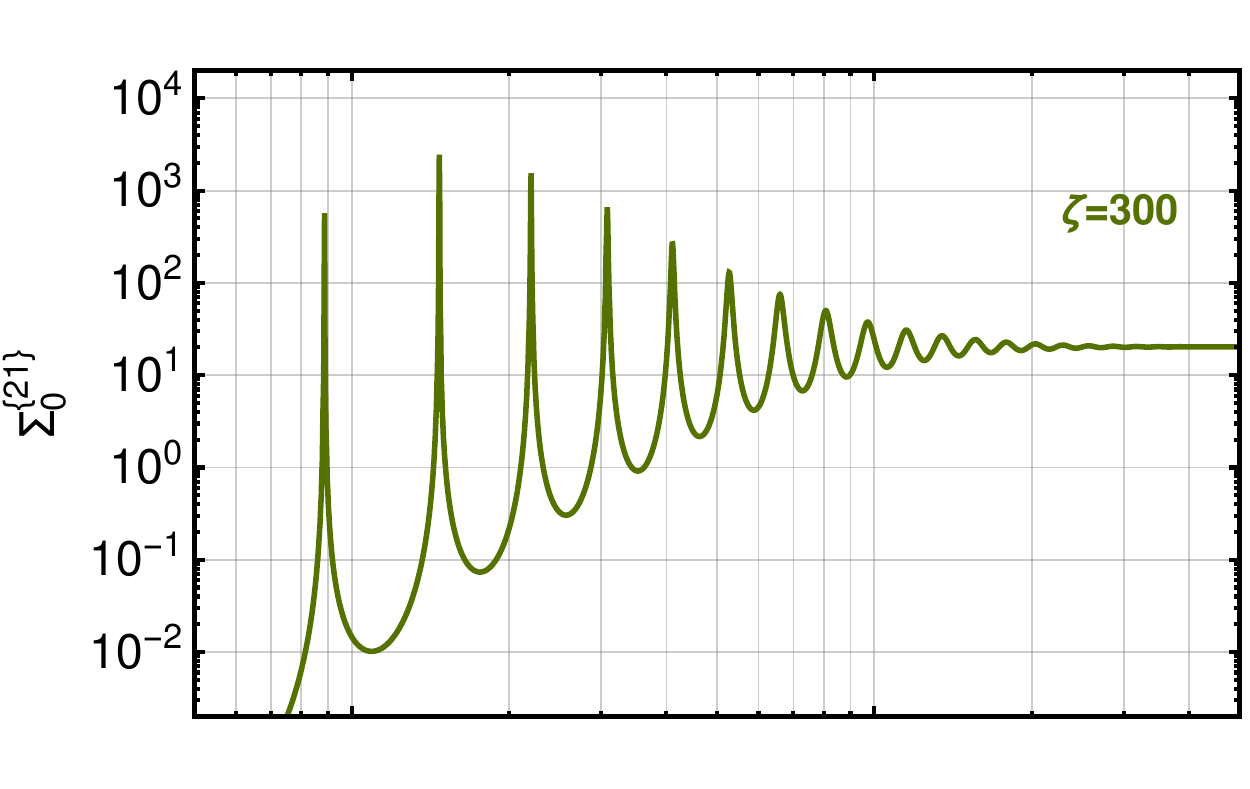}

\vspace{-0.6cm}
\includegraphics[width=\linewidth]{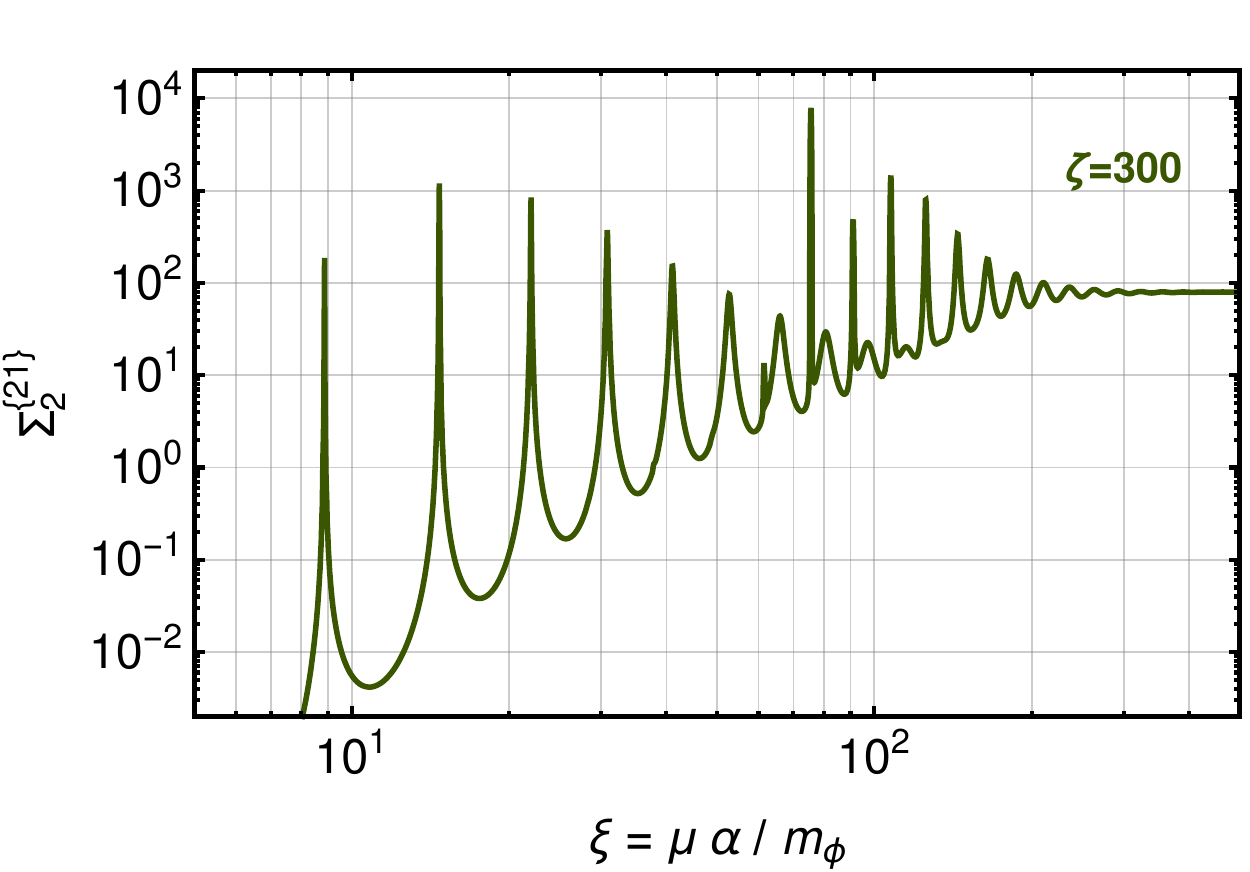}
}
~~~
\parbox[c]{0.42\linewidth}{
\includegraphics[width=\linewidth]{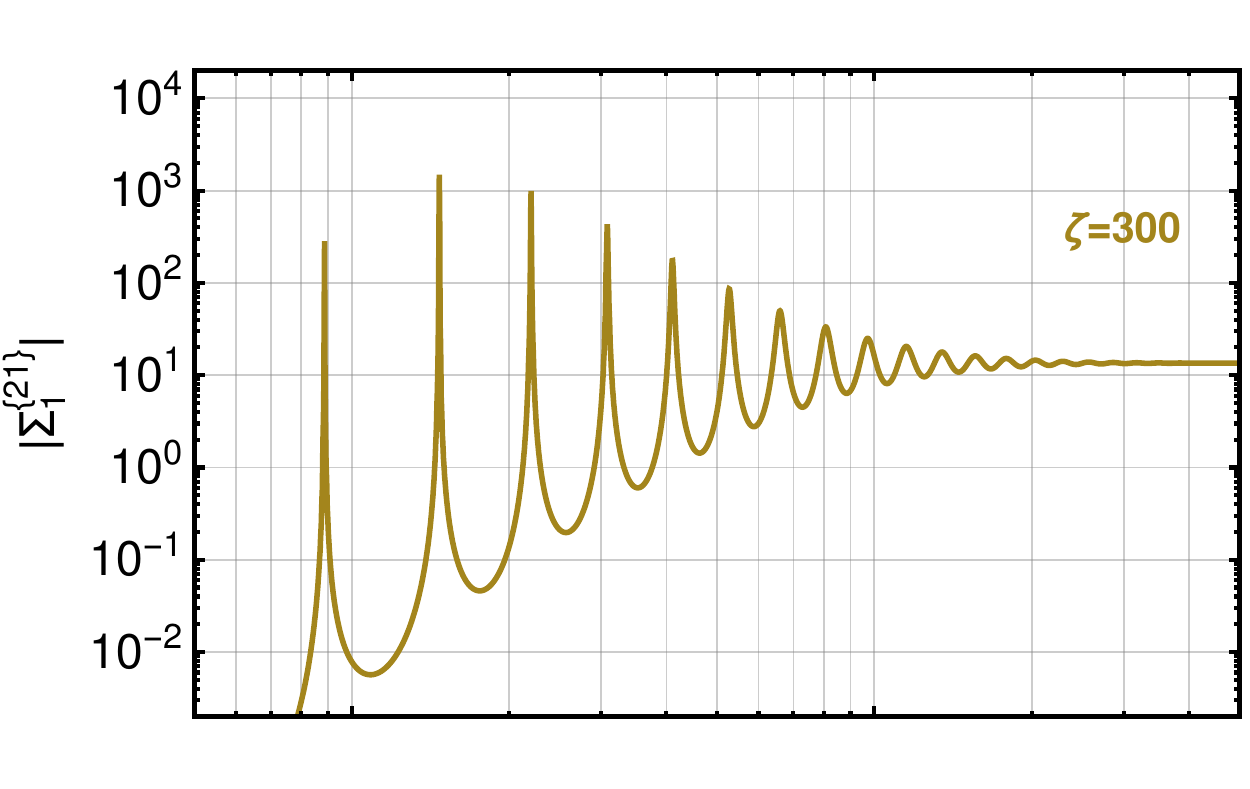}

\vspace{-0.6cm}
\includegraphics[width=\linewidth]{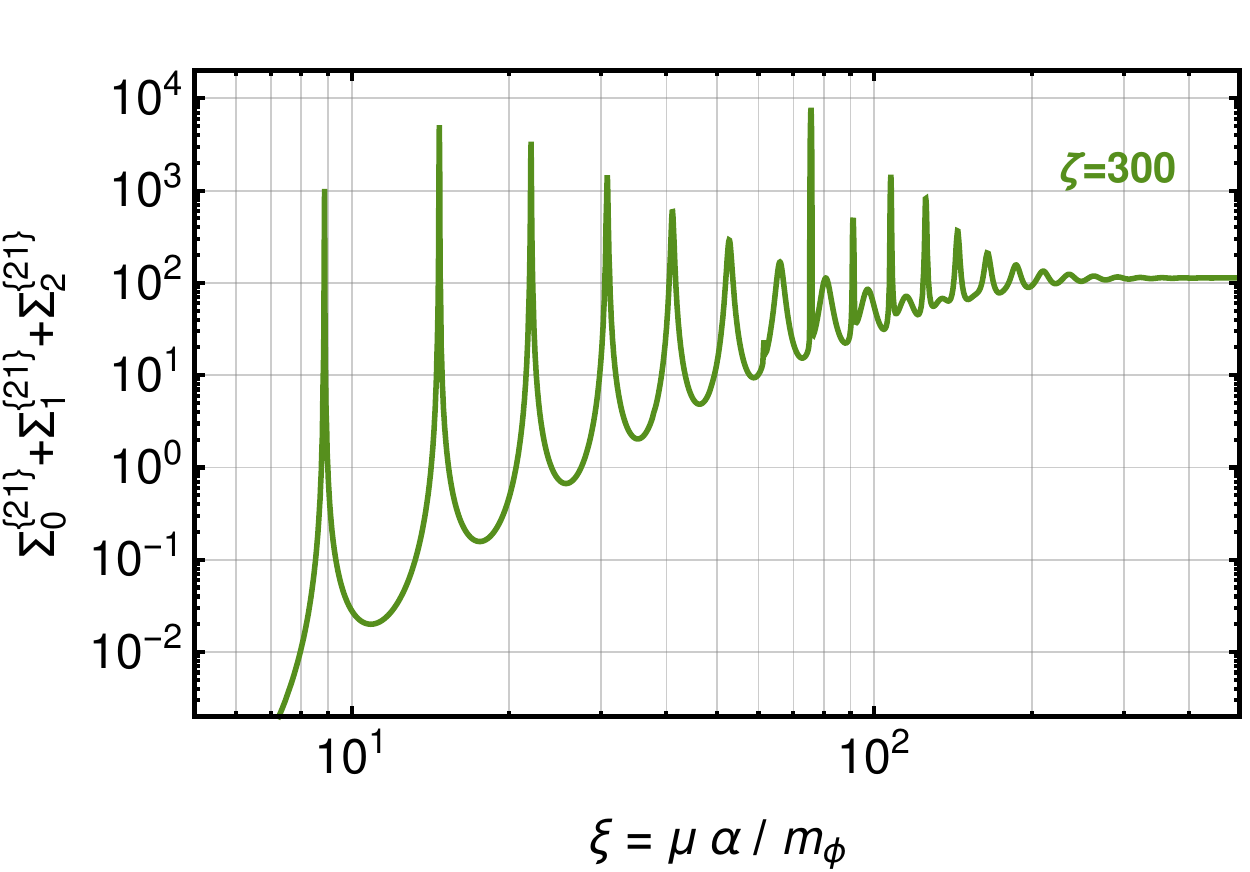}
}

\caption[]{\label{fig:ScalarMed_Degen_Xi_l=1}
The factors $\Sigma_0^{\{21\}}(\zeta,\xi)$, $\Sigma_1^{\{21\}}(\zeta,\xi)$ and $\Sigma_2^{\{21\}}(\zeta,\xi)$ that contribute to the cross-section for the radiative capture of a particle-antiparticle pair into any $n=2, \ell=1$ state, with emission of a scalar force mediator [cf.~eqs~\eqref{eqs:ScalMed_Degen_BSF_l=1}]. The bottom right panel shows their sum, which corresponds to $S_{_{\rm BSF}}^{\{21\}}$, provided that the phase-space suppression is negligible. 
$\Sigma_0^{\{21\}}$ and $\Sigma_1^{\{21\}}$ involve only the $\ell = 1$ mode of the scattering state wavefunction, while $\Sigma_2^{\{21\}}$ includes contributions both from the $\ell=1$ and $\ell=3$ components. The latter is responsible for the sharper resonances at large $\xi$.  
}

\end{figure}

Similarly to the previous section, we keep the terms of order $(|\vec P_\vf| / \k)^2$ from the ${\cal I}_{\vec k, n1m}$ integrals and the zeroth-order terms from the ${\cal K}_{\vec k, n1m}$ integrals [cf.~\cref{eqs:IJK_nlm_expansion}]. Then, from \cref{eq:DiffSigma,eq:ScalMed_M_extended}, we find the total cross-section for capture to any $\ell=1$ state, for a fixed $n$, to be\footnote{We note that the computation of $\s\BSF^{\{210\}}$ with emission of a (nearly massless) scalar mediator in Ref.~\cite{Petraki:2015hla} is incorrect, due to an error in eq.~(F.29).}
\begin{subequations} 
\label{eq:ScalMed_Degen_BSF_l=1}
\label[pluralequation]{eqs:ScalMed_Degen_BSF_l=1}
\beq
\s\BSF^{\{n1\}} \vrel \equiv 
\sum_{m=-1}^1 \s\BSF^{\{n1m\}} \vrel 
\simeq 
\(\frac{\p\a^2}{\mu^2} \) 
\: \a^2 \, S\BSF^{\{n1\}} (\z,\ks;\a) 
\  pss_{n,1}^{1/2}
\, , 
\label{eq:sigma n1 ScalarMed Degen}
\eeq
where $S\BSF^{\{n1\}}$ includes terms suppressed by different powers of $pss_{n,1}$,
\beq
S\BSF^{\{n1\}} (\z,\ks;\a) =
\Sigma_0^{\{n1\}} (\z,\ks) 
+ pss_{n,1}    \times  \Sigma_1^{\{n1\}} (\z,\ks) 
+ pss_{n,1}^2  \times  \Sigma_2^{\{n1\}} (\z,\ks) \, ,
\label{eq:Sn1 ScalarMed Degen}
\eeq
with
\beq
\Sigma_0^{\{n1\}} (\z,\ks) 
=  \frac{3}{4} \(\frac{1+\z^2 \g_{n,1}^2(\ks)}{\z^2}\) \times
\left|\int_0^\infty dx 
\(\frac{2 e^{-x/\ks}}{x}\) \x_{n,1}^*(x) 
\ \x_{|\vec k|,1}^{}(x)
\right|^2
\, ,
\label{eq:Sigma0 n1 ScalarMed Degen}
\eeq
\begin{multline}
\Sigma_1^{\{n1\}} (\z,\ks) 
= - \frac{1}{2^4} \(\frac{1+\z^2 \g_{n,1}^2(\ks)}{\z^2}\)^3 \times
\\ \times \(
\[\int_0^\infty dx 
\(\frac{2 e^{-x/\ks}}{x}\) \x_{n,1}^*(x) \: \x_{|\vec k|,1}^{}(x) \] 
\[\int_0^\infty dx \: x^2 \, \x_{n,1}^{}(x) \: \x_{|\vec k|,1}^*(x) \]
+ \cc \),
\label{eq:Sigma1 n1 ScalarMed Degen}
\end{multline}
\begin{multline}
\Sigma_2^{\{n1\}} (\z,\ks) =  \frac{1}{2^4 \cdot 5^2} 
\(\frac{1+\z^2 \g_{n,1}^2(\ks)}{\z^2}\)^5 \ \times
\\
\times
\( 11 \, \left| \int_0^\infty dx \: x^2 \, \x_{n,1}^*(x) \, \x_{|\vec k|,1}^{}(x) \right|^2  
  + 4 \, \left| \int_0^\infty dx \: x^2 \, \x_{n,1}^*(x) \, \x_{|\vec k|,3}^{}(x) \right|^2
\) \, .
\label{eq:Sigma2 n1 ScalarMed Degen}
\end{multline}
\end{subequations}
We showcase the resonant features of $S\BSF^{\{21\}} (\z,\ks;\a)$ in \cref{fig:ScalarMed_Degen_Xi_l=1}.

\subsubsection{Annihilation \label{sec:ScalarMed_Ann}}

We now consider the annihilation of a bosonic particle-antiparticle pair into two scalar force mediators, 
$\x^* + \x \to \vf+\vf$. The dominant contribution to the annihilation cross-section arises from the $\ell=0$ component of the scattering state wavefunction, 
\begin{subequations} 
\label{eq:ScalMed_Ann}
\label[pluralequation]{eqs:ScalMed_Ann}
\begin{align}
\s\ann^{sc} \vrel &= \frac{\pi \a^2}{\mu^2} \times S\ann^{(0)} (\z,\ks) \, ,
\label{eq:ScalMed_Ann_Scalars}
\end{align}
where $S\ann^{(0)}$ and its Coulomb limit are given in \cref{eq:Sann0,eq:Sann0_Coul}; we repeat them here for convenience
\begin{align}
S\ann^{(0)} (\z,\ks) 
&\equiv |\f_{\vec k}(\vec r = 0)|^2 
= \lim_{x\to 0} \[ \frac{\x_{|\vec k|,\ell=0}^{}(x)}{x} \] \, ,
\label{eq:Sann0_re} 
\\
S\annC^{(0)} (\z) &\equiv \lim_{\ks\to\infty} S\ann^{(0)}(\z,\ks) = \frac{2\p\z}{1-e^{-2\pi\z}} \, . 
\label{eq:Sann0_Coul_re} 
\end{align}
\end{subequations}

\subsection{Identical particles \label{sec:ScalarMed_Identical}}

If the interacting particles are identical, then their total wavefunction is either symmetric or antisymmetric in their interchange, depending on whether the particles are bosons or fermions, respectively. For a pair of fermions, the spatial wavefunction depends on their spin state. A pair of spin-$1/2$ particles may be either in the antisymmetric spin-singlet state, or in the symmetric spin-triplet state. Their spatial wavefunction should then be symmetric or antisymmetric, respectively. Thus, for a pair of identical particles, the scattering-state spatial wavefunctions are 
\begin{subequations}
\label{eq:WFs_Sym&Antisym}
\label[pluralequation]{eqs:WFs_Sym&Antisym}
\begin{align}
\text{Bosons, Fermions with total spin 0: }& \qquad
\frac{1}{\sqrt{2}} \left[ \f_{\vec k} (\vec r) + \f_{-\vec k} (\vec r) \right] \, ,
\label{eq:WFs_Sym}
\\
\text{Fermions with total spin 1: }& \qquad
\frac{1}{\sqrt{2}} \left[ \f_{\vec k} (\vec r) - \f_{-\vec k} (\vec r) \right] \, .
\label{eq:WFs_Antisym}
\end{align}
\end{subequations}
The wavefunction \eqref{eq:WFs_Sym} implies that the contribution of the even-$\ell$ modes participating in a process is doubled, while the contribution of the odd-$\ell$ modes vanishes. The opposite is true for the wavefunction \eqref{eq:WFs_Antisym}. 

For a pair of bosonic identical particles (IP), the BSF and annihilation cross-sections are related to those for distinguishable particles (DP), computed in \cref{sec:ScalarMed_Degen}, as follows
\begin{subequations}
\label{eq:IdenticalBosons}
\label[pluralequation]{eqs:IdenticalBosons}
\begin{align}
\s\BSF^{\{n00\}} \text{ for IP}
~&=~ 2 \times [\s\BSF^{\{n00\}} \text{ for DP],~~~cf.~\cref{eq:ScalMed_Degen_sigma_n00}} \, ,
\label{eq:IdenticalBosons_BSF_n00}
\\
\s\BSF^{\{n1\}} \text{ for IP} ~&=~ 0  \, ,
\label{eq:IdenticalBosons_BSF_n1}
\\
\s\ann \text{ for IP} 
~&=~ 2 \times [\s\ann \text{ for DP],~~~cf.~\cref{eq:ScalMed_Ann_Scalars}} \, .
\label{eq:IdenticalBosons_ann}
\end{align}
\end{subequations}
Note that the vanishing result in \cref{eq:IdenticalBosons_BSF_n1} holds to working order in $\a$; contributions of higher order in $\a$, which we have not computed here, will yield a non-zero cross-section.

\subsection{Discussion \label{sec:ScalarMed_Disc}}

The general aspects discussed in \cref{sec:VecMed_Discussion} in the context of BSF via emission of a vector boson, are pertinent also for BSF with emission of a scalar boson. Here, we point out some features that are specific to the latter.

For BSF via emission of a scalar boson, the dominant transition modes are different for particle-antiparticle or identical-particle pairs, than for pairs of particles with different masses and couplings to the emitted scalar boson. The capture to the ground state is dominated by the monopole and quadrupole modes in the first case [$\ell=0$ and $\ell=2$ modes of the scattering-state wavefunction, respectively, cf.~\cref{eq:ScalMed_Degen_BSF}], and by the dipole mode in the second case [$\ell=1$ mode, cf.~\cref{eq:ScalMed_NonDegen_BSF}]. The monopole and quadrupole modes contribute also to the latter case, but at higher order in the coupling than the dipole transition.

For bosonic particle-antiparticle pairs and pairs of annihilating identical bosons, BSF via scalar emission is significantly slower than annihilation into two scalar bosons. In the Coulomb regime, using  \cref{eq:ScalMed_Degen_BSF_Coul,eq:ScalMed_Ann,eq:IdenticalBosons}, we find, for both self-conjugate and non-self-conjugate species,
\beq
\frac{\s\BSFCgr}{\s\annC}
= \a^2 \: \frac{2^6 \z^2 (3+2\z^2)}{15 (1+\z^2)^2} \: e^{-4\z {\rm arccot}\z} \, .
\label{eq:ScalMed_BSFoverANN}
\eeq
At $\z \gtrsim 1$, this ratio becomes $\sim 0.16\a^2$. 
Away from the Coulomb regime, the relative significance of BSF with respect to annihilation is further diminished. Indeed, as we have seen, the contribution of an $\ell$ mode of the scattering-state wavefunction to an inelastic process scales as $\s_{\rm inel}\vrel \propto \vrel^{2\ell}$ at low velocities.
For BSF, the $\ell=2$ contribution will thus diminish, leaving ultimately only the $\ell=0$ mode at sufficiently low $\vrel$. \Cref{eq:ScalMed_BSFoverANN} and the above discussion imply that the formation and decay of unstable bound states via emission of a scalar mediator cannot deplete significantly the DM density in the early universe, or enhance the indirect detection signals today, in contrast to the case of BSF via vector emission~\cite{vonHarling:2014kha,An:2016gad}. 

Nevertheless, BSF may be important for the capture of asymmetric DM into stable bound states.
Moreover, because of the different velocity scaling of the various $\ell$ modes, even for pairs of particles with different masses/couplings to the radiated scalar boson, the $\ell=0$ mode may dominate the formation of zero-angular momentum bound states at low enough velocities. This is despite the contribution of the $\ell=0$ mode being suppressed by a higher order in the coupling with respect to that of the $\ell=1$ mode.


\clearpage

\begin{subfigures}

\label{fig:ScalarMed_Degen}
\label[pluralfigure]{figs:ScalarMed_Degen}

\begin{figure}[p]
\centering

{\bf Scalar mediator, non-self-conjugate bosonic particle-antiparticle pairs} 
\\
{\bf Capture into the ground state: Resonances}

\parbox[c]{0.4\linewidth}{
\centering
\includegraphics[width=\linewidth]{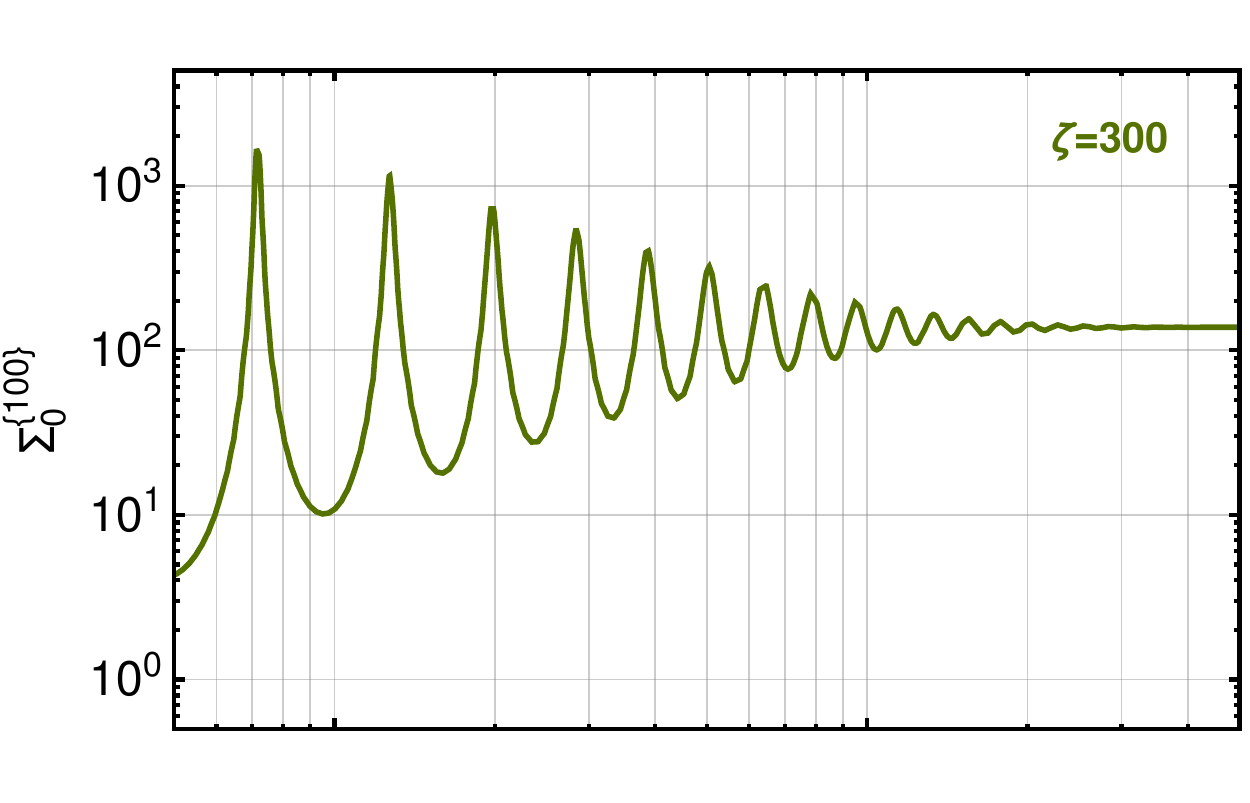}

\vspace{-0.6cm}
\includegraphics[width=\linewidth]{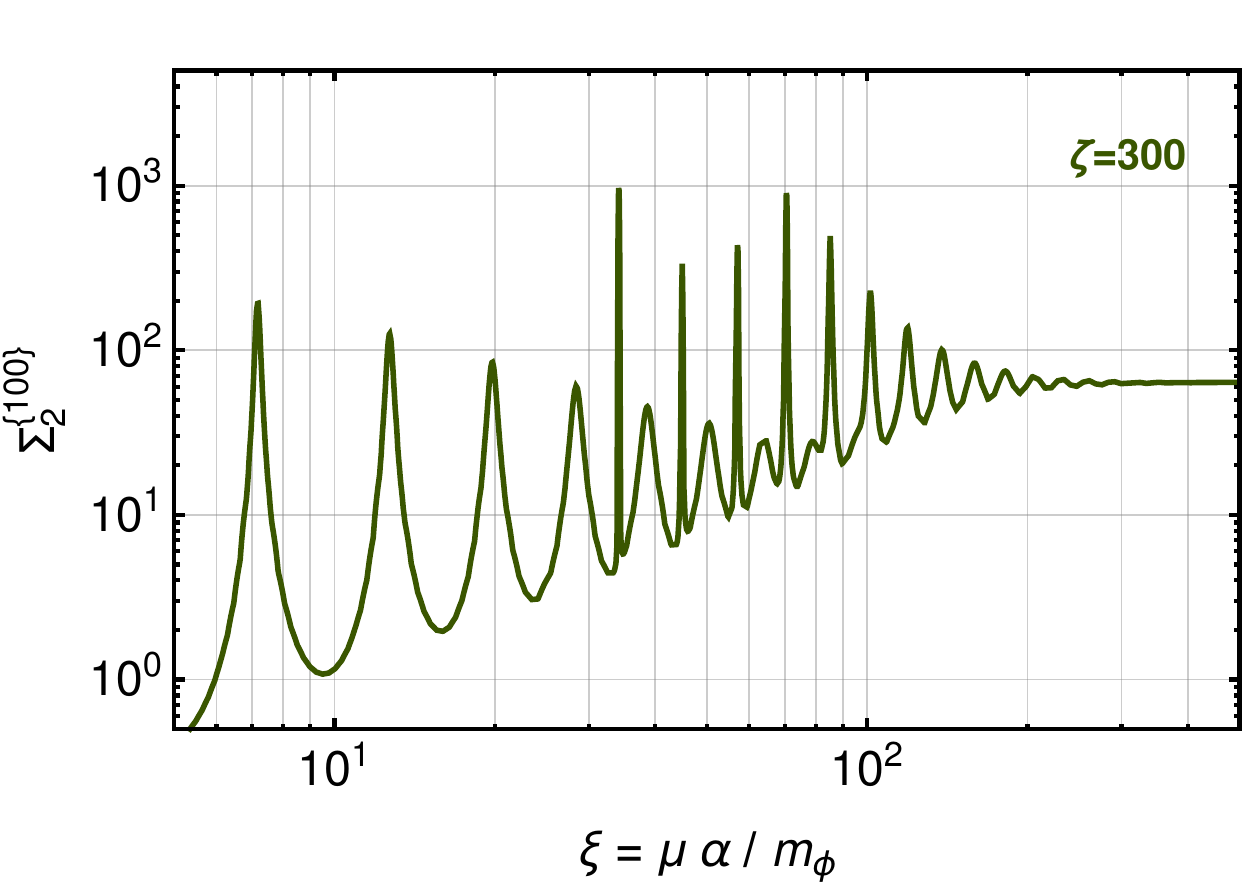}
}
~~~~~
\parbox[c]{0.4\linewidth}{
\includegraphics[width=\linewidth]{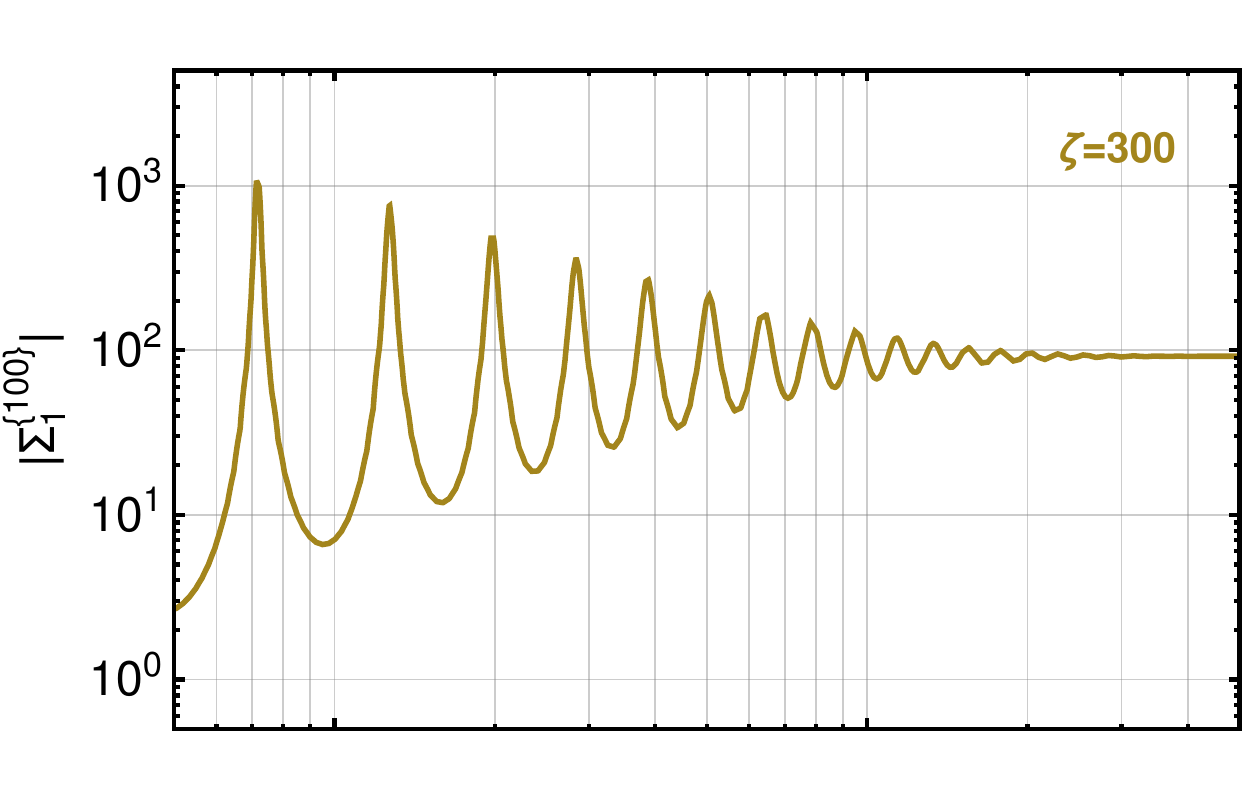}

\vspace{-0.6cm}
\includegraphics[width=\linewidth]{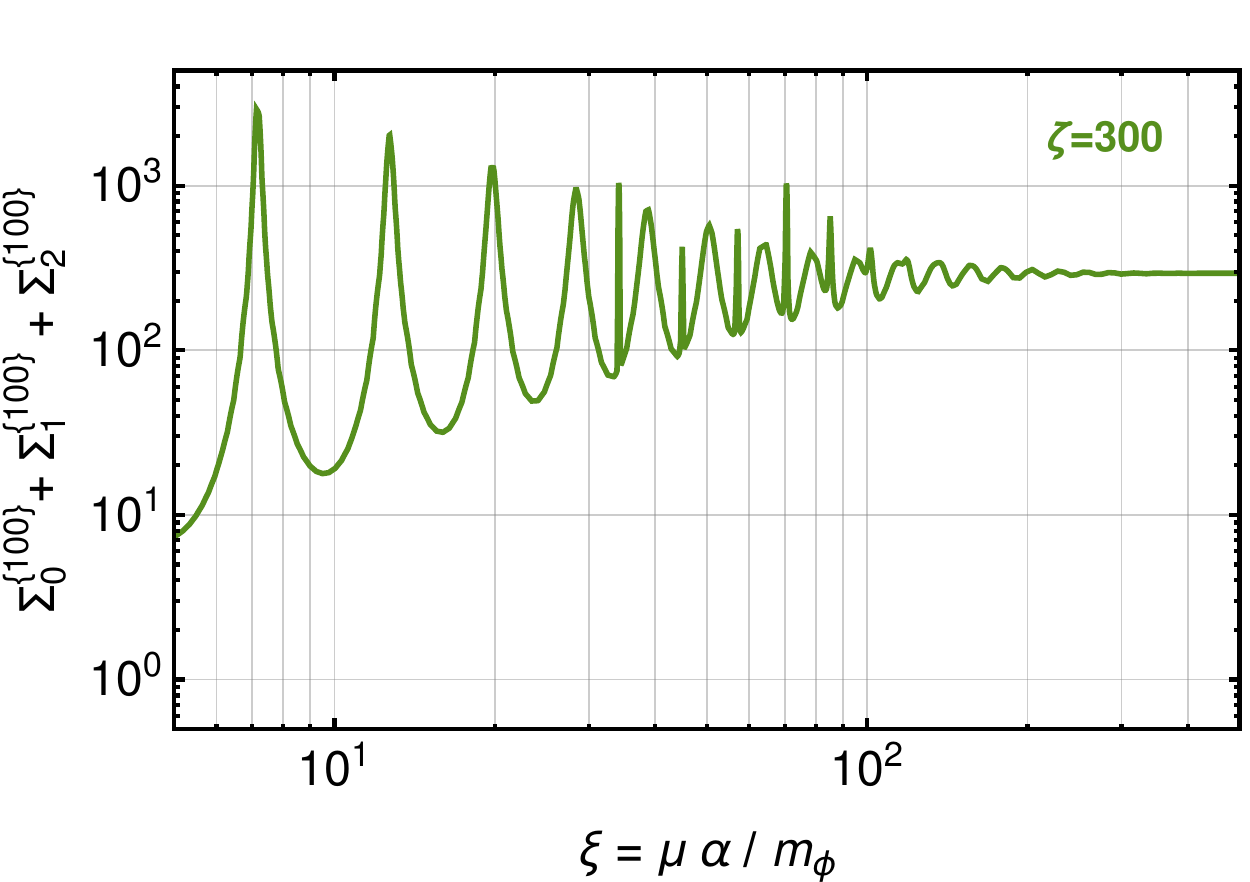}
}

\caption[]{\label{fig:ScalarMed_Degen_Xi}
The factors $\Sigma_0^{\{100\}}(\zeta,\xi)$, $\Sigma_1^{\{100\}}(\zeta,\xi)$ and $\Sigma_2^{\{100\}}(\zeta,\xi)$ that contribute to the cross-section for the radiative capture of a particle-antiparticle pair, into the ground state, with emission of a scalar force mediator [cf.~eqs.~\eqref{eqs:ScalMed_Degen_BSF}]. The bottom panel shows their sum, which corresponds to $S_{_{\rm BSF}}^{\{100\}}$ if the phase-space suppression is negligible. 
$\Sigma_0^{\{100\}}$ and $\Sigma_1^{\{100\}}$ involve only the $\ell = 0$ mode of the scattering-state wavefunction, while $\Sigma_2^{\{100\}}$ includes contributions both from the $\ell=0$ and $\ell=2$ components. The $\ell$ modes determine the resonances.  

}

\bigskip

{\bf Capture into the ground state: Velocity dependence}

\includegraphics[height=6.7cm]{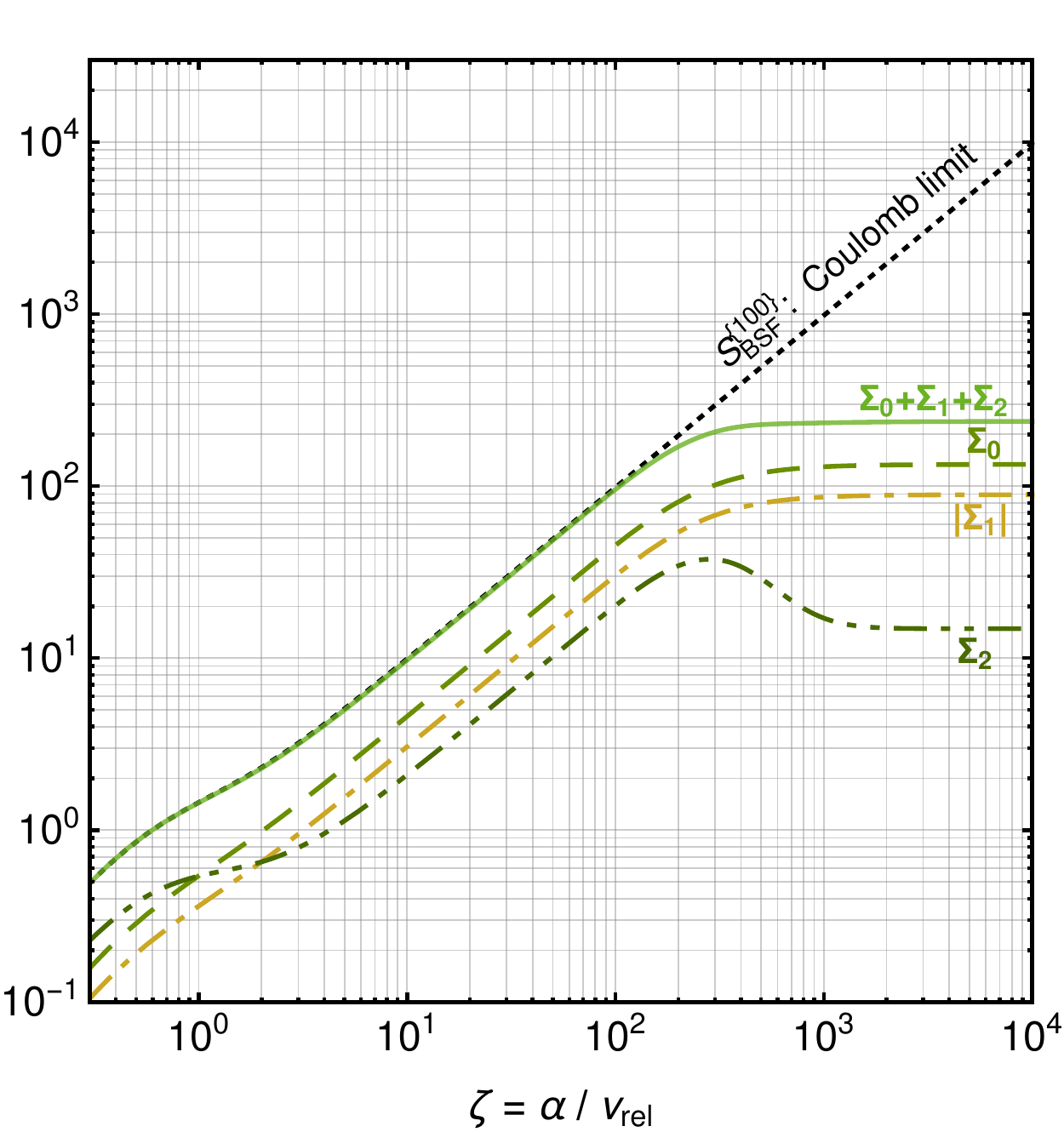}
~~~~
\includegraphics[height=6.7cm]{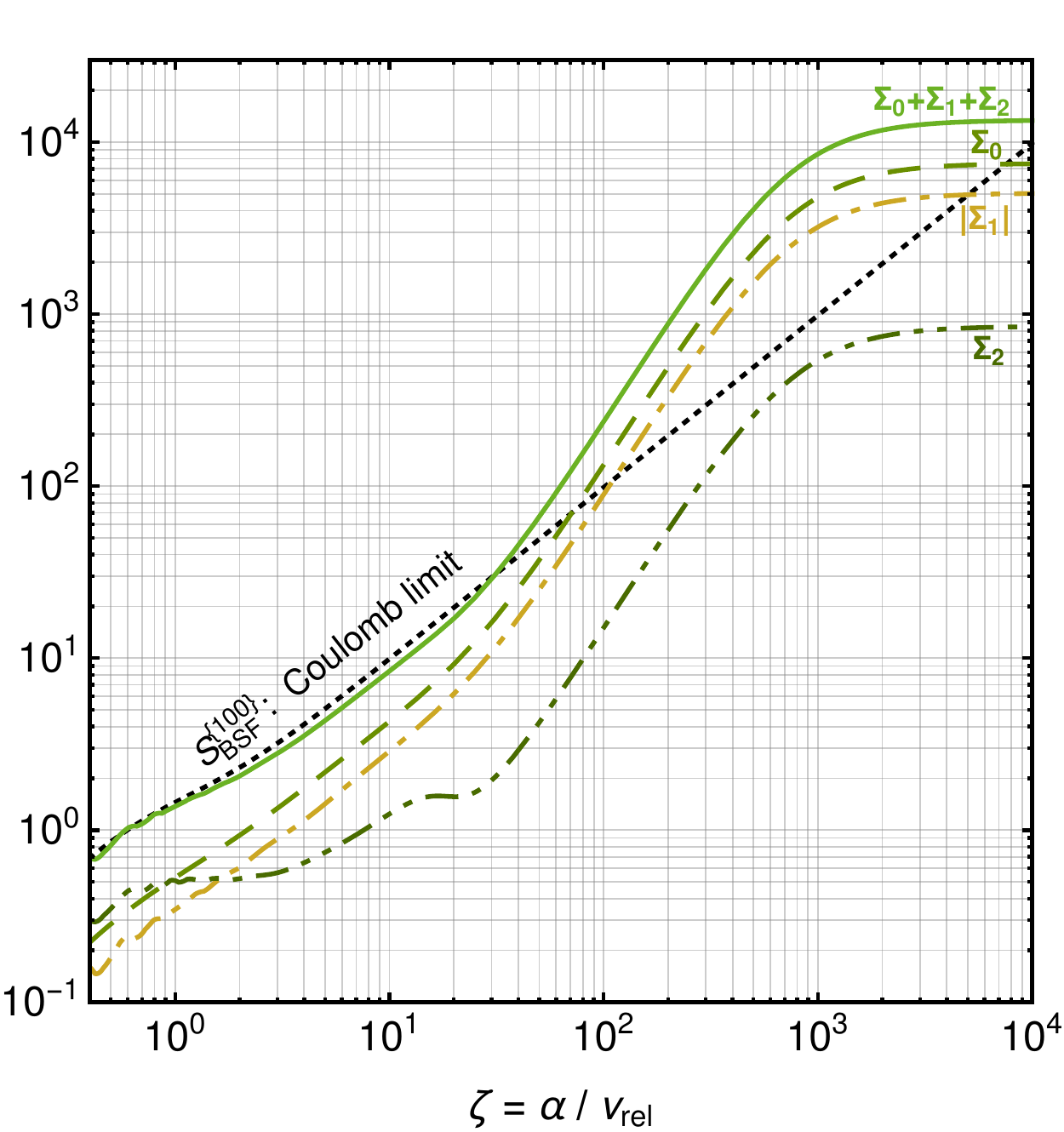}

\caption[]{\label{fig:ScalarMed_Degen_Zeta}
Velocity dependence of the various contributions to the cross-section for the radiative capture of a particle-antiparticle pair into the ground state, with emission of a scalar force mediator: 
$\Sigma_0^{\{100\}}(\zeta,\xi)$ (dashed lines), 
$\Sigma_1^{\{100\}}(\zeta,\xi)$ (dot-dashed lines), 
$\Sigma_2^{\{100\}}(\zeta,\xi)$ (dot-dot-dashed lines),
and their sum (solid lines)   
[cf.~eq.~\eqref{eqs:ScalMed_Degen_BSF}].  
Because $\Sigma_0^{\{100\}}$, $\Sigma_1^{\{100\}}$ and $\Sigma_2^{\{100\}}$ involve the $\ell=0$ mode of the scattering-state wavefunction, they saturate to a constant value at low velocities. 
$\Sigma_2^{\{100\}}$ contains also a contribution from the $\ell=2$ mode; away from the Coulomb regime, this contribution decreases with decreasing velocity, and eventually becomes subdominant, as can been seen in the left panel.
}

\end{figure}
\end{subfigures}

\clearpage


\section{Conclusion \label{Sec:Conc}}

We have computed the cross-sections for the radiative formation of bound states by particles whose interaction is described in the non-relativistic regime by a Yukawa potential. We considered capture processes via emission of either a vector or a scalar boson, and inspected in detail the features of the cross-sections in the entire parametric regime where bound states exist. Bound-state effects can be important both for hidden-sector scenarios in which DM couples directly to light force mediators, as well as for TeV-scale WIMP models where non-perturbative effects due to the long-range nature of the interactions have already been shown to be significant~\cite{Cirelli:2005uq,  Cirelli:2007xd, Cirelli:2009uv, Hryczuk:2010zi, Hryczuk:2014hpa, Beneke:2014hja, Cirelli:2014dsa}. 

The formation of DM bound states has multifaceted implications.  The formation of unstable bound states in the early universe, and their subsequent decay, can deplete the density of symmetric or self-conjugate thermal-relic DM, and therefore affect the predictions for its mass and couplings~\cite{vonHarling:2014kha, Ellis:2015vaa, Kim:2016zyy, Kim:2016kxt}. The same chain of processes taking place in the dense environment of haloes today~\cite{Pospelov:2008jd, MarchRussell:2008tu, An:2016gad, Cirelli:2016rnw}, or in the interior of stars where DM may be captured~\cite{Kouvaris:2016ltf}, enhances the expected rate of the indirect detection signals and results in stronger constraints~\cite{Cirelli:2016rnw}. Moreover, it gives rise to correlated spectral features; besides the high-energy radiation produced in the decay of the bound states, their formation is accompanied by the emission of low-energy radiation that dissipates the binding energy, and which may be detectable.

\begin{figure}[t]
\centering
\includegraphics[height=0.45\textwidth]{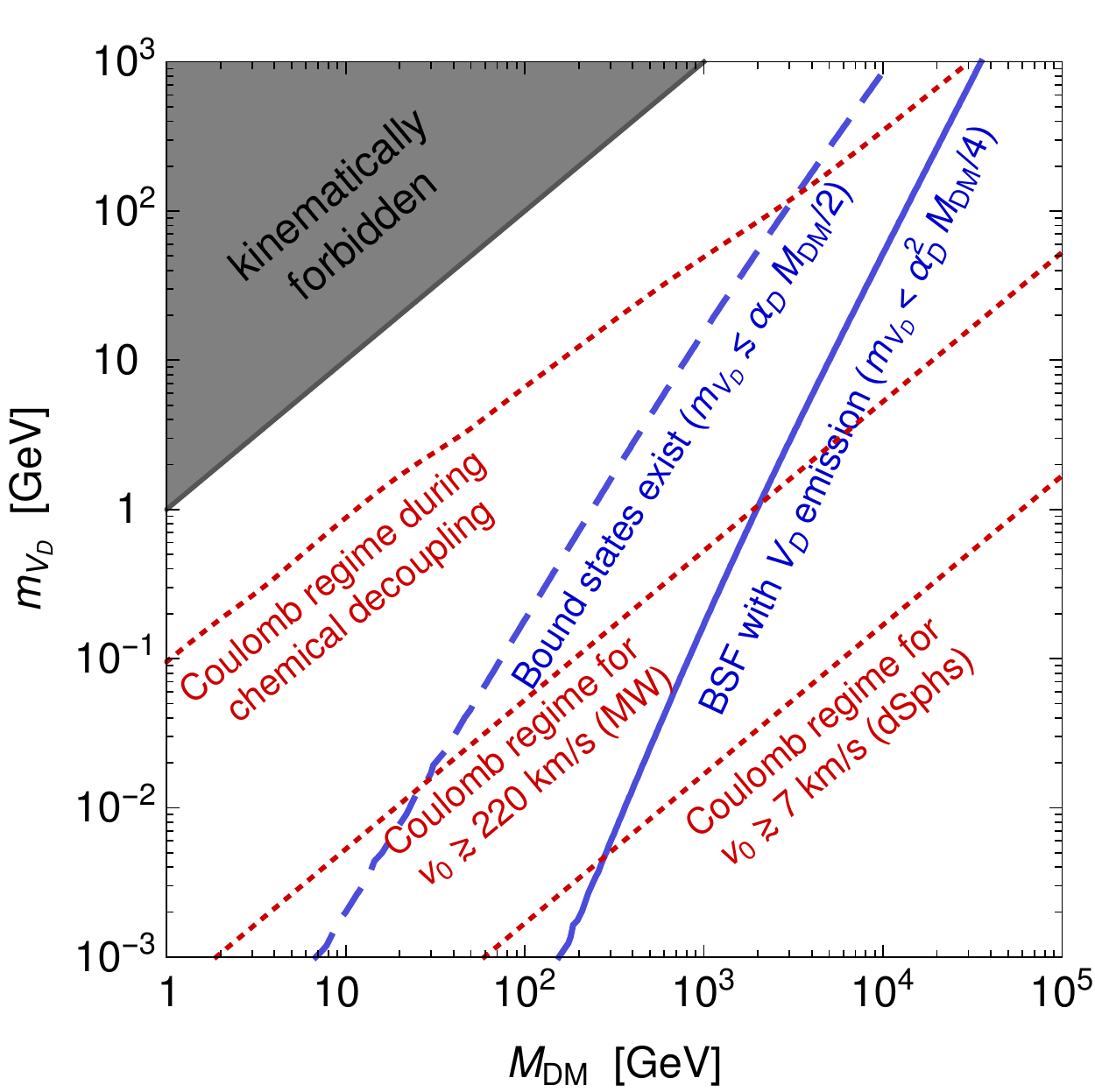}~~~~
\includegraphics[height=0.45\textwidth]{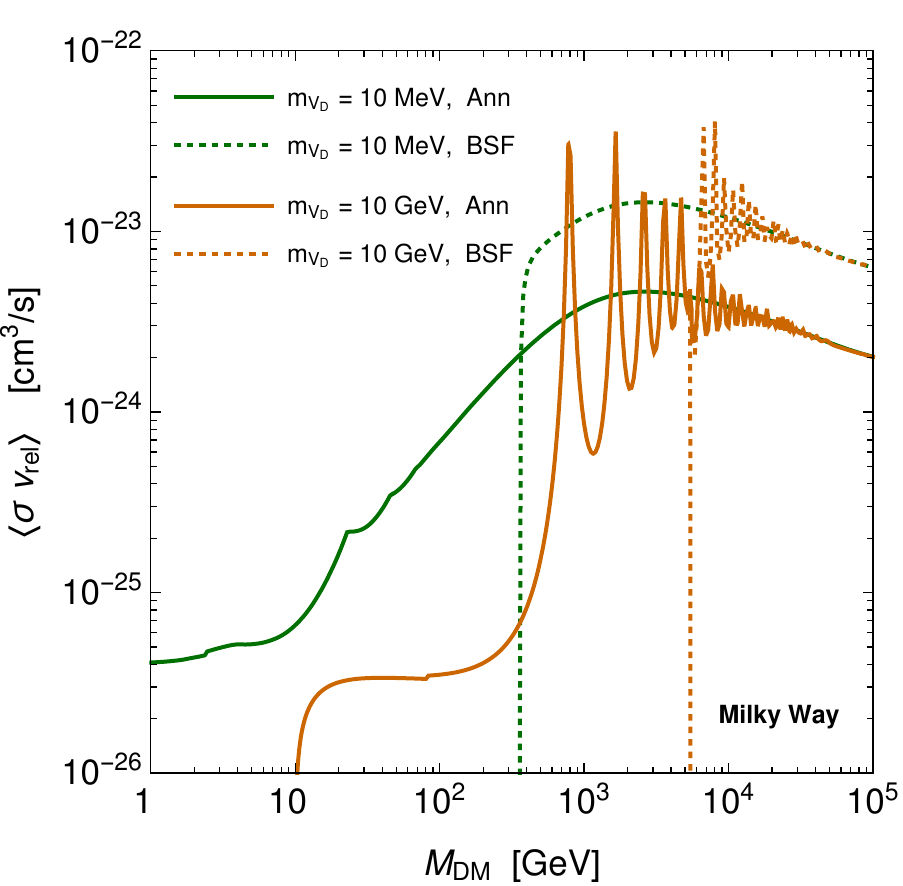}
\caption[]{\label{fig:U1model}
(Plots reproduced from Ref.~\cite{Cirelli:2016rnw}.)  
We consider dark matter consisting of Dirac Fermions $X,\,\bar{X}$, of mass $M_{\rm DM}$, that couple to a dark photon of mass $m_{V_D}$. We assume that the DM relic density arose from thermal freeze-out in the hidden sector, and determine the dark fine-structure constant $\alpha_D$ as a function of $M_{\rm DM}$ and $m_{V_D}$, taking into account both the direct DM annihilation into radiation and the formation and decay of unstable bound states, wherever applicable, according to Ref.~\cite{vonHarling:2014kha}.

\smallskip
\emph{Left:} Below the blue dashed line, the Bohr momentum of the $X-\bar{X}$ pair is larger than the dark photon mass ($\xi \gtrsim 1$); bound states exist (cf.~appendix~\ref{app:BoundState WF num}) and the Sommerfeld effect is significant. 
However, bound states can form with emission of a dark photon only in the parameter space below the blue solid line, where the mediator mass is less than the binding energy.

\smallskip 
Below the red dotted lines, the average momentum transfer between $X$ and $\bar{X}$ is larger than the mediator mass, $(M_{\rm DM}/2)v_{\rm rel} \gtrsim m_{V_D}$ [or $\xi \gtrsim \zeta$,  cf.~eq.~\eqref{eq:CoulombRegime}], for typical velocities during the DM chemical decoupling in the early universe, in the Milky Way, and in the Dwarf Spheroidal galaxies, as indicated in the plot. (We take $v_{\rm rel} = \sqrt{2}v_0$.) In this regime, the interaction cross-sections are well approximated by their Coulomb limit. In the Coulomb limit, bound-state formation is faster than annihilation, $\sigma_{\rm BSF}^{\{100\}} \simeq 3.13 \, \sigma_{\rm ann}$ [cf.~fig.~\ref{fig:VecMed_Coulomb} and eqs.~\eqref{eq:VecMed_S100_Coul}, \eqref{eq:Sann0_Coul}].

\smallskip
\emph{Right:} 
The cross-sections times relative velocity, for annihilation (solid lines) and radiative capture to the ground state (dotted lines), averaged over the DM velocity distribution in the Milky Way ($v_0 = 220$~km/s, $v_{\rm esc} = 533$~km/s), for two different values of the dark photon mass,  $m_{V_D} = 10$~MeV (green smooth lines) and  $m_{V_D} = 10$~GeV (orange lines with resonances). 

\smallskip
For $m_{V_D} = 10$~MeV, the Coulomb limit pertains to $M_{\rm DM} \gtrsim 10$~GeV, as can be seen in the left panel. Bound-state formation is faster than annihilation wherever it is kinematically allowed, $M_{\rm DM} \gtrsim 400$~GeV. 
For $M_{\rm DM} \lesssim 1~{\rm TeV}$, the Sommerfeld effect is not significant during freeze-out, which implies roughly $\alpha_D \propto M_{\rm DM}$. However, in the Milky Way, where the average velocity is lower, the Sommerfeld effect enhances the cross-sections proportionally to $\alpha_D/v_{\rm rel}$, for $M_{\rm DM} \gtrsim 10$~GeV.
Then, in the mass range $M_{\rm DM} \sim 10~{\rm GeV} - 1~{\rm TeV}$, 
the annihilation cross-section scales roughly as $\langle \sigma v_{\rm rel}\rangle \propto M_{\rm DM}$. 
For $M_{\rm DM} \gtrsim$~TeV, the Sommerfeld enhancement is operative both during freeze-out and in the Milky Way, which causes the annihilation and BSF cross-sections to vary only slowly with $M_{\rm DM}$.

\smallskip
For $m_{V_D} = 10$~GeV, the Coulomb limit pertains only to $M_{\rm DM} \gtrsim 30$~TeV. For lower values of $M_{\rm DM}$, both the annihilation and the BSF cross-sections exhibit resonances, which are however located at different values of the DM mass. Bound-state formation may still dominate over annihilation in the mass range where it is kinematically possible. 
At the BSF threshold ($M_{\rm DM} \simeq 5.5$~TeV), a disruption appears in the annihilation cross-section, and the resonances are shifted to lower values of the DM mass; this is due to the effect of BSF on the DM relic density, which reduces the estimated $\alpha_D$.
}
\end{figure}

We showcase some of the above in \cref{fig:U1model}, for a minimal model of fermionic DM coupled to a light but massive dark photon that mixes kinetically with hypercharge~\cite{Holdom:1985ag,Foot:1991kb}. Models of this kind are frequently invoked in the literature~\cite{Kors:2004dx, Feldman:2006wd, Pospelov:2007mp, Fayet:2007ua, Goodsell:2009xc, Fayet:2016nyc}, for example in the context of self-interacting DM~\cite{Spergel:1999mh,Feng:2009mn,Loeb:2010gj}, as well as a plausible explanation of various astrophysical anomalies~\cite{ArkaniHamed:2008qn, Pospelov:2008jd, Cholis:2008qq, Abdullah:2014lla, Berlin:2014pya}. 
A thorough investigation of its phenomenology --- including, for first time in the literature, a self-consistent treatment of bound-state effects both in the DM relic density determination and the indirect detection signals --- has been recently carried out in Ref.~\cite{Cirelli:2016rnw}, where it was demonstrated that the formation and decay of bound states strengthen the constraints derived from $\gamma$-ray observations of the Milky Way and its Dwarf Spheroidal galaxies.

Beyond symmetric or self-conjugate DM, bound-state effects can be even more significant in asymmetric DM models. Asymmetric DM with long-range self-interactions can form stable bound states.
Inside haloes today, the low-energy radiation emitted during capture into a bound state, or in various level transitions between bound-state energy levels, can give rise to signals observable by indirect searches~\cite{Pearce:2013ola, Pearce:2015zca, Cline:2014eaa, Boddy:2014qxa, Detmold:2014qqa}. 
The cosmological formation of stable bound states typically screens or curtails the DM self-interactions, and has to be properly accounted for in any consistent phenomenological study~\cite{Petraki:2014uza}. This is particularly important in the context of the self-interacting DM scenario~\cite{Spergel:1999mh, Kusenko:2001vu,Feng:2008mu, Loeb:2010gj}, as well as in scenarios that feature a dissipative hidden sector~\cite{Foot:2013nea, Fan:2013tia, Foot:2013lxa, Foot:2013uxa, Foot:2014uba, Foot:2014osa, Foot:2015mqa,Boddy:2016bbu}. Moreover, stable bound states of asymmetric DM can give rise to distinct signatures in direct detection experiments~\cite{Laha:2013gva,Laha:2015yoa,Butcher:2016hic}. 
Finally, DM bound states may result in detectable collider signals~\cite{Shepherd:2009sa, An:2015pva, Bi:2016gca, Nozzoli:2016coi}.

\section*{Acknowledgments}

We thank Franz Herzog, Bira van Kolck, and Andreas Nogga for useful discussions. 
This work was supported by the Netherlands Foundation for Fundamental Research of Matter (FOM) and the Netherlands Organisation for Scientific Research (NWO). J.d.V. and K.P. acknowledge support by NWO in the form of, respectively, the VENI and VIDI grants.
K.P. was supported by the European Research Council (ERC) under the EU Seventh Framework Programme (FP7/2007-2013)/ERC Starting Grant (agreement n. 278234 -- `NewDark' project), and by the ANR ACHN 2015 Grant (`TheIntricateDark' project).


\clearpage
\appendix

\noindent
{\Large \bf Appendices}

\section{Wavefunctions \label{App:WaveFun}}

\subsection{The Schr\"{o}dinger equation \label{app:Schrodinger eq}}

The Schr\"{o}dinger equations for the bound and scattering states are
\begin{subequations} \label[pluralequation]{eqs:SchrEq}
\begin{align}
\[-\frac{\nabla^2}{2\mu} + V(\vec r)\] \psi_{n\ell m}(\vec r) 
&= {\cal E}_{n\ell} \, \psi_{n\ell m}(\vec r) \, ,
\label{eq:SchrEq Bound}
\\
\[-\frac{\nabla^2}{2\mu} + V(\vec r)\] \f_{\vec k}(\vec r) 
&= {\cal E}_{\vec k} \, \f_{\vec k}(\vec r) \, ,
\label{eq:SchrEq Scatt}
\end{align}
\end{subequations}
with the wavefunctions normalised as follows
\begin{subequations} 
\label{eq:SchrWF norm}
\label[pluralequation]{eqs:SchrWF norm}
\begin{gather}
\int d^3r \: \psi_{n\ell m}^*(\vec r) \, \psi_{n\ell m}(\vec r) = 1 \, , \label{eq:psi norm}
\\
\int d^3r \: \f_{\vec k}^*(\vec r) \, \f_{\vec k'}(\vec r) 
= (2\p)^3 \d^3(\vec k - \vec k') \, . \label{eq:phi norm}
\end{gather}
\end{subequations}
For a central potential $V(\vec r) = V(r)$ potential, we perform the standard separation of variables
\begin{subequations} 
\label{eq:WFs var separ}
\label[pluralequation]{eqs:WFs var separ}
\begin{align}
\psi_{n\ell m}(\vec r) 
&= \k^{3/2} \[ \frac{\x_{n\ell}^{}(\k r)}{\k r} \] Y_{\ell m} (\W_{\vec r}) \, ,
\label{eq:psi var separ}
\\
\f_{\vec k}(\vec r) &= 
\sum_{\ell = 0}^\infty (2\ell+1) 
\left[ \frac{\x_{|\vec k|,\ell}^{} (\k r)}{\k r} \right]
\: P_{\ell} (\hat{\vec k} \cdot\hat{\vec r}) \, ,
\label{eq:phi var separ}
\end{align} 
\end{subequations}
with
\begin{subequations} 
\label{eq:chi norm}
\label[pluralequation]{eqs:chi norm}
\begin{gather}
\int_0^\infty dx \: |\x_{n\ell}^{}(x)|^2 = 1 \, ,
\label{eq:chi bound norm}
\\Σ
\int_0^\infty dx \: \x_{n\ell}^*(x) \, \x_{|\vec k|,\ell}^{}(x) = 0 \, .
\label{eq:chi bound-scatt ortho}
\end{gather}
\end{subequations} 
We set $x=\k r$ and 
\beq 
\g \equiv \sqrt{-2\mu {\cal E}}/\k \, .
\label{eq:gamma def}
\eeq 
Then, for the Yukawa potential of \cref{eq:Yukawa}, the radial Schr\"{o}dinger equations read
\beq
\x''(x) + 
\[-\frac{\ell(\ell+1)}{x^2} - \g^2 
+\frac{2e^{-x/\ks}}{x} \] \x (x) = 0 \, ,
\label{eq:chi diff}
\eeq
where we temporarily dropped the indices in the wavefunctions and energy eigenvalues for generality.

At $x\to 0$, and for $\ell>0$,  the second term of \cref{eq:chi diff} is dominated by the centrifugal contribution. In this region, the two independent solutions of \cref{eq:chi diff}  scale as $x^{\ell+1}$ (regular) and $x^{-\ell}$ (irregular). Here, we are interested in regular solutions, for ${\cal E} = {\cal E}_n <0$ and ${\cal E} = {\cal E}_{\vec k} >0$, representing bound and scattering states respectively. This implies the boundary condition
\beq
\lim_{x\to 0} \x'(x) = (\ell+1) \lim_{x\to 0} \, [\x(x)/x] \, .
\label{eq:BC derivative@0}
\eeq
The condition~\eqref{eq:BC derivative@0} will be valid also for $\ell = 0$.

\subsection{Bound states \label{app:BoundState WF num}}

\begin{figure}[t]
\centering
\includegraphics[width=6.7cm]{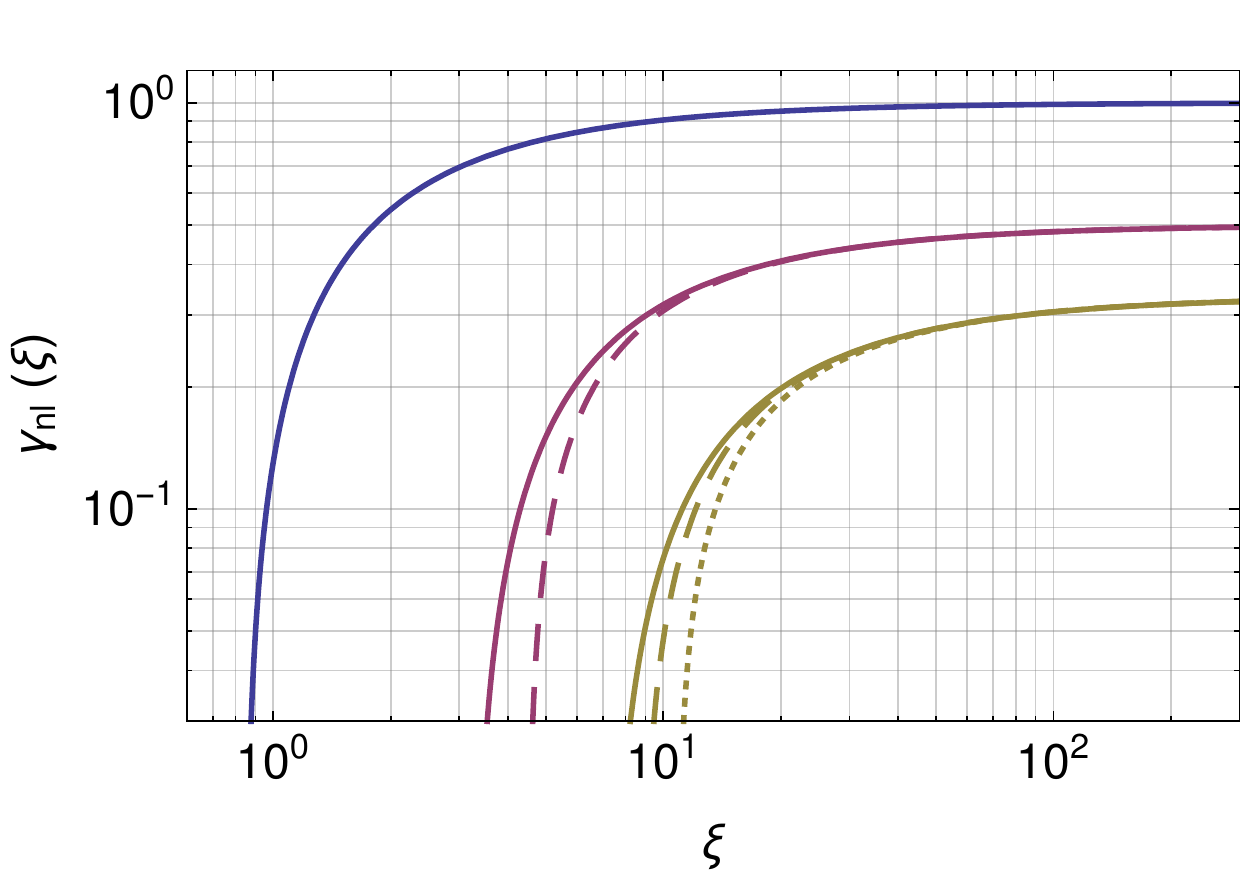}
\caption{$\gamma_{n\ell}(\xi) = \sqrt{-2\mu{\cal E}_{n\ell}}/\kappa$, for $n =1$~(blue), $n=2$ (purple), $n=3$~(yellow), and $\ell = 0$ (solid), $\ell = 1$ (dashed) and $\ell = 2$ (dotted), with $0 \leqslant \ell \leqslant n-1$.}
\label{fig:discrete spectrum}
\end{figure}

For ${\cal E} < 0$, we seek solutions of \cref{eq:chi diff} that vanish at infinity,
\beq
\lim_{x\to\infty} \x(x) = 0 \, .
\label{eq:BC 0@infinity}
\eeq
The boundary conditions \eqref{eq:BC derivative@0} and \eqref{eq:BC 0@infinity}, and the normalisation condition \eqref{eq:chi bound norm} completely specify the discrete spectrum of wavefunctions and energy eigenvalues. For a Yukawa potential, the discrete energy eigenvalues depend on the principal quantum number $n$, as well as on $\ell$,  
\beq
{\cal E} = {\cal E}_{n\ell}(\ks) \equiv 
- \gamma_{n\ell}^2(\xi) \times \frac{\kappa^2}{2\mu} \, .
\label{eq:E_nl}
\eeq 
The lifting of the well-known $\ell$-degeneracy of the energy eigenvalues of the Coulomb limit, is due to the non-conservation of the Laplace-Runge-Lenz vector by the Yukawa potential. We determine $\g_{n\ell}(\ks)$ numerically, and present it in \cref{fig:discrete spectrum}, for $n = 1,2,3$.  We find that it can be well fit by the formula
\beq
\gamma_{n\ell}(\ks) \simeq  \frac{1}{n} \(1-\frac{n^2 \ks_c}{\ks}\)^\r \, .
\label{eq:gamma fit}
\eeq
The best fit parameters $\ks_c$ and $\r$ are given in \cref{tab:gamma fit}, for $1 \leqslant n \leqslant 3$.\footnote{Note that we are using a higher precision numerical fit for the computation of cross-sections of bound-state related processes.} 
Equation~\eqref{eq:gamma fit} reproduces the Coulomb limit, $\lim_{\ks \to \infty} \g_{n\ell}(\ks) = 1/n$. Away from the Coulomb limit, the existence of bound states implies an $n$- and $\ell$-dependent lower bound on $\ks$.

\begin{table}[t!]
\centering
\begin{tabu}{|[1.2pt] c|[1.2pt]  c|[1.2pt]  c|c|[1.2pt]  c|c|c|[1.2pt]}
\tabucline[1.2pt]{-}  
$n$ 		& 1 & \multicolumn{2}{c|[1.2pt]}{2} & \multicolumn{3}{c|[1.2pt]}{3}
\\  \hline  
$\ell$ 	& 0 		& 0 		& 1 			& 0 			& 1 			& 2
\\  \tabucline[1.2pt]{-}  
$\xi_c$ 	& 0.8399	& 0.8059	& 1.1195		& 0.79678	& 0.96883	& 1.1991
\\  \hline 
$\rho$ 	& 1.1129	& 1.1597	& 0.81847	& 1.1746		& 0.94518	& 0.7638
\\ \tabucline[1.2pt]{-}  
\end{tabu} 
\caption{The fit parameters $\xi_c$ and $\rho$, in the fitting formula $\gamma_{n\ell}(\xi) \simeq  (1/n) (1- n^2 \xi_c/\xi)^\rho$, for the energy eigenvalues ${\cal E}_{n\ell} = - \gamma_{n\ell}^2(\xi) \times\kappa^2/(2\mu)$. In the Coulomb limit, $\lim_{\xi\to \infty} \gamma_{n\ell}(\xi) = 1/n$. 
\label{tab:gamma fit}}
\end{table}

\begin{figure}[t]
\centering
\includegraphics[width=0.45\textwidth]{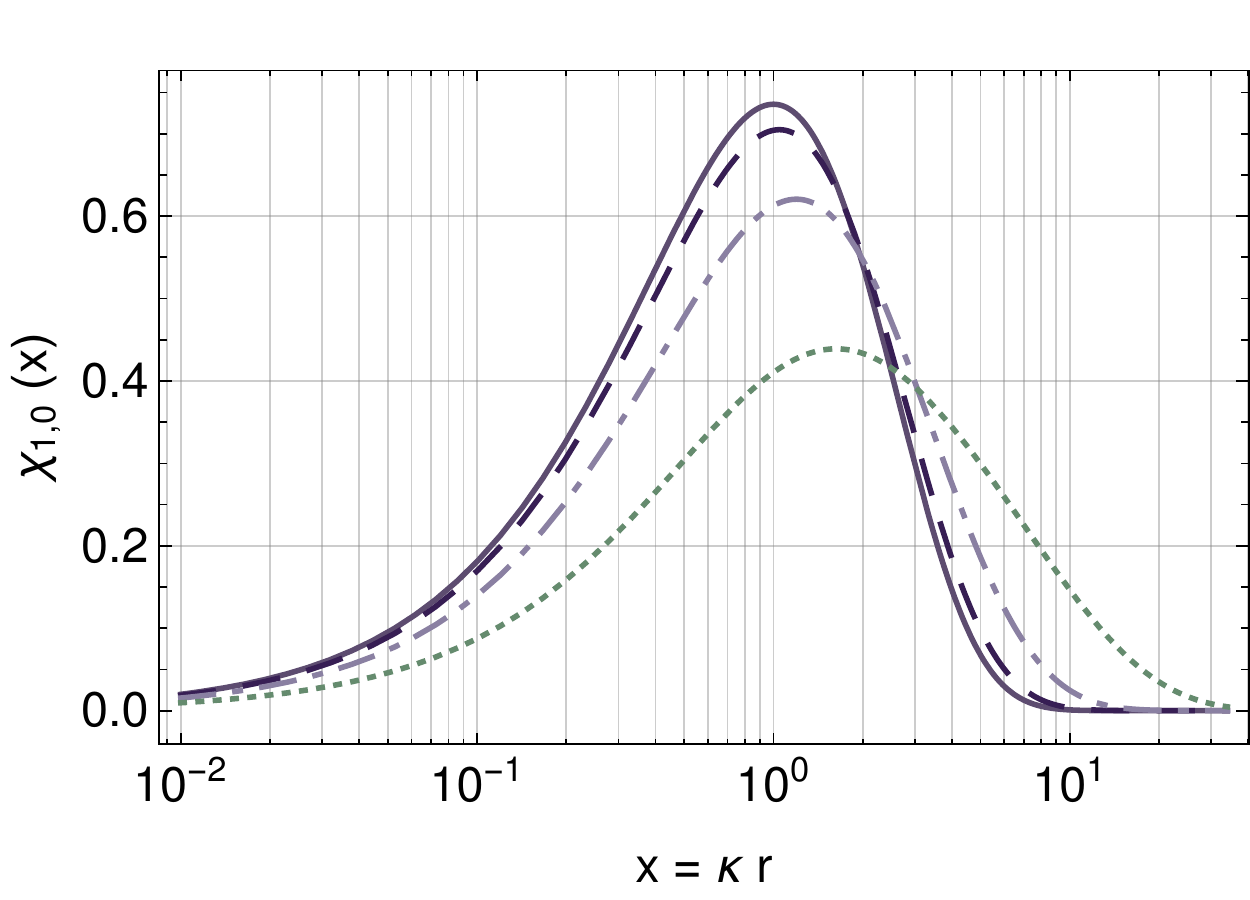}~~~  
\includegraphics[width=0.45\textwidth]{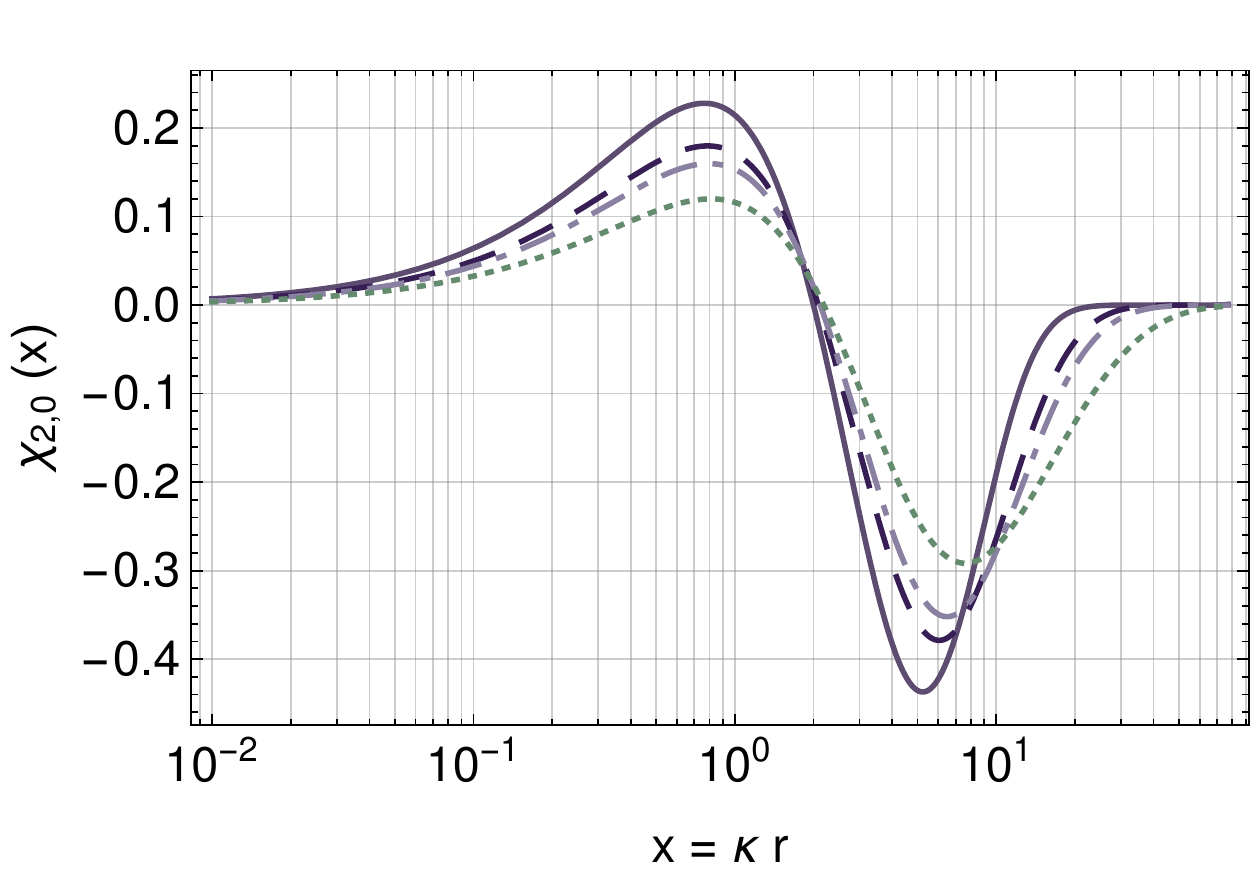}
\caption[]{
\emph{Left:} The bound-state wavefunction $\chi_{n=1,\ell=0}^{} (x)$, for various values of $\xi \equiv \mu \alpha/m_\vf$: Analytical Coulomb limit $\xi \to \infty$ (solid), $\xi = 3$ (dashed), $\xi = 1.5$ (dotdashed), and $\xi = 1$ (dotted).  
\emph{Right:} The bound-state wavefunction $\chi_{n=2,\ell=0}^{} (x)$, in the Coulomb limit $\xi \to \infty$ (solid), and for $\xi = 6$ (dashed), $\xi = 5$ (dotdashed), and $\xi = 4$ (dotted).
}
\label{fig:WF_bound}
\end{figure}

In \cref{fig:WF_bound}, we show the wavefunctions $\x_{1,0}^{}(x)$ and $\x_{2,0}^{}(x)$, for various values of $\ks$.

\paragraph{Coulomb limit.} 
In the limit $\ks \to \infty$, \cref{eq:chi diff} admits analytic solutions with
\beq
\g_{n\ell}^C = \lim_{\ks \to \infty} \g_{n\ell}(\ks) = 1/n \, ,
\label{eq:gamma nl Coul}
\eeq
%
%
\beq
\x_{n\ell}^C (x) = \lim_{\ks\to\infty} \x_{n\ell} (x)
= \frac{1}{n} \[\frac{(n-\ell-1)!}{(n+\ell)!}\]^{1/2}
\ e^{-x/n} \ (2x/n)^{\ell+1} \ L_{n-\ell-1}^{2\ell+1} (2x/n)
\label{eq:WF BS Coul}
\eeq
where $L_n^a$ are the generalised Laguerre polynomials of degree $n$. (We assume the normalisation condition $\int_0^\infty x^a e^{-x} L_n^{(a)} L_m^{(a)} dx = [\G(n+a+1)/n!] \, \d_{n,m}$.)

\subsection{Two-particle scattering state \label{app:ScattState WF num}}

For the continuous spectrum with ${\cal E} = {\cal E}_{\vec k} > 0$, we set
\beq
{\cal E}_{\vec k} = \frac{\vec k^2}{2\mu} = \frac{1}{2} \mu \vrel^2 \, ,
\label{eq:E_k}
\eeq
or equivalently $\g^2 = -1/\z^2$. The wavefunctions are specified by the boundary condition~\eqref{eq:BC derivative@0}, and the asymptotic behaviour at $x \to \infty$.
At large $x$, the wavefunction $\x_{|\vec k|, \ell}^{}$ behaves as (see e.g.~Ref.~\cite[chapter 7]{Sakurai_QMbook})
\begin{subequations}
\beq
\x_{|\vec k|,\ell}^{} (x) 
\ \stackrel{x\to\infty}{\longrightarrow} \ 
\frac{\zeta}{2i}\[e^{i (x/\z+\d_\ell)} - e^{-i(x/\z -\ell \pi)}\] \, ,
\label{eq:SS WF large x}
\eeq
where the phase shifts $\d_\ell$ depend on $\zeta$ and $\ks$. This implies that 
\beq
\left|\x_{|\vec k|,\ell}^{} (x) \right|^2 + 
\left|\x_{|\vec k|,\ell}^{} (x -\p\z/2) \right|^2 
= \z^2 \, .
\label{eq:SS WF large x norm}
\eeq
\end{subequations}

\paragraph{Coulomb limit.} At $\ks \to \infty$, the analytical solutions of \cref{eq:chi diff} with ${\cal E}={\cal E}_{\vec k}>0$, are
\begin{subequations}
\beq
\chi_{|\vec k|,\ell}^C (x) 
= e^{\pi \zeta/2} \ \frac{\Gamma(1+\ell-i\zeta)}{(2\ell+1)!}
\ x(2ix/\zeta)^\ell \ e^{-i x/\zeta}
\ {}_1F_1 (1+\ell+i\zeta; \ 2\ell+2; \ 2ix/\zeta) \, .
\label{eq:chi_k Coul}
\eeq
The sum \eqref{eq:phi var separ} over the $\ell$ modes can be expressed in closed form,
\beq
\f_{\vec k}^C (\vec r) =  \lim_{\ks \to \infty} \f_{\vec k} (\vec r) =  
\ e^{\pi \zeta/2} \ \Gamma (1-i \zeta) 
\ {}_1F_1 [i\z;\ 1; \ i(kr - \vec k \cdot \vec r)] 
\ e^{i \vec k \cdot \vec r} \, .
\label{eq:phi_Coul}
\eeq
\end{subequations}

\clearpage
\section{Convolution integrals \label{App:ConvIntegr}}

\subsection{Definition}
The cross-sections for radiative BSF depend on the following integrals~\cite{Petraki:2015hla}
\begin{subequations} 
\label{eq:ConvIntegrals_Def_re}
\label[pluralequation]{eqs:ConvIntegrals_Def_re}
\begin{align}
{\cal I}_{\vec k, n\ell m} (\vec b) 
&\equiv 
\int \frac{d^3p}{(2\p)^3} 
\: \tilde \psi_{n\ell m}^* (\vec p) 
\: \tilde \f_{\vec k}  (\vec p + \vec b)
= \int d^3 r 
\: \psi_{n\ell m}^* (\vec r) 
\: \f_{\vec k}  (\vec r) 
\: e^{-i \vec b \cdot \vec r} \: ,
\label{eq:Ical_Def_re}
\\
\boldsymbol{\cal J}_{\vec k, n\ell m} (\vec b) 
&\equiv 
\int \frac{d^3p}{(2\p)^3} \: \vec p 
\: \tilde \psi_{n\ell m}^* (\vec p) 
\: \tilde \f_{\vec k}  (\vec p +\vec b) 
= i \int d^3 r 
\: [\nabla \psi_{n\ell m}^* (\vec r)] 
\:  \f_{\vec k}  (\vec r) 
\: e^{-i \vec b \cdot \vec r}
\: ,
\label{eq:Jcal_Def_re}
\\
{\cal K}_{\vec k, n\ell m} (\vec b) 
&\equiv 
\int \frac{d^3p}{(2\p)^3} \: \vec p^2 
\: \tilde \psi_{n\ell m}^* (\vec p) 
\: \tilde \f_{\vec k}  (\vec p + \vec b) 
= - \int d^3 r 
\: [\nabla^2 \psi_{n\ell m}^* (\vec r)] 
\: \f_{\vec k}  (\vec r) 
\: e^{-i \vec b \cdot \vec r}
\: ,
\label{eq:Kcal_Def_re}
\end{align}
where $\vec b$ is proportional to the momentum of the radiated particle, $\vec b \propto \vec P_\vf$. Using the Schr\"odinger \cref{eq:SchrEq Bound}, and \cref{eq:E_nl}, ${\cal K}_{\vec k, n\ell m} (\vec b)$ takes also the form
\beq
{\cal K}_{\vec k, n\ell m} (\vec b)
= 
-\kappa^2 \gamma_{n\ell}^2(\ks) \: {\cal I}_{\vec k, n\ell m} (\vec b) 
- 2\mu \int d^3 r 
\: V(\vec r) \psi_{n\ell m}^* (\vec r) 
\: \f_{\vec k}^{}  (\vec r) 
\: e^{-i \vec b \cdot \vec r}
\: .
\label{eq:Kcal NoDeriv}
\eeq
\end{subequations}
The first term in \cref{eq:Kcal NoDeriv} yields a higher order correction to the contribution from ${\cal I}_{\vec k, n\ell m} (\vec b)$ in a given process, and we shall typically ignore it. The second term above is the leading relativistic correction.

\subsection{Useful identities for angular integration \label{app:Identities}}

\begin{subequations} 
\label{eq:Identities}
\label[pluralequation]{eqs:Identities}

The plane waves $e^{-i\vec{b}\cdot \vec{r}}$, which appear in the integrals~\eqref{eqs:ConvIntegrals_Def_re} and essentially stand for the wavefunction of the radiated boson, can be expanded in terms of Legendre polynomials using the identity
\beq
e^{-i\vec{b}\cdot \vec{r}} 
= \sum_{\ell = 0}^\infty (2\ell+1) \, (-i)^\ell 
\, j_\ell^{} (b\,r) 
\, P_\ell (\hat{\vec{b}} \cdot \hat{\vec{r}}) \, ,
\label{eq:exp expansion}
\eeq
where $j_\ell^{}$ is the spherical Bessel function. Derivatives of $j_\ell^{}$ that will arise in the following, can be re-expressed in terms of Bessel functions using the identity
\beq
\frac{dj_\ell^{} (z)}{dz} = \frac{\ell j_\ell^{} (z)}{z} - j_{\ell+1}^{} (z) \, .
\label{eq:Bessel deriv}
\eeq
The Legendre polynomials can be expanded in spherical harmonics as follows
\beq
P_\ell (\hat{\vec{x}}\cdot\hat{\vec{y}}) 
= \frac{4\p}{2\ell+1} \: \sum_{m=-\ell}^\ell 
Y_{\ell m}^*(\W_\vec{x}) Y_{\ell m}(\W_\vec{y}) \, ,
\label{eq:P ell expansion}
\eeq
Using the expansion~\eqref{eq:exp expansion} and the identity~\eqref{eq:P ell expansion}, will give rise to angular integrals that involve three spherical harmonics, and which can be expressed in terms of the Wigner-$3j$ symbol,
\beq
\int \! d\W \: Y_{\ell_1 m_1}^{}\!(\W)  \: Y_{\ell_2 m_2}^{}\!(\W)  \: Y_{\ell_3 m_3}^{}\!(\W) 
=  \sqrt{\frac{(2\ell_1+1)(2\ell_2+1)(2\ell_3+1)}{4\pi}}
\begin{pmatrix}
\ell_1 & \ell_2 & \ell_3
\\
0  & 0 & 0
\end{pmatrix}
\begin{pmatrix}
\ell_1 & \ell_2 & \ell_3
\\
m_1 & m_2 & m_3
\end{pmatrix}
.
\label{eq:3 Harmonics int}
\eeq
We always assume that the spherical harmonics are normalised according to
\beq
\int d\W \: Y_{\ell m}^*(\W) \, Y_{\ell' m'}^{}(\W) = \d_{\ell\ell'} \, \d_{m m'} \, .
\label{eq:Harmonics norm}
\eeq

\end{subequations}

\subsection{Expansion in the momentum of the emitted radiation \label{app:expansion RadMom}}

The exponential decay of the bound-state wavefunction $\ps_{n\ell m}(\vec r)$ at large $r$ ensures that the integrands in \cref{eqs:ConvIntegrals_Def_re} are significant only for $\k r \, \gamma_{n\ell}^{}(\ks) \lesssim$~few (cf.~section~\ref{app:BoundState WF num}). In this range, the argument of the Bessel function in \cref{eq:exp expansion} is 
\beq
b \,r = (\h |\vec{P}_\vf|) \, r 
\ \lesssim \
{\rm few} \times  \frac{\h |\vec{P}_\vf|}{\kappa \gamma_{n\ell}(\ks)}
= {\rm few} \times \frac{\h \, \a}{2\gamma_{n\ell}(\ks)} 
\ \frac{1+ \z^2 \g_{n\ell}^2(\ks)}{\z^2} \, pss_{n\ell}^{1/2} \, ,
\nn
\eeq
where here $\h = \h_1$ or $\h_2$, and we used \cref{eq:ForceMed_momentum} for $|\vec{P}_\vf|$. From this we deduce that $b r \ll 1$, if
\beq
\vrel^2 \ll \a \ \g_{n\ell}^{}(\ks) \, .
\label{eq:expansion condition}
\eeq
The condition \eqref{eq:expansion condition} covers the range of interest. Provided that it is satisfied, we may expand the Bessel function of \cref{eq:exp expansion}, and keep only leading-order terms. For this purpose, we shall use the expansion
\begin{align}
j_\ell (z) = \sum_{s=0}^\infty \frac{(-1)^s z^{\ell + 2s}}{2^s s! (2s+2\ell+1)!!} \, ,
\label{eq:Bessel_expansion}
\end{align}
where for our purposes, $z=b r$. 

In computing the integrals~\eqref{eqs:ConvIntegrals_Def_re}, we express the wavefunctions as in \cref{eqs:WFs var separ}, and carry out the radial integration in the variable $x = \kappa r$. The expansion of the Bessel function over $z = (b/\k) x $ amounts thus to an expansion over $b/\kappa$. Since $b \propto |\vec P_\vf|$, with $|\vec P_\vf|$ given in \cref{eq:ForceMed_momentum}, this is ultimately an expansion in $\a$.

\subsection{Capture into bound states of arbitrary angular momentum \label{app:ConvIntegr_angular}}

In order to evaluate the integrals \eqref{eqs:ConvIntegrals_Def_re}, we first perform the angular integration using the identities~\eqref{eqs:Identities}, and then expand in powers of $b/\kappa$, using \cref{eq:Bessel_expansion}. We find
\begin{subequations} 
\label{eq:IJK_nlm_expansion}
\label[pluralequation]{eqs:IJK_nlm_expansion}
\begin{align}
{\cal I}_{\vec k, n\ell m} (\vec b) &= 
\(\frac{4\p}{\kappa}\)^{3/2}
\ \sum_{s=0}^\infty 
\ \sum_{\lR=0}^\infty 
\(\frac{b}{\kappa}\)^{\lR+2s}
\frac{(-1)^{\lR+s} \ i^{\lR} }{2^s s!\,(2s+2\lR+1)!!}
\nn   \\   &\times 
  \sum_{\mR = -\lR}^{\lR}
\ \sum_{\lI=0}^\infty 
\ \sum_{\mI = -\lI}^{\lI} (-1)^\mI
\ Y_{\lI \mI}^*  (\W_{\vec k})
\ Y_{\lR \mR}^{} (\W_{\vec b})
\nn   \\   &\times
\sqrt{(2\ell+1)(2\lR+1)(2\lI+1)}
\begin{pmatrix}
\ell& \lR 	& \lI
\\
0 & 0 & 0
\end{pmatrix}
\begin{pmatrix}
\ell & ~ \lR 	& ~ \lI
\\
-m 	 & ~-\mR 	& ~ \mI
\end{pmatrix}
\nn   \\   &\times
\int_0^\infty dx 
\: x^{\lR+2s}
\: \chi_{n\ell}^* (x) 
\: \chi_{|\vec{k}|,\lI}^{} (x) \, ,
\label{eq:Inlm general}
\end{align}
\begin{align}
\boldsymbol{\cal J}_{\vec k, n\ell m} (\vec b) 
&= \frac{(4\pi)^2 i}{\kappa^{1/2}} 
\ \sum_{s=0}^{\infty} 
\ \sum_{\lR = 0}^{\infty} 
\(\frac{b}{\kappa}\)^{\lR+2s}
\ \frac{(-1)^{\lR+s} \ i^{\lR}}{2^s s!\,(2s + 2\lR +1)!!}
\nn \\
&\times
  \sum_{\mR=-\lR}^{\lR}
\ \sum_{\ell_{\rm I}=0}^\infty
\ \sum_{\mI=-\ell_{\rm I}}^{\ell_{\rm I}}
\  Y_{\lI \mI}^*  (\W_{\vec k})
\  Y_{\lR \mR}^{} (\W_{\vec b})
\nn \\
&\times 
\int d\W \ Y_{\lI \mI}^{}(\W) \ Y_{\lR \mR}^*(\W)
\int_0^\infty dx 
\ x^{1+\lR+2s} 
\ \nabla_{\vec x} \left[ \frac{\chi_{n\ell}^*(x)}{x} \ Y_{\ell m}^*(\W)\right]
\ \chi_{|\vec k|, \lI}^{}(x)
\, ,
\label{eq:Jnlm general}
\end{align}
\begin{align}
{\cal K}_{\vec k, n\ell m} (\vec b) &= 
\sqrt{(4\p)^3 \kappa}
\ \sum_{s=0}^\infty 
\ \sum_{\lR=0}^\infty 
\(\frac{b}{\kappa}\)^{\lR+2s}
\frac{(-1)^{\lR+s} \ i^{\lR} }{2^s s!\,(2s+2\lR+1)!!}
\nn   \\   &\times 
  \sum_{\mR = -\lR}^{\lR}
\ \sum_{\lI=0}^\infty 
\ \sum_{\mI = -\lI}^{\lI} (-1)^\mI
\ Y_{\lI \mI}^*  (\W_{\vec k})
\ Y_{\lR \mR}^{} (\W_{\vec b})
\nn   \\   &\times
\sqrt{(2\ell+1)(2\lR+1)(2\lI+1)}
\begin{pmatrix}
\ell& \lR 	& \lI
\\
0 & 0 & 0
\end{pmatrix}
\begin{pmatrix}
\ell & ~ \lR 	& ~ \lI
\\
-m 	 & ~-\mR 	& ~ \mI
\end{pmatrix}
\nn   \\   &\times
\int_0^\infty dx 
\[-\gamma_{n\ell}^2(\ks)  +  \frac{2e^{-x / \ks}}{x} \] 
\: x^{\lR+2s}
\: \chi_{n\ell}^* (x) 
\: \chi_{|\vec{k}|,\lI}^{} (x) \, .
\label{eq:Knlm general}
\end{align}
\end{subequations}
In the above, $\{\lI, \mI\}$ and $\{\lR, \mR\}$ are the orbital angular momentum quantum numbers of the incoming state and the radiated particle, respectively. In the $\nabla_{\vec x}$ operator of \cref{eq:Jnlm general}, the radial coordinate should be understood to be $x \equiv \kappa r$. We also note that $(-1)^m Y_{\ell m}^*(\W) = Y_{\ell, -m}(\W)$.

The expansions~\eqref{eqs:IJK_nlm_expansion} can be used to evaluate the amplitudes for the capture processes of interest, by keeping the leading order terms, as appropriate.

\subsection[Capture into $\ell=0$ bound states]{Capture into $\boldsymbol{\ell=0}$ bound states \label{app:n00 states}}

We may evaluate the integrals needed for capture into zero angular momentum bound states, directly from \cref{eqs:IJK_nlm_expansion}. Instead, here we shall perform the angular integration independently, and then expand in powers of the radiated momentum.

\subsubsection{Angular integration}

We will need the following angular integrals
\begin{subequations}
\begin{align}
\int d\W_{\vec{r}} 
\ P_\ell (\hat{\vec{k}}\cdot\hat{\vec{r}}) 
\ e^{-i \vec{b}\cdot \vec{r}} 
\ &\equiv \ \varpi 
\, , \\
\int d\W_{\vec{r}} \: \hat{\vec{r}}
\: P_\ell (\hat{\vec{k}}\cdot\hat{\vec{r}}) 
\: e^{-i \vec{b}\cdot \vec{r}}
\ &= \ 
-\frac{\nabla_{\vec b} \varpi}{i r} 
\, .
\end{align}
\end{subequations}
Using the identities~\eqref{eqs:Identities}, we find
\begin{subequations} \label[pluralequation]{eqs:Angular Integrals}
\beq
\varpi =  
\int d\W_{\vec{r}} \: P_\ell (\hat{\vec{k}}\cdot\hat{\vec{r}}) \: e^{-i \vec{b}\cdot \vec{r}} 
=  4\p (-i)^\ell \, j_\ell (b\,r) \, P_\ell (\hat{\vec{k}}\cdot\hat{\vec{b}}) \, ,
\label{eq:ang int I}
\eeq
and
\begin{multline}
\int d\W_{\vec{r}} \: \hat{\vec{r}} \: P_\ell (\hat{\vec{k}}\cdot\hat{\vec{r}}) \: e^{-i \vec{b}\cdot \vec{r}} 
= -4\p (-i)^{1+\ell} \ \times
\\
\left\{
\hat{\vec{k}}  \ \frac{j_\ell (br)}{br} \, \frac{dP_\ell(y)}{dy} 
+\hat{\vec{b}} \[\frac{j_\ell (br)}{br} \(\ell P_\ell (y) -y \frac{dP_\ell (y)}{dy} \) - j_{\ell+1}(br) \, P_\ell (y) \] 
\right\}_{y=\hat{\vec{k}}\cdot\hat{\vec{b}}} .
\label{eq:ang int J}
\end{multline}
\end{subequations}

Combining the wavefunction decompositions \eqref{eqs:WFs var separ} and the angular integrals \eqref{eqs:Angular Integrals}, the integrals \eqref{eqs:ConvIntegrals_Def_re} for capture into $\{n00\}$ bound states, become
\begin{subequations} 
\label{eq:Integrals k-n00}
\label[pluralequation]{eqs:Integrals k-n00}
\beq
{\cal I}_{\vec{k},n00} (\vec{b}) 
= \sqrt{\frac{4\pi}{\kappa^3}}
\ \sum_{\ell=0}^\infty P_\ell (\hat{\vec{k}}\cdot\hat{\vec{b}}) \: (2\ell+1) (-i)^\ell
\int_0^\infty dx \ \x_{n,0}^* (x) \: \x_{|\vec k|,\ell}^{} (x) 
\ j_\ell (b x/\kappa)
\, ,
\label{eq:I k-n00}
\eeq
\begin{multline}
\boldsymbol{\cal J}_{\vec{k},n00}  (\vec{b}) 
=-\sqrt{\frac{4\pi}{\kappa}} 
\ \sum_{\ell=0}^\infty (2\ell+1) (-i)^\ell
\, \int_0^\infty dx  
\ \[\frac{d\x_{n,0}^*(x)}{dx} - \frac{\x_{n,0}^*(x)}{x}\] 
\x_{|\vec k|,\ell}^{}(x)
\, \times 
\\
\times
\left\{ 
 \hat{\vec{k}} \ \frac{dP_\ell}{dy} \ \frac{j_\ell(b x/\k)}{b x/\k}  
+\hat{\vec{b}} \[ \(\ell P_\ell(y) - y\, \frac{dP_\ell(y)}{dy} \) \, \frac{j_\ell(b x/\k)}{b x/\k} 
- P_\ell(y) \ j_{\ell+1}(b x/\k) \]
\right\}_{y = \hat{\vec{b}}\cdot\hat{\vec{k}} } 
\label{eq:J k-n00}
\end{multline}
and
\beq
{\cal K}_{\vec{k},n00} (\vec{b}) 
= - \sqrt{4\p\k} 
\ \sum_{\ell=0}^\infty P_\ell (\hat{\vec{k}}\cdot\hat{\vec{b}}) 
\: (2\ell+1) (-i)^\ell 
\ \int_0^\infty  d x
\ \frac{d^2 \x_{n,0}^*(x)}{dx^2} 
\ \x_{|\vec k|,\ell}^{} (x)
\ j_\ell (b x/\k) \, .
\label{eq:K k-n00}
\eeq
\end{subequations}

\subsubsection{Leading-order contributions}

Using \cref{eq:Bessel_expansion}, we now expand the integrals~\eqref{eqs:Integrals k-n00} in powers of $b/\kappa$, and keep the leading-order terms. For ${\cal I}_{\vec{k},n00}$, the zero-th order contribution vanishes, due to the orthogonality of the $\ps_{n\ell m}$ and $\f_{\vec k}$ (or particularly, the $\x_{n\ell}$ and $\x_{|\vec k|,\ell}$) wavefunctions. In our computations in \cref{Sec:VecMed,Sec:ScalMed}, we shall need terms up to ${\cal O}[(b/\kappa)^2]$. For ${\cal J}_{\vec k, n00}$ and ${\cal K}_{\vec k, n00}$, the zero-th order terms in $b/\kappa$ suffice.
\begin{subequations} \label[pluralequation]{eqs:IJK n00}
\begin{multline}
{\cal I}_{\vec{k},n00} (\vec b) 
\simeq
- \sqrt{\frac{4\pi}{\kappa^3}} \,
\left\{
\(\frac{b}{\kappa}\) \ i \, P_1(\hat{\vec k} \cdot \hat{\vec b})
\int_0^\infty dx \: x \, \chi_{n,0}^*(x) \, \chi_{|\vec k|,1}^{}(x)
\right. \\ \left.
+ 
\(\frac{b}{\kappa}\)^2
\[
\frac{P_0(\hat{\vec k} \cdot \hat{\vec b})}{6} 
\int_0^\infty dx \: x^2 \, \chi_{n,0}^*(x) \, \chi_{|\vec k|,0}^{}(x)
+
\frac{P_2(\hat{\vec k} \cdot \hat{\vec b})}{3} 
\int_0^\infty dx \: x^2 \, \chi_{n,0}^*(x) \, \chi_{|\vec k|,2}^{}(x)
\] \right\} ,
\label{eq:In00 expansion}
\end{multline}
\begin{align}
\boldsymbol{\cal J}_{\vec k, n00} (\vec b) 
&\simeq 
-\hat{\vec k} \ \sqrt{\frac{4\p}{\k}}
\int_0^\infty dx \[\frac{d\chi_{n,0}^*(x)}{dx} -\frac{\chi_{n,0}^*(x)}{x} \] \chi_{|\vec k|,1}^{}(x) ,
\label{eq:Jn00 expansion}
\\
{\cal K}_{\vec k, n00} (\vec b) 
&\simeq 
-\sqrt{4\pi \kappa}
\int_0^\infty dx 
\ \frac{d^2\chi_{n,0}^*(x)}{dx^2}
\ \chi_{\vec k, 0}^{}(x) 
\nn \\
&= \sqrt{4\pi \kappa} \int_0^\infty dx 
\( \frac{2e^{-x/\ks}}{x} \)
\chi_{n,0}^*(x) \ \chi_{\vec k, 0}^{}(x) 
\, .
\label{eq:Kn00 expansion}
\end{align}
\end{subequations}

\subsection[Capture into $\ell=1$ bound states]{Capture into $\boldsymbol{\ell=1}$ bound states \label{app:n_l=1_m states}}

It is straightforward to obtain the leading order terms of the ${\cal I}_{\vec k, n1m}$ and ${\cal K}_{\vec k, n1m}$ integrals, using \cref{eqs:IJK_nlm_expansion}. 
Here, we give explicitly only the leading order contributions to the ${\cal J}_{\vec k, n1m}$ integrals. To zero-th order in $b$,
\begin{subequations} \label[pluralequation]{eqs:Jn1m}
\begin{multline}
\boldsymbol{\cal J}_{\vec k, n10} (\vec b) 
\simeq i \sqrt{\frac{12 \p}{\kappa}} 
\left\{
\( \vec{\hat{k}} \: \cos \theta_{\vec k}-\frac{\vec{\hat{e}}_z}{3}\)
\int_0^\infty dx \[ \chi_{n,1}'(x) - \frac{2\chi_{n,1}^{}(x)}{x}\]^* 
\chi_{|\vec k|,2}^{}(x) 
\right. \\ \left.
+ \frac{\vec{\hat{e}}_z}{3}
 \int_0^\infty \! dx  \[\chi_{n,1}'(x) + \frac{\chi_{n,1}^{}(x)}{x}\]^* 
\chi_{|\vec k|,0}^{}(x)
\right\} \, ,
\label{eq:Jn10} 
\end{multline}
\begin{multline}
\boldsymbol{\cal J}_{\vec k, n11} (\vec b) 
\simeq -i \sqrt{\frac{6\pi}{\kappa}}
\left\{
\(\vec{\hat{k}} \: \sin \theta_{\vec k} \: e^{i\phi_{\vec k}} 
-\frac{\vec{\hat{e}}_x + i \vec{\hat{e}}_y}{3} \)
\int_0^\infty dx \: \[\chi_{n,1}'(x) - \frac{2\chi_{n,1}^{}(x)}{x}\]^* \chi_{|\vec k|,2}^{}(x)
\right. \\ \left.
+ \frac{\vec{\hat{e}}_x + i \vec{\hat{e}}_y}{3}
\int_0^\infty \! dx  
\[\chi_{n,1}'(x) + \frac{ \chi_{n,1}^{}(x)}{x}\]^* 
\chi_{|\vec k|,0}^{}(x) \right\} \, ,
\label{eq:Jn11} 
\end{multline}
and
\beq
\boldsymbol{\cal J}_{\vec k, n1-1} (\vec b) = -
\boldsymbol{\cal J}_{\vec k, n11} (\vec b) \, ,
\eeq
\end{subequations}
where 
$\vec{\hat{k}} = 
\vec{\hat{e}}_x \, \sin\theta_{\vec k} \cos\phi_{\vec k} +
\vec{\hat{e}}_y \, \sin\theta_{\vec k} \sin\phi_{\vec k} +
\vec{\hat{e}}_z \, \cos\theta_{\vec k}$.

\clearpage
\section{Coulomb limit for capture into $\boldsymbol{\ell=0}$ bound states \label{App:Coulomb ell=0}}

\subsection{Wavefunctions}

For our analytical computations in the Coulomb limit, we shall use the closed form of the scattering state wavefunction [cf.~\cref{eq:phi_Coul}],
\begin{subequations} 
\label{eq:WFs Coul}
\label[pluralequation]{eqs:WFs Coul}
\beq
\f_{\vec k}^C (\vec r) = 
\sqrt{S_0(\z)}  
\ {}_1F_1 [i\z;\ 1; \ i(kr - \vec k \cdot \vec r)] 
\ e^{i \vec k \cdot \vec r} \, ,
\label{eq:phi_Coul_re}
\eeq
where\footnote{The factor $\sqrt{S_0(\z)}$ in \cref{eq:phi_Coul} often appears in the literature as $e^{\p\z/2} \G(1-i\z)$.} 
\beq
S_0(\z) \equiv \frac{2\p\z}{1-e^{-2\p\z}} \, .
\label{eq:S0_Def}
\eeq
We will consider only capture to zero angular momentum bound states. The $\ell=0$ bound-state wavefunctions and their derivatives, are [cf.~\cref{eq:WF BS Coul}]
\begin{align}
\ps_{n00}^C(\vec r) 
&= \sqrt{\frac{\k^3}{\p n^5}}
\ e^{-\k r/n} \ \sum_{s=0}^{n-1} 
\frac{n! \: (-2 \k r/n)^s}{(n-s-1)! \: (s+1)! \: s!} \, , 
\label{eq:psi_n00_Coul}
\\
\nabla \ps_{n00}^C(\vec r) 
&= - \hat{\vec r} \ \sqrt{\frac{\k^5}{\p n^7}} \ e^{-\k r/n} 
\: \sum_{s=0}^{n-1} \frac{n! \: (2n-s)}{(n-s-1)!}
\: \frac{(-2\k r/n)^s }{(s+2)! \: s!} \, , 
\label{eq:psiNabla_n00_Coul}
\\
\nabla^2 \ps_{n00}^C(\vec r) 
&= \sqrt{\frac{\k^7}{\p n^9}} \ e^{-\k r/n} 
\: \sum_{s=0}^{n-1} \frac{n! \: (2n-s)}{(n-s-1)!}
\: \frac{(-2\k r/n)^s }{(s+2)! \: s!} \(1-\frac{2+s}{\kappa r/n}\)
\, ,
\label{eq:psiNabla2_n00_Coul}
\end{align}
where we expanded the Laguerre polynomials for later convenience.
\end{subequations}

\subsection{Identities} 

We shall use the identity~\cite{AkhiezerMerenkov_sigmaHydrogen}
\beq
\int d^3r \ \frac{e^{-\bar{\k} r}}{4\pi r} 
\ {}_1F_1 [i\z;1;i(kr-\vec k \cdot \vec r)] 
\ e^{i(\vec k - \vec b) \cdot \vec r} 
\ = \
\frac{[\vec b^2 + (\bar{\kappa} - i k)^2]^{-i\z}}
{[(\vec k-\vec b)^2 + \bar{\kappa}^2]^{1-i\z}}
\ \equiv \ f_{\vec k, \vec b}(\bar{\kappa}) \, ,
\label{eq:Identity Coul}
\eeq
where ${}_1F_1$ is the confluent hypergeometric functions of the first kind. From the \cref{eq:Identity Coul}, we find
\begin{subequations} \label[pluralequation]{eqs:Identities Coul expansion}
\begin{gather}
f_{\vec k, \vec b= \vec 0}(\bar{\kappa}) 
= \frac{1}{k^2} 
\: \frac{e^{-2\z {\rm arccot} (\bar{\kappa}/k)}}{1+(\bar{\kappa}/k)^2} \, , 
\label{eq:f @ b=0}
\\
\[\nabla_{\vec b} f_{\vec k, \vec b}(\bar{\kappa})\]_{\vec b = \vec 0} 
= \hat{\vec{k}} \ \frac{2(1-i\zeta)}{k^3} \ 
\frac{e^{-2\z {\rm arccot} (\bar{\kappa}/k)}}{[1+(\bar{\kappa}/k)^2]^2} \, ,
\label{eq:df/db @ b=0}
\end{gather}
and, keeping up to $b^2$ terms,
\begin{multline}\label{eq:df/dkappa}
\frac{d}{d\bar{\kappa}} f_{\vec k, \vec b}(\bar{\kappa}) \simeq
\frac{2}{k^3} \: \frac{e^{-2\z {\rm arccot} (\bar{\kappa}/k)}}{[1+(\bar{\kappa}/k)^2]^2} 
\: \[ \z - (\bar{\k}/k) 
+ \frac{2 b \cos(\hat{\vec k}\cdot\hat{\vec b})}{k} 
\frac{(1-i\z)(\z-2\bar{\kappa}/k)}{1+(\bar{\kappa}/k)^2}
\right. \\ \left.
+ \frac{2 b^2}{k^2} 
\frac{\cos^2(\hat{\vec k}\cdot\hat{\vec b}) 
\ (\z-3\bar{\kappa}/k)(1-i\z)(2-i\z) +i(1-i\bar{\kappa}/k) \[(\z-\bar{\kappa}/k)^2-i(\bar{\kappa}/k) (1-i\z)\]}{[1+(\bar{\kappa}/k)^2]^2}
\] .
\end{multline}
\end{subequations}

\subsection{Convolution integrals} 

Starting from the definitions~\eqref{eqs:ConvIntegrals_Def_re}, and using the Coulomb wavefunctions~\eqref{eqs:WFs Coul} and the identity~\eqref{eq:Identity Coul}, we find
\begin{subequations} 
\label{eq:Conv Integrals n00 Coul}
\label[pluralequation]{eqs:Conv Integrals n00 Coul}
\beq
{\cal I}_{\vec k \to n00}^C (\vec b) 
=-\sqrt{\frac{16 \pi \k^3}{n^5} \: S_0(\z)} 
\[\sum_{s=0}^{n-1} \frac{n! \: (2 \bar{\k})^s}{(n-s-1)!\: (s+1)! \: s!} 
\: \frac{d^{s+1}}{d\bar{\kappa}^{s+1}} 
f_{\vec k, \vec b}(\bar{\kappa}) \]_{\bar{\kappa} = \kappa/n} ,
\eeq
\beq
\boldsymbol{\cal J}_{\!\!\vec k \to n00}^C (\vec b) 
=-\sqrt{\frac{16\pi \k^5}{n^7} \: S_0(\z)} 
\ \nabla_{\vec b}
\[\sum_{s=0}^{n-1} \frac{n! \: (2n-s)}{(n-s-1)!}
\: \frac{(2\bar{\k})^s }{(s+2)! \: s!}
\: \frac{d^s}{d\bar{\kappa}^s} 
f_{\vec k, \vec b}(\bar{\kappa}) \]_{\bar{\kappa} = \kappa/n} \!,
\eeq
\begin{multline}
{\cal K}_{\vec k \to n00}^C (\vec b) 
= \sqrt{\frac{16 \pi \k^7}{n^9} \: S_0(\z)} 
\[\sum_{s=0}^{n-1} \frac{n! \: (2n-s)}{(n-s-1)!}
\: \frac{(2\bar{\k})^s}{(s+2)! \: s!} \right. 
\ \times  \\   
\left. \times \: 
\( \frac{d^{s+1}}{d\bar{\kappa}^{s+1}} f_{\vec k, \vec b}(\bar{\kappa})
+ \frac{n(2+s)}{\k}
\: \frac{d^s}{d\bar{\kappa}^s} f_{\vec k, \vec b}(\bar{\kappa}) \)\]_{\bar{\kappa} = \kappa/n} \, .
\end{multline}
\end{subequations}

\subsection{Convolution integrals: Expansion in the momentum of emitted radiation.}

Using \cref{eqs:Identities Coul expansion}, we expand the integrals~\eqref{eqs:Conv Integrals n00 Coul}, 
keeping terms up to order $(b/\kappa)^2$ for ${\cal I}_{\vec k, n00}$, and only zero-th order terms for ${\cal J}_{\vec k, n00}$ and ${\cal K}_{\vec k, n00}$. We obtain the following\footnote{
Note that the contribution from the $b$-independent term of \cref{eq:df/dkappa} to the ${\cal I}_{\vec k \to n00}^C$ integral vanishes (as expected) when the summation over $s$ is performed, i.e. 
\beq
\sum_{s=0}^{n-1} \frac{n! 2^s \z_n^s}{(n-s-1)!(s+1)!s!}
\frac{d^s}{d\z_n^s}\[\frac{e^{-2\z {\rm arccot} \z_n}}{(1+\z_n^2)^2}(\z-\z_n)\]_{\z_n = \z/n}
=0 \, . \nn
\eeq
}
\begin{subequations} 
\label{eq:ConvIntegr_n00_Coul_expansion}
\label[pluralequation]{eqs:ConvIntegr_n00_Coul_expansion}
\begin{align}\label{eq:In00 Coul expansion}
&{\cal I}_{\vec k \to n00}^C (\vec b) 
\simeq 
-\sqrt{\frac{n^3\pi}{\k^3} \: S_0(\z)}  
\ \frac{b}{\k}
\ \sum_{s=0}^{n-1} \frac{n! \: (2 \z_n)^{4+s}}{(n-s-1)!\: (s+1)! \: s!} \: \times
\nn \\ 
&\times \frac{d^s}{d\z_n^s} \left\{
\: \frac{e^{-2\z {\rm arccot} \, \z_n}}{(1+\z_n^2)^3} 
\: \[ 
\cos(\hat{\vec k}\cdot\hat{\vec b}) \ (1-i\z)(\z-2\z_n)
\right.\right. \nn \\  &\left.\left.
+ \frac{b \, \z}{\k} 
\ \frac{\cos^2(\hat{\vec k}\cdot\hat{\vec b}) 
\ (\z-3\z_n)(1-i\z)(2-i\z) +i(1-i\z_n) \[(\z-\z_n)^2-i\z_n (1-i\z)\]}
{1+\z_n^2}\] \right\}_{\z_n = \z/n} ,
\end{align}
\beq\label{eq:Jn00 Coul expansion}
\boldsymbol{\cal J}_{\vec k \to n00}^C (\vec b) 
\simeq - \hat{\vec{k}} \ 
\sqrt{\frac{\pi}{\kappa n} \: S_0(\z)} 
\  (1-i\zeta) 
\:  \sum_{s=0}^{n-1} \frac{n! \: (2n-s)}{(n-s-1)!}
\: \frac{(2 \z_n)^{3+s} }{(s+2)! \: s!}
\: \frac{d^s}{d\z_n^s} 
\[\frac{e^{-2\z \: {\rm arccot} \, \z_n}}{\(1+\z_n^2\)^2}\]_{\z_n = \z/n} ,
\eeq
\begin{multline}\label{eq:Kn00 Coul expansion}
{\cal K}_{\vec k \to n00}^C (\vec b) 
\simeq  
\sqrt{\frac{\pi \k}{n^3} \: S_0(\z)} 
\  \sum_{s=0}^{n-1} \frac{n! \: (2n-s)}{(n-s-1)!}
\: \frac{(2 \z_n)^{3+s}}{(s+2)! \: s!} \ \times  
\\   
\times \frac{d^s}{d\z_n^s} 
\[\frac{e^{-2\z\,{\rm arccot}\,\z_n}}{1+\z_n^2}
\(\frac{\z-\z_n}{1+\z_n^2} + \frac{n(2+s)}{2\z}\)\]_{\z_n = \z/n} \, .
\end{multline}
\end{subequations}
We use \cref{eqs:ConvIntegr_n00_Coul_expansion} in our computations of the Coulomb limit of BSF cross-sections in 
\cref{eq:VecMed_BSF_n00_Coul,eq:ScalMed_NonDegen_BSF_Coul,eq:ScalMed_Degen_S100_Coul}.

\clearpage
\section{Bound-state formation in momentum space \label{App:BSF_MomemtumSpace}}

In this appendix, we describe a momentum-space procedure to compute BSF cross-sections, that is based on methods developed originally for few-body problems, by the nuclear-physics community. We have adjusted codes that were written for low-energy proton-proton collisions \cite{deVries:2013fxa} and deuteron formation via neutron capture on a proton target~\cite{deVries:2015pza}, in order to calculate the BSF cross sections discussed in the rest of this paper. 
Below, we outline the procedure for the formation of scalar DM bound states via a vector mediator. More details can be found in Ref.~\cite{Glockle,Epelbaum:2004fk, deVries:2013fxa}. All BSF cross sections obtained in this work have been checked by both the coordinate- and momentum-space routine. We point out that the notation and some conventions in this appendix are somewhat disjoint from those in the main text, and some symbols are used here for different purposes. 

\subsection{Solution of the Lippmann-Schwinger and bound-state equation}

The starting point is the non-relativistic Lippmann-Schwinger (LS) equation which in its general form is written as
\begin{equation}
T^{l^\prime l\, s^\prime s}_j (p^\prime, p, E) 
= V^{l^\prime l\, s^\prime s}_j (p^\prime, p) +\sum_{l^{\prime \prime} \, s^{\prime \prime}} 
\int_0^\infty dp^{\prime \prime} \, V^{l^\prime l^{\prime \prime}\, s^\prime s^{\prime
    \prime}}_j (p^\prime, p^{\prime \prime})\left(\frac{p^{\prime \prime\,2}}
{E- p^{\prime \prime\,2}/(2\mu) + i\epsilon}\right) T^{l^{\prime \prime} l\, 
s^{\prime \prime} s}_j (p^{\prime \prime}, p, E)~,
\nonumber
\end{equation}
where $E$ is the center-of-mass energy, $p=\abs{\vec p}$ and $p'=\abs{\vec p'}$ are the relative momenta of the incoming and outgoing DM particles in the center-of-mass frame, and $T^{l^\prime l\, s^\prime s}_j$ denotes the $T$-matrix (scattering matrix) element  corresponding to conserved total angular momentum $j$ for states with initial and final orbital angular momentum (spin) $l$ ($s)$ and $l^\prime$ ($s^\prime)$. $V^{l^\prime l\, s^\prime s}_j (p^\prime, p)$ denotes a partial-wave-decomposition of the DM potential. For scalar DM we remove the spin indices and use $j=l=l'$, such that 
\begin{equation}\label{LS}
T^{l}(p^\prime, p, E) = V^{l}
(p^\prime, p)+ \int_0^\infty 
dp^{\prime \prime}\, V^{l}(p^\prime, p^{\prime \prime})\left(\frac{p^{\prime \prime\,2}}
{E- p^{\prime \prime\,2}/(2\mu) + i\epsilon}\right) T^{l}(p^{\prime \prime}, p, E)~.
\end{equation}

The potential for a vector mediator in momentum space is given by
\begin{equation}
V(\vec p, \vec p')
= - \frac{4\pi \alpha}{(\vec p - \vec p')^2+m_\varphi^2} \, .
\label{onepion}
\end{equation}
The partial-wave-decomposed potential that appears in \cref{LS} is defined as
\begin{eqnarray}\label{pwd}
V^{l}(p^\prime, p) &=& \frac{1}{(2\pi)^3}\langle p^\prime\, l || V(\vec p, \vec p') || p\,l\rangle\nonumber\\
&=& - \frac{4\pi \alpha}{(2\pi)^2} \int_{-1}^{+1} dx\,P_l(x) \frac{1}{p^2+p^{\prime\,2}-2 p p^\prime x + m_\varphi^2}\ ,
\end{eqnarray}
where $P_l(x)$ denotes the Legendre polynomials.

To numerically solve the LS equation, we need to deal with the $i\epsilon$ in the numerator of \cref{LS}. We write
\begin{equation}
\frac{1}{E- p^{\prime \prime\,2}/(2\mu) + i\epsilon} 
= \frac{2\mu}{q_0^2- p^{\prime \prime\,2} + i\epsilon} = 
\frac{2\mu}{q_0 + p^{\prime \prime}} \left(\frac{\mathcal P}{q_0 - p^{\prime
      \prime}} - i\pi \delta(q_0 -p^{\prime \prime})\right)~,
\end{equation}
where $\mathcal P$ denotes  the principal value integral, and we introduced $E \equiv q_0^2/(2\mu)$. The LS equation can then be written a
\begin{eqnarray}
T^l (p^\prime, p, E) &=& V^l (p^\prime, p) +(2\mu)\mathcal P\!\!
\int_0^{p_{\mathrm{max}}} 
dp^{\prime \prime}\, V^l(p^\prime, p^{\prime \prime})\left(\frac{p^{\prime \prime\,2}}{q_0^2
- p^{\prime \prime\,2} }\right)T^l (p^{\prime \prime}, p, E)\nonumber\\
&& - i\frac{\pi (2\mu) q_0}{2}  
V^l(p^\prime, q_0)
T^l(q_0, p, E)~,
\end{eqnarray}
where we introduced $p_{\mathrm{max}}$ which corresponds to the maximum momentum of the momentum grid that is applied in the actual numerical solution. The main problem is the divergence at $p'' = q_0$ which we therefore subtract and add to get
\begin{eqnarray}
T^l(p^\prime, p, E) 
&=& V^l(p^\prime, p) +(2\mu) 
\int_0^{p_{\mathrm{max}}} dp^{\prime \prime} \, 
\bigg[V^l(p^\prime, p^{\prime \prime})
\left(\frac{p^{\prime \prime\,2}}{q_0^2- p^{\prime \prime\,2} }\right) T^l
(p^{\prime \prime}, p, E)\nonumber\\
&& - V^l
(p^\prime, q_0)\left(\frac{q_0^2}{q_0^2- p^{\prime \prime\,2} }\right) 
T^l(q_0, p, E)\bigg]
\nonumber\\
&&+ (2\mu)  q_0^2 V^l
(p^\prime, q_0) T^l (q_0, p, E) 
\mathcal P\!\!  \int_0^{p_{\mathrm{max}}} dp^{\prime \prime} \frac{1}{q_0^2- p^{\prime \prime\,2} }
\nonumber\\
&& - i\frac{\pi (2\mu)  q_0}{2}\sum_{l^{\prime \prime}\,s^{\prime \prime}}  
V^l(p^\prime, q_0)
T^l(q_0, p, E)~,
\end{eqnarray}
such that  the first integral is no longer singular. The second integral can be done analytically and does not depend on the form of the potential
\begin{equation}
\mathcal P\!\!  \int_0^{p_{\mathrm{max}}} dp^{\prime \prime} 
\frac{1}{q_0^2- p^{\prime \prime\,2} } = \frac{1}{2q_0} 
\ln\left(\frac{ p_{\mathrm{max}} + q_0}{ p_{\mathrm{max}}-q_0}\right)~.
\end{equation}
The LS equation can now be discretized on a momentum grid and written as a complex eigenvalue equation which we solve using the LAPACK library \cite{LAPACK}. 

To obtain the momentum-space bound-state wave function we solve the homogeneous part of the LS equation
\begin{equation}\label{BS}
\Psi_{nlm}(p) = \frac{1}{E^b_{nlm} - \frac{p^2}{(2\mu)}} \int dp^\prime\,p^{\prime\, 2} V^l(p,p') \Psi_{nlm}(p')\, ,
\end{equation}
where $E^b_{nlm}$ is the (negative) binding energy and the real wave function is normalized as
\begin{equation}
\int dp \,p^{2}\Psi_{nlm}(p)^2 =1\, . 
\end{equation}
\Cref{BS} can be immediately discretized and written as an eigenvalue equation which we again solve with a LAPACK routine \cite{LAPACK}. The binding energy is varied until we find a consistent solution. 

Although for simplicity we discussed only scalar DM and vector mediator, the above routines can be easily extended to solve coupled-channel LS and bound-state equations.

\subsection{The BSF cross sections}

The goal is to obtain the amplitude for the BSF process $X_1 + X_2 \rightarrow B_{nlm}(X_1 X_2) + \varphi$. We write this amplitude as
\begin{equation}
\label{amplitude}
A^{nlm,\lambda}(\vec p_{in},\vec P_\varphi) = \langle  \Psi_{nlm}\,|\, O^\lambda(\vec P_\varphi)\,|\, \vec p_{in}\,\rangle^{(+)},
\end{equation}
where $\lambda$ denotes the polarization of the vector mediator and $\vec p_{in}$ $(p_{in} = | \vec p_{in}| \simeq \mu v_{rel} $) and $|\vec P_\varphi|=P_\varphi$ are, respectively, the incoming relative momentum of the DM pair in the c.o.m. frame and the outgoing mediator momentum. $O^\lambda$ describes the current. The ${}^{(+)}$ superscript on the incoming state implies that this is the fully scattered state
obtained from applying the $T$-matrix to a free state.

We insert a complete set of states $1=\sum_{l=0}^{\infty} \sum_{m=-l}^{l} \int dp\,p^2\,|p\,lm\rangle \langle  p\, lm | $ and use
\begin{equation} 
| \vec p_{in}\rangle = \sum_{l'=0}^{\infty} \sum_{m'=-l'}^{l'}  Y^*_{l' m'}(\hat p_{in}) |p_{in}\, l' m' \rangle,
\end{equation}
to write
\begin{equation}
A^{nlm, \lambda}(\vec p_{in},\vec P_\varphi) = A^{nlm, \lambda} _{\mathrm{free}}(\vec p_{in},\vec P_\varphi) + A_{\mathrm{scat}}^{nlm, \lambda}(\vec p_{in},\vec P_\varphi),
\end{equation}
where
\begin{eqnarray}
\label{Afree_0}
A^{nlm, \lambda}_{\mathrm{free}}(\vec p_{in},\vec P_\varphi) &=& \sum_{l'=0}^{\infty} \sum_{m'=-l'}^{l'}\langle  \Psi_{nlm}\,|\, O^\lambda(\vec P_\varphi) \,|\, p^{in}\, l'm_l' \rangle Y^*_{l' m'}(\hat p^{in}) \,,
\nonumber\\
A^{nlm, \lambda}_{\mathrm{scat}}(\vec p_{in},\vec P_\varphi) &=& \sum_{l'=0}^{\infty} \sum_{m'=-l'}^{l'} \int dp\,p^2 \langle \Psi_{nlm} \,|\, O^\lambda(\vec P_\varphi)\,|\, p\,l'm_l'\rangle Y^*_{l^\prime m'}(\hat p_{in}) \frac{2\mu}{p_{in}^{2} - p^{2}+i\epsilon} T^{l'}(p,p_{in})\ ,\nonumber
\end{eqnarray}
where $T$ denotes the $T$-matrix obtained above. The integral appearing in $A^{nlm, \lambda}_{\mathrm{scat}}$ is numerically solved in the same way as was done for the $T$-matrix by adding and subtracting the divergence at $p=p_{in}$.

The next step requires the calculation of $O_{nlm}^{\lambda l'm'}(p, \vec P_\varphi)\equiv \langle \Psi_{nlm} \,|\, O^\lambda(\vec P_\varphi)\,|\, p\,l'm' \rangle$. For simplicity we investigate the current for two DM scalars with equal mass and opposite charge $c_1 = -c_2 =c$ such that $O^\lambda = cg / (4 \mu)\,(P + P')^\lambda \tau_3$ where $P$ and $P'$ are, respectively, the in- and outgoing momentum of the scalar interacting with the outgoing mediator. We have introduced an 'isospin' operator, $\tau_3 = \mathrm{diag}(1,\,-1)$, that indicates that the two DM scalars carry opposite charge. Of course, different charge configurations can be considered as well. At the same time the bound- and scattering state have been assigned\footnote{We have not written these isospin factors in the potential in \cref{pwd} for simplicity. We could have replaced $-4\pi \alpha \rightarrow (4\pi \alpha) \tau^{(1)}_3\tau^{(2)}_3$ where $\tau^{(i)}_3$ indicates the isospin of DM particle $(i)$. As $\langle t\ m_t =0| \tau^{(1)}_3\tau_3^{(2)}| t' \ m_t'=0\rangle = -\delta^{t t'}$, we obtain the same scattering equations for $t=t'=0$ and $t=t'=1$ and we therefore dropped the isospin indices.} a total isospin ($t$ and $t'$) and third component of total isospin ($m_t$ and $m_t'$) which are useful bookkeeping devices.  For scalars with opposite charge we have $m_t=m_t'$=0.

Using momentum conservation and the fact that the currents under consideration only couple to one of the DM particles (two-body current appear at higher order and can be included along the same lines) we obtain
\begin{equation}
\label{O1}
O_{nlm}^{\lambda l'm'}(p, \vec P_\varphi) 
= \int d\Omega(\hat p)  \Psi_{nlm} (k) Y^\star_{l m}(\hat k) \, Y_{l' m'}(\hat p)* 
\frac{cg}{4\mu}(2k^\lambda)*2\langle t m_t| \tau_3| t' m_t'\rangle
\end{equation}
where $\vec k = \vec p - \vec P_\varphi/2$ ($k= | \vec k |$ and $\hat k = \vec k/k$). As each scalar carries isospin $1/2$ the combination of 2 scalars gives total isospin $t={0,1}$ and $m_t = {0,\pm 1}$. For the case at hand, we have $m_t =0$ such that the total isospin can be both $t=0$ or $t=1$. The Pauli principle requires a symmetric wave function for two scalars such that  $t=0$ ($t=1$) implies odd (even) orbital angular momentum. Since $\langle t\ m_t=0| \tau_3| t' \ m_t'=0\rangle = (1-\delta^{t t'})$, we see that the total isospin flip requires $|l-l'|$ to be odd.\footnote{For identical scalars, however, we have $m_t = \pm 1$ so that only $t=1$ is allowed. This implies that $|l-l'|$ must be even, as discussed in \cref{sec:ScalarMed_Identical}.} 

We can either solve \cref{O1} numerically or, as we do here, perform the angular integral analytically. Using standard angular momentum techniques we can write
\begin{eqnarray}
\label{O1PWD}
O_{nlm}^{\lambda l'm'}(p, \vec P_\varphi) 
&=&  
\frac{cg}{\mu}(1-\delta^{t t'})\sqrt{\pi}(-1)^{m+m'} \sqrt{\hat l} \sum_{G=0}^{\infty} \sum_{m_G=-G}^{G} (l 1 G;000)( l 1 G; -m\, \lambda\, m_G)
\nonumber \\
&&\times  \sum_{\lambda_1=0}^G\sum_{\lambda_2=0}^G\delta^{\lambda_1+\lambda_2,G} \sqrt{\hat\lambda_1 \hat \lambda_2}\sqrt{\frac{\hat G !}{\hat \lambda_1! \hat \lambda_2!}}\,( p )^{\lambda_1} (-P_\varphi/2)^{\lambda_2}\, \sum_{f=0}^{\infty} (-1)^f \hat f^{3/2} \,g_{fG}^{nlm}(p,P_\varphi)
\nonumber \\
&&\times \sum_{g=0}^{\infty} \ \sum_{m_g=-g}^{m_g}  
\bma f &f&0 \\ \lambda_1 & \lambda_2 & G \\ l'&g&G \ema  
\left( f \lambda_1 l' ; 000 \right) \left( f \lambda_2 g ; 000 \right)
\nonumber\\
&&\times \left(l' g G ; -m' \,m_g\,m_{G}\right)\,Y_{g m_g}(\hat P_\varphi)\ , 
\end{eqnarray}
in terms of Clebsch-Gordan coefficients $(j_1\,j_2\,j;m_1\,m_2\,m_1+m_2)$, we introduced a nine-J symbol, and  $\hat l = 2l+1$. The function $g_{fG}^{nlm}(p, P_\varphi)$ denotes a single numerical angular integral
\begin{equation}
\label{gkG}
g_{fG}^{nlm}(p,P_\varphi) \equiv \int_{-1}^1 dx^\prime  
\frac{|k(x^\prime)|}{|k(x^\prime)|^{G}} \Psi_{nlm}(k(x^\prime))\,P_f(x^\prime)\ ,
\end{equation}
where $k(x^\prime) = p^2 +P_\varphi^2/4 - p P_\varphi x'$.

We are now in the position to evaluate the scattering amplitude in \cref{amplitude}. Although the obtained expressions are valid in any coordinate frame, it is convenient to specify $\vec P_\varphi =P_\varphi \hat z$. In this frame, the differential BSF cross section is given by
\begin{equation}
\frac{d\sigma^{nlm}_{\mathrm{BSF}}}{d\Omega}=  \frac{\pi \mu P_\varphi}{p_{in}}\left( \sum_{\lambda =\pm 1}|A^{nlm, \lambda}(\vec p_{in},\vec P_\varphi =P_\varphi \hat z)|^2 + \frac{m_{\varphi}^2 }{P_\varphi^2 +m_{\varphi}^2} |A^{nlm, \lambda=0}(\vec p_{in},\vec P_\varphi =P_\varphi \hat z)|^2\right)\, ,
\end{equation}
such that only the transverse polarizations contribute in the Coulomb limit.

\clearpage
\def\bibfont{\small}
\bibliography{Bibliography.bib}

\end{document}